\documentclass[a4paper,11pt]{article}
\pdfoutput=1 % if your are submitting a pdflatex (i.e. if you have
\usepackage{jheppub} % for details on the use of the package, please
\usepackage[utf8]{inputenc}
\usepackage[T1]{fontenc} % if needed
\usepackage{booktabs} % if needed
\usepackage{slashed}
\usepackage{placeins}
\usepackage{float}
\usepackage{changes}
\usepackage{tikz}
\usetikzlibrary{arrows}
\usetikzlibrary{shapes,arrows}
\usetikzlibrary{positioning}
\usetikzlibrary{fit}
\usetikzlibrary{shadows}

\usepackage{amsmath,amssymb,amsfonts}	% DON'T use cite, screws with bibtex + hyperref
\usepackage{color}
\usepackage{amsmath}
\usepackage{amsfonts}
\usepackage{amssymb}
\usepackage{graphicx}
\usepackage{slashed}            % for slashed characters in math mode
\usepackage{bbm}                % for \mathbbm{1} (unit matrix)
\usepackage{xspace}				% For spacing after commands
\usepackage{tikz}
\usetikzlibrary{arrows,shapes}
\usetikzlibrary{trees}
\usetikzlibrary{matrix,arrows} 				% For commutative diagram
\usetikzlibrary{positioning}				% For "above of=" commands
\usetikzlibrary{calc,through}				% For coordinates
\usetikzlibrary{decorations.pathreplacing}  % For curly braces
\usepackage{pgffor}							% For repeating patterns
\usetikzlibrary{decorations.pathmorphing}	% For Feynman Diagrams
\usetikzlibrary{decorations.markings}
\tikzset{
    vector/.style={decorate, decoration={snake}, draw},
	provector/.style={decorate, decoration={snake,amplitude=2.5pt}, draw},
	antivector/.style={decorate, decoration={snake,amplitude=-2.5pt}, draw},
    fermion/.style={draw=black, postaction={decorate},
        decoration={markings,mark=at position .55 with {\arrow[draw=black]{>}}}},   
    fermionbar/.style={draw=black, postaction={decorate},
        decoration={markings,mark=at position .55 with {\arrow[draw=black]{<}}}},
    fermionnoarrow/.style={draw=black},
    gluon/.style={decorate, draw=black,
        decoration={coil,aspect=0.5,amplitude=4pt, segment length=4pt}},
    scalar/.style={dashed,draw=black, postaction={decorate},
        decoration={markings,mark=at position .55 with {\arrow[draw=black]{>}}}},
    scalarbar/.style={dashed,draw=black, postaction={decorate},
        decoration={markings,mark=at position .55 with {\arrow[draw=black]{<}}}},
    scalarnoarrow/.style={dashed,draw=black},
    electron/.style={draw=black, postaction={decorate},
        decoration={markings,mark=at position .55 with {\arrow[draw=black]{>}}}},
	bigvector/.style={decorate, decoration={snake,amplitude=4pt}, draw},
	ghost/.style={dotted,draw=black, postaction={decorate},
        decoration={markings,mark=at position .55 with {\arrow[draw=black]{>}}}},
	ghostbar/.style={dotted,draw=black, postaction={decorate},
        decoration={markings,mark=at position .55 with {\arrow[draw=black]{<}}}},
         gluonloop/.style={decorate, draw=black,
        decoration={coil,aspect=0.6,amplitude=3pt, segment length=4pt}},   
}
\usepackage[]{units}

\usepackage{pdflscape}
\usepackage{./gnuplot-lua-tikz}
\usepackage{subfig}
\usepackage[tight-spacing=true]{siunitx}
\newcommand{\myparagraph}[1]{\paragraph{#1}\mbox{}\\}
\newcommand*\circled[1]{\tikz[baseline=(char.base)]{
            \node[shape=circle,draw,inner sep=2pt] (char) {#1};}}          
\newcommand{\MG}{\texttt{MG5aMC@NLO}}
\title{Squark production in R-symmetric SUSY with Dirac gluinos: NLO corrections}

\author[b,1]{Philip Diessner,\note{Corresponding author.}}
\author[a,c]{Wojciech Kotlarski,}
\author[a]{Sebastian Liebschner,}
\author[a]{and Dominik St\"ockinger}

\affiliation[a]{Institut f\"ur Kern- und Teilchenphysik, TU Dresden\\
 01069 Dresden, Germany}
\affiliation[b]{Deutsches Elektronen-Synchrotron DESY, Hamburg, Germany}
\affiliation[c]{Faculty of Physics, University of Warsaw,\\ Pasteura 5, 02093 Warsaw, Poland}

\emailAdd{philip.diessner@desy.de}
\emailAdd{wojciech.kotlarski@fuw.edu.pl}
\emailAdd{sebastian.liebschner@tu-dresden.de}
\emailAdd{dominik.stoeckinger@tu-dresden.de}
\preprint{DESY 17-100}
\abstract{
R-symmetry leads to a distinct realisation of SUSY with a significantly modified coloured sector featuring a Dirac gluino and a scalar colour octet (sgluon).
We present the impact of R-symmetry on squark production at the 13 TeV LHC.
We study the total cross sections and their NLO corrections from all strongly interacting states, their dependence on the Dirac gluino mass and sgluon mass as well as their systematics for selected benchmark points.
We find that tree-level cross sections in the R-symmetric model are reduced compared to the MSSM but the NLO K-factors are generally larger in the order of ten to twenty per cent.
In the course of this work we derive the required DREG $\to$ DRED transition counterterms and necessary on-shell renormalisation constants.
The real corrections are treated using FKS subtraction, with results cross checked against an independent calculation employing the two cut phase space slicing method.
}

\begin{document} 
\maketitle
\flushbottom

\section{Introduction}

The Minimal Supersymmetric Standard Model (MSSM) is one of the most studied extensions of 
the SM. Often, an unbroken R-parity is assumed which implies that 
supersymmetric particles can only be produced in pairs. In this paper we 
consider a distinct realisation of supersymmetry (SUSY) based on an unbroken, continuous R-symmetry.
The basic feature of R-symmetry is that particles and superpartners have different R-charges where the differences are unambiguously prescribed by the SUSY algebra.

For definiteness we focus on the Minimal R-symmetric Supersymmetric Standard Model  (MRSSM) \cite{Kribs:2007ac} but our discussion will apply also to more general R-symmetric models.

From the phenomenological point of view, the MRSSM is an appealing model, with immediate restrictions following from R-symmetry.
The model contains Dirac instead of Majorana gauginos and adjoint scalar superpartners for all gauge fields.
This also leads to significant changes in the Higgs sector due to the presence of additional singlet and triplet scalars.
The $\mu$- and $A$-terms of the MSSM are forbidden; mixing between left and right handed squarks or sleptons is forbidden, and large contributions to flavour and CP violating observables are suppressed \cite{Kribs:2007ac,Dudas:2013gga}. 

In a recent series of papers, the electroweak sector has been investigated
and it has been shown that the MRSSM can accommodate the experimentally measured 
$W$ and Higgs boson mass as well as the electroweak precision variables and is compatible with direct detection searches for dark matter and LHC searches of electroweak particles \cite{Diessner:2014ksa,Diessner:2015yna}. 
For related studies, see \cite{Braathen:2016mmb,Nakano:2015gws,Ding:2015epa}.

In this paper we focus on the strongly interacting sector of R-symmetric SUSY. 
It is characterised by Dirac gluinos, scalar gluons and left and right handed squarks which are mass eigenstates and have opposite R-charges. 
The phenomenology of this SUSY QCD sector of the MRSSM has already been studied at  the tree level \cite{Heikinheimo:2011fk,Kribs:2013oda}, where it was shown that conservation of R-charge is 
responsible for a rather drastic suppression of the inclusive squark production cross 
section leading to lower bounds on squark masses. This suppression is even 
more amplified when the gluino mass is increased.\footnote{Interestingly, Dirac 
gauginos can be heavier than Majorana gauginos without being less natural in 
view of the hierarchy problem \cite{Hardy:2013ywa}.}
Corrections to this 
observable at the next-to-leading order (NLO) have only been approximately estimated in ref.~\cite{Kribs:2012gx} by using global MSSM instead of MRSSM K-factors.
% using \texttt{Prospino}~\cite{Beenakker:1996ed}. 
It is the purpose of this paper to describe an exact NLO SUSY-QCD calculation in the MRSSM
at the example of squark-squark ($\tilde{q}\tilde{q}$) and squark-antisquark ($\tilde{q}\tilde{q}^\dagger$) production at the LHC. We will 
expose and explain differences to the analogous calculation in the MSSM.

The paper is structured as follows. 
The next section describes the strongly interacting sector of the MRSSM.
In section 3 we evaluate the leading order cross sections for the production of colour charged MRSSM particles.
Section 4 describes the evaluation and renormalisation of the virtual amplitudes. 
Our results have been obtained with two independent codes which use different regularisation schemes \cite{Signer:2008va}.
We provide a list of counterterms including the transition counterterm from dimensional regularisation to dimensional reduction.
Section 5 is devoted to the description of real corrections.
We present two alternative ways of dealing with infrared singularities used in this work: the two cut phase space slicing method and the FKS subtraction.
We also discuss the treatment of the on-shell resonances.
In section 6 we proceed with a phenomenological analysis.
We give a detailed comparison of K-factors in the MSSM and the MRSSM, explaining the physical origins of their differences.
Then we give an overview of the NLO corrected cross sections in different regions of parameter space, including an uncertainty discussion. 
Section 7 contains our conclusions, and the appendix collects a list of Feynman rules, implementation details, and numerical results for verification purposes.

\section{Details of the model}\label{sec:model}

The field content %of the SUSY QCD sector 
of the MRSSM is enlarged in comparison to the one of the MSSM. The necessity of 
this arises from R-symmetry. As can be seen from table~\ref{tab:fieldcontent} the gluino $\tilde g_L$, as superpartner of 
the gluon,
has R-charge $+1$. In order to account for a non-vanishing Dirac gluino mass it needs to be partnered with a new field
$\tilde g_R$ with R-charge $-1$. 
The Dirac nature of gluinos manifests itself in an $N=2$ supersymmetric gauge sector 
including scalar gluons $O$, called sgluons, which transform in the same representation as the gluon under gauge transformations.
\begin{table}[ht] 
\begin{center}
\begin{tabular}{c cc|cc|c}
\toprule
\multicolumn{2}{c}{superfield} & \multicolumn{2}{c}{boson} & \multicolumn{2}{c}{fermion} \\
\midrule
 left-handed (s)quark &$\widehat{Q}_L$ &  $\tilde{q}_L$ &  $1$ & $q_L$ & $0$\\
 right-handed (s)quark &$\widehat{Q}_R$ &  $\tilde{q}_R^\dagger$ &  $1$ &  $\bar q_R$ &  $0$\\
 gluon vector superfield &$\widehat{V}$ &  $g$ & 0 &  $\tilde{g}_L$ & $+1$\\ 
 adjoint chiral superfield &$\widehat{O}$ & $O$ &  $0$ & $\tilde{g}_R$ & $-1$\\
 \bottomrule
\end{tabular}
\caption{The table shows the strongly interacting field content of the
  MRSSM, together with the R-charges of the component fields. The
  superfield in the last line is absent in the MSSM. It comprises of the right-handed component of the Dirac gluino and two real sgluons.}\label{tab:fieldcontent}
\end{center}
\end{table}
The corresponding Lagrangian of the strongly interacting part of the 
MRSSM, including one massless quark of arbitrary flavour reads
\begin{align}
\mathcal{L}_{\mathrm{RSQCD}} = &\int\mathrm{d}^4\theta\ \left( \widehat{\overline{Q}}_L \mathrm{e}^{2 g_s\widehat{V}_s} \widehat{Q}_L + \widehat{\overline{Q}}_R \mathrm{e}^{-2g_s\widehat{V}^T_s} \widehat{Q}_R + \widehat{\overline{O}} \mathrm{e}^{2 g_s\widehat{V}^{\mathrm{adj.}}_s} \widehat{O}\right)\nonumber\\
+& \left( \int \mathrm{d}^2\theta \frac{1}{16g_s^2} \widehat{W}_s^{a\alpha}\widehat{W}^a_{s\alpha} + h.c. \right) + \mathcal{L}_{\mathrm{soft}}\,.\label{eq:L_super_RSQCD}
\end{align}
Note that the vector superfield of the gluon $\widehat{V}_s$ in the first line of eq.~\eqref{eq:L_super_RSQCD} transforms for each term in the appropriate representation of $SU(3)_C$, i.e. the fundamental, the antifundamental and the adjoint one.\\
The soft breaking Lagrangian accounts for squark, gaugino and sgluon masses. These mass terms arise from a hidden sector spurion. 
For gauginos the D-type spurion is given by $\widehat{W}_\alpha^\prime = \theta_\alpha D$ and mediates 
supersymmetry breaking at the mediation scale $M$: $\int\mathrm{d}\theta^2\frac{\widehat{W}_\alpha^\prime}{M}\widehat{W}_s^\alpha \widehat{O}$. 
After integrating out the spurion one obtains~\cite{Fox:2002bu,Diessner:2014ksa}
\begin{align}
\mathcal{L}_{\mathrm{soft}} =\ & -\frac{m_{\tilde{q}_L}^2}{2} |\tilde{q}_L|^2 - \frac{m_{\tilde{q}_R}^2}{2} |\tilde{q}_R|^2 \nonumber\\
& -m_{O}^2\left|O^{a}\right|^2 - m_{\tilde{g}}\left(\overline{\tilde{g}_R}\tilde{g}_L -\sqrt{2}D^a O^a + \text{h.c.}\right)\;,
\end{align}
where $D^a$ is the usual auxiliary field in the $SU(3)$ sector.
The complex sgluon state has to be decomposed into two real fields:
\begin{align}
O = \frac{O_s + iO_p}{\sqrt{2}}\,.\label{eq:sgluon}
\end{align}
The mass of the CP-even scalar $O_s$ receives an 
additional contribution from the gluino Dirac mass
whereas the mass of the pseudoscalar $O_p$ is solely given by the soft breaking parameter:
\begin{align}
m_{O_s} = \sqrt{m_{O}^2 + 4 m_{\tilde{g}}^2}\;,
\quad m_{O_p} = m_O\,.
\label{eq:octetmasses}
\end{align}
The respective Feynman rules derived from this Lagrangian differ
partly from the ordinary MSSM ones. Those Feynman rules which have no
MSSM counterpart are listed in appendix~\ref{sec:FeynmanRules}. 

To study the relevant SUSY-QCD effects of the MRSSM quantitatively, we define three benchmark points given in table~\ref{tab:bms}.
\begin{table}[ht] 
\begin{center}
\begin{tabular}{lrrrr}
\toprule
& $m_{\tilde{q}}$ & $m_{\tilde{g}}$ & $m_{\tilde{O}_s}$& $m_{\tilde{O}_p}$\\
\midrule
BM1 & 1500 & 1000 & 5385 & 5000\\
BM2 & 1500 & 2000 & 6403 & 5000\\
BM3 & 500 & 2000 & 6403 & 5000\\
\bottomrule
\end{tabular}
\end{center}
\caption{Benchmark points. Assuming unified squark masses and the sgluon masses
apply only to the MSSM. All masses are given in GeV.}
\label{tab:bms}
\end{table}

\section{Squark and gluino production at the leading order}\label{sec:tree-level}
\begin{figure}[!htp]
\begin{center}
\begin{tikzpicture}[line width=1.0 pt, scale=1.3, arrow/.style={thick,->,shorten >=2pt,shorten <=2pt,>=stealth}]
	\draw[fermion] (-1,0.5)--(-0.42,0);
	\draw[fermionbar] (-1,-0.5)--(-0.42,0);
	\node at (-1.2,0.5) {$u$};
	\node at (-1.2,-0.5) {$\overline{u}$};
	\draw[gluon] (-0.42,0)--(0.42,0); 
	\draw[scalar] (0.42,0)--(1,0.5);
	\draw[scalarbar] (0.42,0)--(1,-0.5);
	\node at (1.5,0.5) {$\tilde{u}_L / \tilde{u}_R$};
	\node at (1.5,-0.5) {$\tilde{u}^\dagger_L / \tilde{u}^\dagger_R$};
%	\draw[arrow] (-1,0.7)--(-0.5,0.26);
%	\node at (-0.75,0.8) {$k_a$};
%	\draw[arrow] (-1,-0.7)--(-0.5,-0.26);
%	\node at (-0.75,-0.8) {$k_b$};
%	\draw[arrow] (0.5,0.26)--(1,0.7);
%	\node at (0.75,0.8) {$p_1$};
%	\draw[arrow] (0.5,-0.26)--(1,-0.7);
%	\node at (0.75,-0.8) {$p_2$};
\begin{scope}[shift={(4,0)}]
	\draw[fermion] (-1,0.5)--(0,0.5);
	\draw[fermionbar] (-1,-0.5)--(0,-0.5);
	\draw[fermionnoarrow] (0,0.5)--(0,-0.5);
	\draw[gluon] (0,0.5)--(0,-0.5); 
	\draw[scalar] (0,0.5)--(1,0.5);
	\draw[scalarbar] (0,-0.5)--(1,-0.5);
\end{scope}
\end{tikzpicture}
\end{center}
\begin{center}
\begin{tikzpicture}[line width=1.0 pt, scale=1.3]
    \node at (0,0) {};
\begin{scope}[shift={(0,-1)}]
	\draw[gluon] (-1,0.5)--(-0.42,0);
	\draw[gluon] (-1,-0.5)--(-0.42,0);
	\node at (-1.2,0.5) {$g$};
	\node at (-1.2,-0.5) {$g$};
	\draw[gluon] (-0.42,0)--(0.42,0); 
	\draw[scalar] (0.42,0)--(1,0.5);
	\draw[scalarbar] (0.42,0)--(1,-0.5);
	\node at (1.5,0.5) {$\tilde{u}_L / \tilde{u}_R$};
	\node at (1.5,-0.5) {$\tilde{u}^\dagger_L / \tilde{u}^\dagger_R$};
\begin{scope}[shift={(3.5,0)}]
	\draw[gluon] (-1,0.5)--(0,0.5);
	\draw[gluon] (-1,-0.5)--(0,-0.5);
	\draw[scalarbar] (0,0.5)--(0,-0.5);
	\draw[scalar] (0,0.5)--(1,0.5);
	\draw[scalarbar] (0,-0.5)--(1,-0.5);
\end{scope}
\begin{scope}[shift={(6,0)}]
	\draw[gluon] (-1,0.5)--(0,0.5);
	\draw[gluon] (-1,-0.5)--(0,-0.5);
	\draw[scalar] (0,0.5)--(0,-0.5);
	\draw[scalarnoarrow] (0,0.5)--(0.4,0.1);
	\draw[scalarbar] (0.6,-0.1)--(1,-0.5);
	\draw[scalarnoarrow] (0,-0.5)--(0.5,0);
	\draw[scalar] (0.5,0)--(1,0.5);
\end{scope}
\begin{scope}[shift={(8.5,0)}]
	\draw[gluon] (-1,0.5)--(0,0);
	\draw[gluon] (-1,-0.5)--(0,0);
	\draw[scalarbar] (0,0)--(1,-0.5);
	\draw[scalar] (0,0)--(1,0.5);
\end{scope}
\end{scope}
\end{tikzpicture}
\end{center}
\begin{center}
\begin{tikzpicture}[line width=1.0 pt, scale=1.3]
    \node at (0,0) {};
\begin{scope}[shift={(0,-1)}]
	\draw[fermion] (-1,0.5)--(0,0.5);
	\draw[fermion] (-1,-0.5)--(0,-0.5);
	\node at (-1.2,0.5) {$u$};
	\node at (-1.2,-0.5) {$u$};
	\draw[gluon] (0,0.5)--(0,-0.5);
	\draw[fermionnoarrow] (0,0.5)--(0,-0.5);
	\draw[scalar] (0,0.5)--(1,0.5);
	\draw[scalar] (0,-0.5)--(1,-0.5);;
	\node at (1.3,0.5) {$\tilde{u}_L$};
	\node at (1.3,-0.5) {$\tilde{u}_R$};
\begin{scope}[shift={(3,0)}]
	\draw[fermion] (-1,0.5)--(0,0.5);
	\draw[fermion] (-1,-0.5)--(0,-0.5);
	\draw[gluon] (0,0.5)--(0,-0.5);
	\draw[fermionnoarrow] (0,0.5)--(0,-0.5);
	\draw[scalarnoarrow] (0,0.5)--(0.4,0.1);
	\draw[scalar] (0.6,-0.1)--(1,-0.5);
	\draw[scalarnoarrow] (0,-0.5)--(0.5,0);
	\draw[scalar] (0.5,0)--(1,0.5);
\end{scope}
\end{scope}
\end{tikzpicture}
\end{center}
\begin{center}
\begin{tikzpicture}[line width=1.0 pt, scale=1.3]
    \node at (0,0) {};
\begin{scope}[shift={(0,-1)}]
	\draw[fermion] (-1,0.5)--(-0.42,0);
	\draw[fermionbar] (-1,-0.5)--(-0.42,0);
	\node at (-1.2,0.5) {$u$};
	\node at (-1.2,-0.5) {$\overline{u}$};
	\draw[gluon] (-0.42,0)--(0.42,0); 
	\draw[gluon] (0.42,0)--(1,0.5);
	\draw[fermionnoarrow] (0.42,0)--(1,0.5);
	\draw[gluon] (0.42,0)--(1,-0.5);
	\draw[fermionnoarrow] (0.42,0)--(1,-0.5);
	\node at (1.2,0.5) {$\tilde{g}$};
	\node at (1.2,-0.5) {$\overline{\tilde{g}}$};
\begin{scope}[shift={(3,0)}]
	\draw[fermion] (-1,0.5)--(0,0.5);
	\draw[fermionbar] (-1,-0.5)--(0,-0.5);
	\draw[scalar] (0,0.5)--(0,-0.5);
	\draw[gluon] (0,0.5)--(1,0.5);
	\draw[fermionnoarrow] (0,0.5)--(1,0.5);
	\draw[gluon] (0,-0.5)--(1,-0.5);
	\draw[fermionnoarrow] (0,-0.5)--(1,-0.5);
\end{scope}
\begin{scope}[shift={(6,0)}]
	\draw[fermion] (-1,0.5)--(0,0.5);
	\draw[fermionbar] (-1,-0.5)--(0,-0.5);
	\draw[scalar] (0,0.5)--(0,-0.5);
	\draw[gluon] (0,0.5)--(0.4,0.1);
	\draw[fermionnoarrow] (0,0.5)--(0.4,0.1);
	\draw[gluon] (0.6,-0.1)--(1,-0.5);
	\draw[fermionnoarrow] (0.6,-0.1)--(1,-0.5);
	\draw[gluon] (0,-0.5)--(1,0.5);
	\draw[fermionnoarrow] (0,-0.5)--(1,0.5);
\end{scope}
\end{scope}
\end{tikzpicture}
\end{center}
\begin{center}
\begin{tikzpicture}[line width=1.0 pt, scale=1.3]
    \node at (0,0) {};
\begin{scope}[shift={(0,-1)}]
	\draw[gluon] (-1,0.5)--(-0.42,0);
	\draw[gluon] (-1,-0.5)--(-0.42,0);
	\node at (-1.2,0.5) {$g$};
	\node at (-1.2,-0.5) {$g$};
	\draw[gluon] (-0.42,0)--(0.42,0); 
	\draw[gluon] (0.42,0)--(1,0.5);
	\draw[fermionnoarrow] (0.42,0)--(1,0.5);
	\draw[gluon] (0.42,0)--(1,-0.5);
	\draw[fermionnoarrow] (0.42,0)--(1,-0.5);
	\node at (1.2,0.5) {$\tilde{g}$};
	\node at (1.2,-0.5) {$\overline{\tilde{g}}$};
\begin{scope}[shift={(3,0)}]
	\draw[gluon] (-1,0.5)--(0,0.5);
	\draw[gluon] (-1,-0.5)--(0,-0.5);
	\draw[gluon] (0,0.5)--(0,-0.5);	
	\draw[fermionnoarrow] (0,0.5)--(0,-0.5);
	\draw[gluon] (0,0.5)--(1,0.5);
	\draw[fermionnoarrow] (0,0.5)--(1,0.5);
	\draw[gluon] (0,-0.5)--(1,-0.5);
	\draw[fermionnoarrow] (0,-0.5)--(1,-0.5);
\end{scope}
\begin{scope}[shift={(6,0)}]
	\draw[gluon] (-1,0.5)--(0,0.5);
	\draw[gluon] (-1,-0.5)--(0,-0.5);
	\draw[gluon] (0,0.5)--(0,-0.5);
	\draw[fermionnoarrow] (0,0.5)--(0,-0.5);
	\draw[gluon] (0,0.5)--(0.4,0.1);
	\draw[fermionnoarrow] (0,0.5)--(0.4,0.1);
	\draw[gluon] (0.6,-0.1)--(1,-0.5);
	\draw[fermionnoarrow] (0.6,-0.1)--(1,-0.5);
	\draw[gluon] (0,-0.5)--(1,0.5);
	\draw[fermionnoarrow] (0,-0.5)--(1,0.5);
\end{scope}
\end{scope}
\end{tikzpicture}
\end{center}
\begin{center}
\begin{tikzpicture}[line width=1.0 pt, scale=1.3]
    \node at (0,0) {};
\begin{scope}[shift={(0,-1)}]
	\draw[fermion] (-1,0.5)--(-0.42,0);
	\draw[gluon] (-1,-0.5)--(-0.42,0);
	\node at (-1.2,0.5) {$u$};
	\node at (-1.2,-0.5) {$g$};
	\draw[fermion] (-0.42,0)--(0.42,0); 
	\draw[scalar] (0.42,0)--(1,0.5);
	\draw[gluon] (0.42,0)--(1,-0.5);
	\draw[fermionnoarrow] (0.42,0)--(1,-0.5);
	\node at (1.5,0.5) {$\tilde{u}_R / \tilde{u}_L$};
	\node at (1.5,-0.5) {$\tilde{g} / \overline{\tilde{g}}$};
\begin{scope}[shift={(4,0)}]
	\draw[fermion] (-1,0.5)--(0,0.5);
	\draw[gluon] (-1,-0.5)--(0,-0.5);
	\draw[gluon] (0,0.5)--(0,-0.5);	
	\draw[fermionnoarrow] (0,0.5)--(0,-0.5);
	\draw[scalar] (0,0.5)--(1,0.5);
	\draw[gluon] (0,-0.5)--(1,-0.5);
	\draw[fermionnoarrow] (0,-0.5)--(1,-0.5);
\end{scope}
\begin{scope}[shift={(7,0)}]
	\draw[fermion] (-1,0.5)--(0,0.5);
	\draw[gluon] (-1,-0.5)--(0,-0.5);
	\draw[scalar] (0,0.5)--(0,-0.5);
	\draw[gluon] (0,0.5)--(0.4,0.1);
	\draw[fermionnoarrow] (0,0.5)--(0.4,0.1);
	\draw[gluon] (0.6,-0.1)--(1,-0.5);
	\draw[fermionnoarrow] (0.6,-0.1)--(1,-0.5);
	\draw[gluon] (0.6,-0.1)--(1,-0.5);
	\draw[fermionnoarrow] (0.6,-0.1)--(1,-0.5);
	\draw[scalarnoarrow] (0,-0.5)--(0.5,0);
	\draw[scalar] (0.5,0)--(1,0.5);
\end{scope}
\end{scope}
\end{tikzpicture}
\caption{Tree-level diagrams for squark and gluino production in the MRSSM. For simplicity, only one quark flavour is shown. In the third and last line, also a charge conjugated process exists.
}\label{fig:TreeDiagrams}
\end{center}
\end{figure}
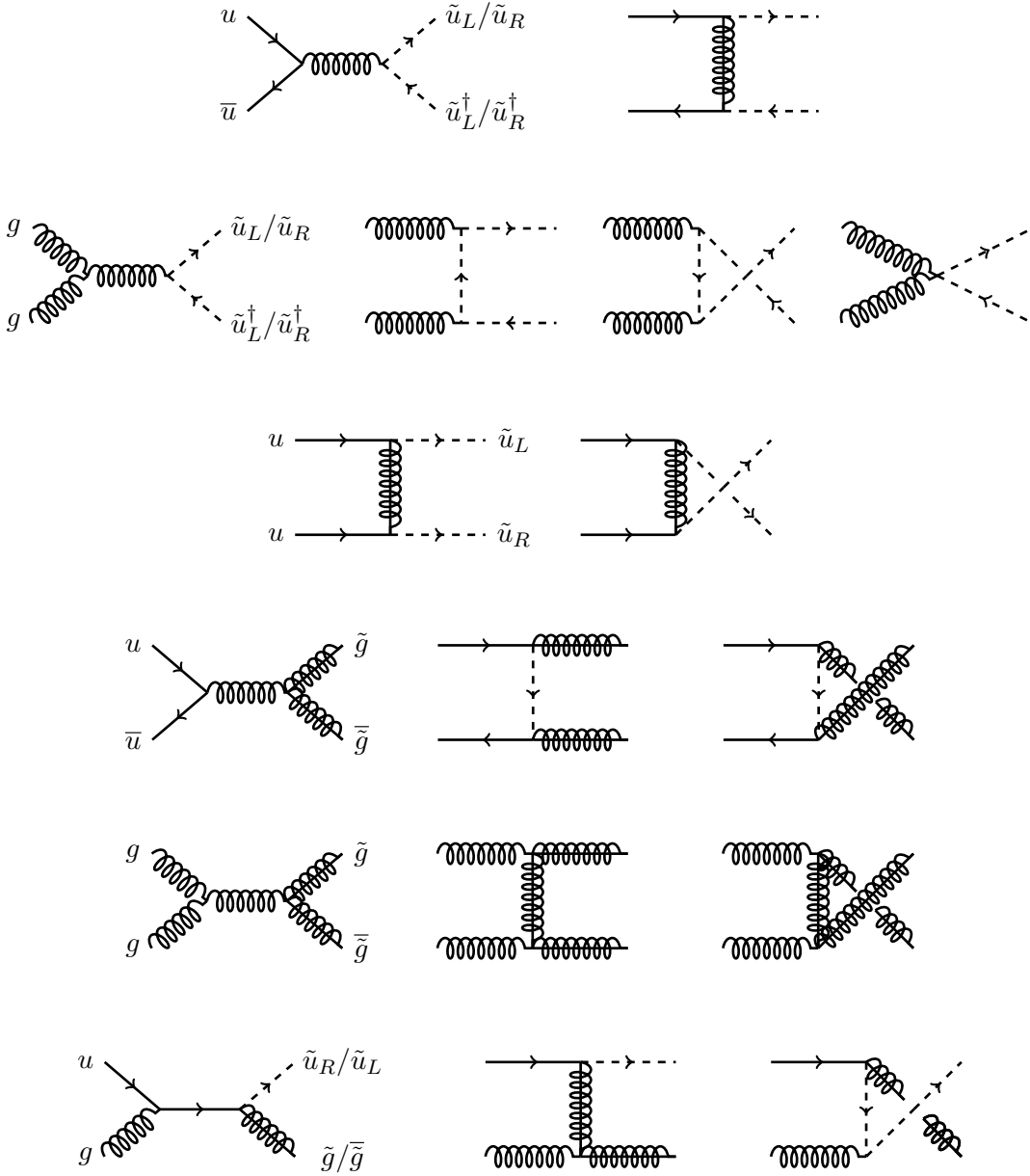

As a starting point for the NLO SUSY-QCD calculation, we recall the tree-level 
production of squarks and gluinos in the MRSSM%
\footnote{For a full tree-level analysis of the production of strongly 
interacting particles, sgluon production should also be
considered. For this we point the reader to refs.~\cite{Choi:2008ub,Plehn:2008ae}.}
and point out differences to the familiar MSSM. 
A detailed study including comparisons of Dirac, Majorana and hybrid gluinos 
can be found in ref.~\cite{Choi:2008pi}. 
As in the MSSM there are six partonic channels contributing to squark and gluino 
production. Three of which for squark-antisquark and squark-squark production
\begin{align}
q_i \overline{q}_j \to \tilde{q}_k\tilde{q}_l^\dagger, \hspace{1cm} 
g g \to \tilde{q}_i\tilde{q}_i^\dagger, \hspace{1cm} 
q_i q_j \to \tilde{q}_i\tilde{q}_j \hspace{0.5cm}.
\end{align} 
and three for gluino-antigluino and squark-(anti)gluino production
\begin{align}
q_i\overline{q}_i \to \tilde{g}\overline{\tilde{g}}, \hspace{1cm} 
g g \to \tilde{g}\overline{\tilde{g}}, \hspace{1cm} 
q_i g \to \tilde{q}_i\tilde{g}\ /\ \tilde{q}_i\overline{\tilde{g}}\hspace{0.5cm}. 
\end{align}
The inclusion of charge conjugated processes is understood if they exist. 
The indices denote quark flavours. The corresponding Feynman diagrams are shown in figure~\ref{fig:TreeDiagrams}.
In the following we give analytic formulae for the partonic cross sections for several channels in the MRSSM and explain if/how they differ to
their respective analogue in the MSSM.
We sum over the $n_f - 1$ first (s)quark flavours (where $n_f \equiv 6$).\footnote{We treat the top squarks separately as they
can only be produced in squark-antisquark pairs and their masses and decay patterns are distinct from other squarks.}
The gauge coupling $g_s$ from the $q q g$-vertex is identical to its supersymmetric analogue $\hat{g}_s$ from the $q\tilde{q}\tilde{g}$-vertex. Taking all squarks as mass degenerate, we obtain the leading order partonic cross sections  
\begin{align}
\hat{\sigma}^B(q_i \overline{q}_j \to \tilde{q}\tilde{q}^\dagger) &= \delta_{ij}  
\frac{g_s^4}{16\pi \hat{s}} (n_f-1) \left[ \frac{4}{27} - \frac{16 m_{\tilde{q}}^2}
{27\hat{s}} \right]\beta_{\tilde{q}}\nonumber\\
&+ \delta_{ij} \frac{g_s^2\hat{g}_s^2}{16\pi \hat{s}}  \left[ \left( \frac{4}{27} + 
\frac{8 m_-^2}{27 \hat{s}} \right)\beta_{\tilde{q}}  + \left( 
\frac{8m_{\tilde{g}}^2}{27\hat{s}} + \frac{8m_-^4}{27\hat{s}^2} \right)L_1 
\right]\nonumber\\
& + \frac{\hat{g}_s^4}{16\pi \hat{s}} \left[ -\frac{8}{9}\beta_{\tilde{q}} + \left( 
-\frac{4}{9} - \frac{8m_-^2}{9\hat{s}} \right)L_1 \right],\label{eq:qqbar_xsec}\\
\hat{\sigma}^B(g g \to \tilde{q}\tilde{q}^\dagger) &= \frac{(n_f-1) g_s^4}{16\pi \hat{s}} 
\left[ \left(\frac{5}{24} + \frac{31 m_{\tilde{q}}^2}
{12\hat{s}}\right)\beta_{\tilde{q}} + \left( \frac{4m_{\tilde{q}}^2}{3\hat{s}} + 
\frac{m_{\tilde{q}}^4}{3\hat{s}^2} \right) \ln \frac{1-\beta_{\tilde{q}}}
{1+\beta_{\tilde{q}}} \right], \label{eq:gg_xsec}\\
\hat{\sigma}^B(q_i q_j \to \tilde{q}\tilde{q}) &= \frac{\hat{g}_s^4}{16\pi \hat{s}} \left[ 
-\frac{8}{9}\beta_{\tilde{q}} +  \left( -\frac{4}{9} - \frac{8m_-^2}{9\hat{s}} 
\right)L_1 \right],\label{eq:qq_xsec}\\
\hat{\sigma}^B(q \overline{q} \to \tilde{g}\overline{\tilde{g}}) &= \frac{g_s^4}
{16\pi \hat{s}} \left[ \frac{16}{9} + \frac{32m_{\tilde{g}}^2}{9\hat{s}} \right] 
\beta_{\tilde{g}}\nonumber\\
& + \frac{\hat{g}_s^2 g_s^2}{16\pi \hat{s}}  \left[ \left( -\frac{4}{3}-\frac{
8m_-^2}{3\hat{s}} \right)\beta_{\tilde{g}} + \left( \frac{8 m_{\tilde{g}}^2}{3\hat{s}} + 
\frac{8m_-^4}{3\hat{s}^2} \right) L_2 \right]\nonumber\\
& + \frac{\hat{g}_s^4}{16\pi \hat{s}} \left[ \left( \frac{32}{27} + \frac{32 m_-^4}
{m_-^4 + m_{\tilde{q}}^2\hat{s}} \right)\beta_{\tilde{g}} - \frac{64m_-^2}{27\hat{s}}L_2 
\right],\label{eq:qqbar_to_sgsgbar}\\
\hat{\sigma}^B(g g \to \tilde{g}\overline{\tilde{g}}) &= \frac{g_s^4}{16\pi \hat{s}} \left[ 
\left( -6 - \frac{51 m_{\tilde{g}}^2}{2\hat{s}} \right)\beta_{\tilde{g}} + \left( 
-\frac{9}{2} - \frac{18 m_{\tilde{g}}^2}{\hat{s}} + \frac{18 m_{\tilde{g}}^4}{\hat{s}^2} 
\right)\ln \frac{1-\beta_{\tilde{g}}}{1+\beta_{\tilde{g}}} \right],\\
\hat{\sigma}^B(q g \to \tilde{q} \tilde{g}) &= \frac{g_s^2\hat{g}_s^2}{16\pi \hat{s}} \left[ 
\frac{\kappa}{\hat{s}}\left( -\frac{7}{9} - \frac{32 m_{-}^2}{9\hat{s}} \right) + \left( 
-\frac{8m_-^2}{9\hat{s}} + \frac{2m_{\tilde{q}}^2m_-^2}{\hat{s}^2} + \frac{8 m_-^4}{9\hat{s}^2} 
\right)L_3\right.\nonumber\\
&+\left. \left( -1-\frac{2m_-^2}{\hat{s}} + \frac{2m_{\tilde{q}}m_-^2}{\hat{s}^2} 
\right)L_4 \right]\,,
\end{align}
where $\hat{s}$ is the squared partonic centre-of-mass energy and the following abbreviations of ref.~\cite{Beenakker:1996ch} are used 
\begin{align}
&\beta_{\tilde{q}} = \sqrt{1-\frac{4 m_{\tilde{q}}^2}{\hat{s}}}, && \beta_{\tilde{g}} = \sqrt{1-\frac{4 m_{\tilde{g}}^2}{\hat{s}}},\nonumber\\
&m_-^2 = m_{\tilde{g}}^2 - m_{\tilde{q}}^2, && \kappa = \sqrt{(\hat{s}-m_{\tilde{g}}^2-m_{\tilde{q}}^2)^2-4m_{\tilde{g}}^2m_{\tilde{q}}^2},
\nonumber\\
& L_1 = \ln \frac{\hat{s}+2m_-^2 - \hat{s}\beta_{\tilde{q}}}{\hat{s}+2m_-^2 + \hat{s}\beta_{\tilde{q}}}, && L_2= \ln \frac{\hat{s} - 2m_-^2 - \hat{s}\beta_{\tilde{g}}}{\hat{s} - 2m_-^2 + \hat{s}\beta_{\tilde{g}}}, \nonumber\\
& L_3 = \ln \frac{\hat{s} - m_-^2 - \kappa}{\hat{s} - m_-^2 + \kappa}, && L_4= \ln \frac{\hat{s} + m_-^2 - \kappa}{\hat{s} + m_-^2 + \kappa}\;.
\end{align}

In comparison to the MSSM there are two main differences. 
Firstly, an overall restriction stemming from an unbroken R-symmetry is that the 
final state particles' R-charges must sum up to zero.
Hence in the MRSSM, only 
diagrams for $q\overline{q} \to \tilde q_L\tilde q_L^\dagger$, 
$q\overline{q} \to \tilde q_R\tilde q_R^\dagger$  and $qq \to \tilde q_L\tilde q_R$ 
can be drawn as shown in figure~\ref{fig:TreeDiagrams}. 
Processes with a squark-antisquark (squark-squark) pair of different (same) ``chiralities''
in the final state are forbidden by R-symmetry.

For the allowed channels in the MRSSM the chirality projectors lead
to the replacement
\begin{align}
\frac{\slashed{p} + m_{\tilde{g}}}{p^2 - m_{\tilde{g}}^2} \to 
\frac{\slashed{p}}{p^2 - m_{\tilde{g}}^2}\label{eq:propagator}
\end{align}
for the gluino propagator in the first and third line of figure~\ref{fig:TreeDiagrams}.
On the level of the cross section, this manifests in the substitution
\begin{align}
\frac{\hat{g}_s^4}{16\pi^2 \hat s}\left( -\frac{4}{9} - \frac{4m_-^4}
{9(m_{\tilde{g}}^2 \hat s + m_-^4)} \right)\beta_{\tilde{q}} 
\to 
-\frac{\hat{g}_s^4}{16\pi^2 \hat s} \cdot \frac{8}{9} \beta_{\tilde{q}}\label{eq:substitution}
\end{align}
and the vanishing of the term proportional to $\delta_{ij}$ for $\tilde{q}\tilde{q}$ production in ref.~\cite{Beenakker:1996ch},
compared to the MSSM expressions. % and leads to stronger suppression by heavy gluino masses.
Expanding $\hat{\sigma}^B(q_i q_j \to \tilde{q}\tilde{q})$ appropriately for 
large values of $m_{\tilde{g}}$, shows that the leading term in the MSSM is 
proportional to $m_{\tilde{g}}^{-2}$, whereas in the MRSSM it is 
$m_{\tilde{g}}^{-4}$, as expected from eq.~\eqref{eq:propagator}\footnote{Note that the expansion of the left hand side of eq.~\eqref{eq:substitution} gives $\mathcal{O}(m_{\tilde{g}}^0)$ terms which cancel against terms from the expansion of the logarithm. Terms of $\mathcal{O}(m_{\tilde{g}}^{-2})$ are than left as the leading ones.}.

A second feature of the MRSSM is that gluino and antigluino are no longer 
indistinguishable particles, 
which manifests in twice as much degrees of freedom available for the gluino 
in the final state. In contrast to the above mentioned feature, this  
characteristic induces an increase of some cross sections. It strikes very 
clearly in the cross section of $g g \to \tilde{g}\overline{\tilde{g}}$, which 
is doubled in comparison to the MSSM. For the process $q\overline{q} \to 
\tilde{g}\overline{\tilde{g}}$, both features appear: The first line of 
eq.~\eqref{eq:qqbar_to_sgsgbar} is twice as much as its MSSM analogue in 
ref.~\cite{Beenakker:1996ch} but the following two are not. 
This is because those originate from $t$- and $u$-channel diagrams where only 
one instead of two squark ``chiralities'' occur. In addition, the MRSSM 
result misses the $t$- and $u$-channel interference. 
In the last channel, i.e. 
$q g \to \tilde{q}\tilde{g}$, both features are present and cancel exactly. On 
the one hand, R-charge allows only the production of ``right-handed'' squarks 
in association with the gluino. On the other hand, there is a distinct 
antigluino which can be produced with a ``left-handed'' squark. As for $\tilde{q}\tilde{q}$ production, the charge conjugated process exists.
\begin{figure}[!htbp]
\begin{center}
\includegraphics[scale=0.35]{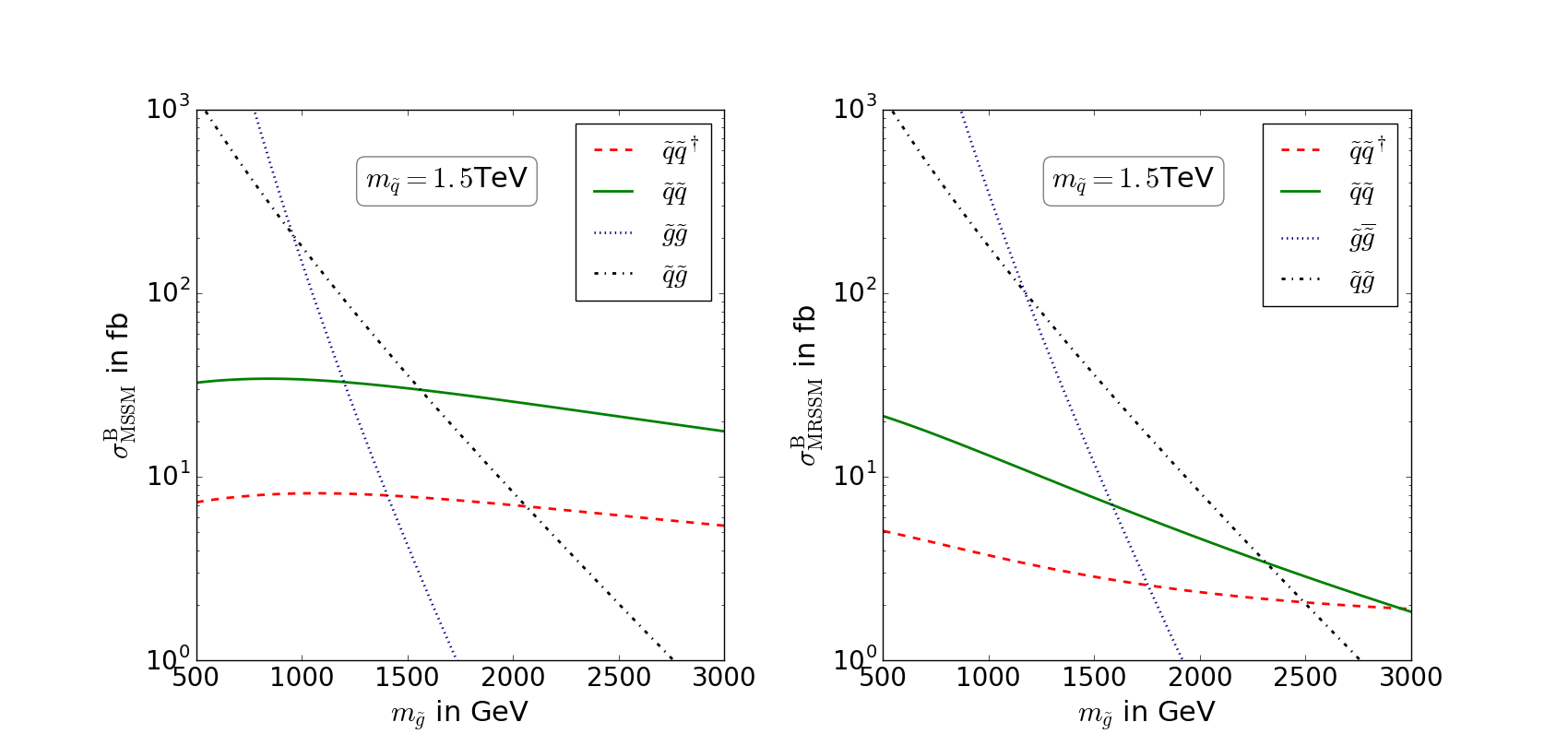}
\includegraphics[scale=0.35]{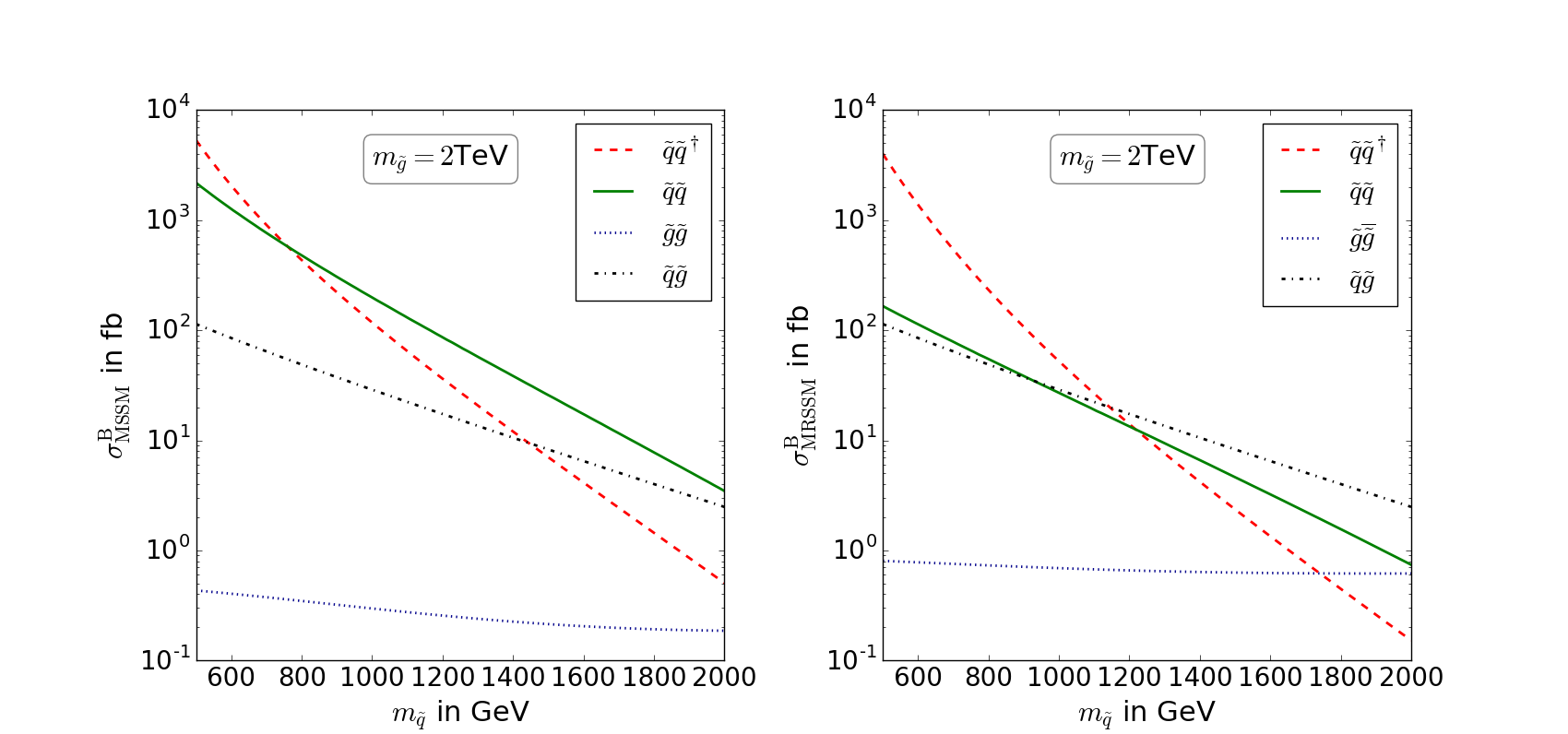}
\caption{The production cross sections for squarks and gluinos at the LHC with 
$\sqrt{S} = \unit[13]{TeV}$ in the MSSM (left) and the MRSSM (right). 
It is summed over five flavours, all possible squark ``chiralities'' and for 
squark-squark and squark-gluino production also the charge conjugated 
processes are taken into account. 
The top row contains results in function of the gluino mass (with squark masses fixed to 1.5 TeV), the bottom one in function of squark masses with the gluino mass fixed to 2 TeV.
The PDF set used is $\mathtt{MMHT2014LO}$~\cite{Harland-Lang:2014zoa}. As renormalisation and factorisation scale $\mu_R = \mu_F = \frac{m_1 + m_2}{2}$ has been chosen, where $m_i$ are the final state's particle masses.
}\label{fig:tree_msq_1}
\end{center}
\end{figure}

In conclusion, we obtain both suppressing as well as amplifying effects in the MRSSM. 
In contrast to the Dirac gluino, the presence of 
sgluons has no effect on the discussed tree-level processes. 
Figure~\ref{fig:tree_msq_1} shows the convolution of the partonic cross sections with parton distribution functions (PDFs) in the MSSM and the MRSSM for the 13 TeV LHC.
For details regarding the parameters used, see the caption of figure~\ref{fig:tree_msq_1}. We sum over all ``chiralities'' 
and five flavours as well as charge conjugated processes, when distinct. 
Due to the prohibition of the above mentioned ``chirality'' states in the MRSSM, we obtain a suppression in the $\tilde{q}\tilde{q}^\dagger$ and $\tilde{q}\tilde{q}$ production (by a factor of 1.3 and 13.2 at BMP2, respectively), 
which increases with the gluino mass. Furthermore, an amplification of gluino production, due to its Dirac nature, is visible.
In combination with experimental results, the absence of detected gluinos 
translates into a larger exclusion limit of the gluino mass than in SUSY-QCD~\cite{Heikinheimo:2011fk}.
We will thus focus on the region of parameter space with a rather large gluino 
mass when discussing our NLO results.

\section{Virtual corrections}\label{sec:virt}
%\marginpar{notation: ``NLO SUSY-QCD'' --- unify!}

The present and the subsequent sections describe the calculation of the
NLO SUSY-QCD corrections to squark production processes in the
MRSSM.\footnote{By NLO SUSY-QCD corrections we always refer to
  next-to-leading order corrections involving the entire coloured
  sector of the respective model. In the MRSSM this includes
  corrections involving Dirac gluinos and sgluons.}
The inclusion of gluino final states we postpone for future
work, since they are not as important in the motivated scenario where
squarks are significantly lighter than gluinos.
Thus, on hadron level we consider squark-squark production
$p p \to \tilde{q}\tilde{q}$ and squark-antisquark production
$p p \to \tilde{q}\tilde{q}^\dagger$. On the partonic level we have to
consider NLO corrections to $q q\to \tilde{q}\tilde{q}$ for the first
process and NLO corrections to $q\overline{q} \to \tilde{q}\tilde{q}^\dagger$ and
$g g \to \tilde{q}\tilde{q}^\dagger$ for the second process.

In the present section we focus on the virtual corrections. 
They involve ultraviolet and soft and collinear
infrared divergences. These divergences are removed after 
renormalisation and combination
with real corrections.

In intermediate steps, the divergences need to be regularised. However,
computations of SUSY-QCD corrections suffer from the fact that no
regularisation is at the same time directly compatible with SUSY and
the standard definition of PDFs, see also ref.~\cite{Signer:2008va}
and ref.~\cite{Gnendiger:2017pys} for a recent review. Hence we have done the
calculation in two complementary ways.
\begin{itemize}
\item The first calculation uses dimensional regularisation
  (version HV in the notation of ref.~\cite{Signer:2008va}) and
  Passarino-Veltman reduction of one-loop integrals. In the HV scheme
  all four-dimensional quantities are treated in
  $D=4-2\epsilon$ dimensions, except that particles outside
  loops are kept unregularised. In this calculation
  SUSY is broken in intermediate steps and SUSY-restoring counterterms
  must be added.
\item The second calculation uses dimensional reduction/the
  four-dimensional helicity scheme (version FDH in the notation of
  ref.~\cite{Signer:2008va}), helicity methods and integrand reduction
  techniques. In the FDH scheme, only space-time and momenta are
  treated in $D$ dimensions, while gluons are kept
  quasi-four-dimensional. Like in HV, only particles inside loops are
  regularised. In this calculation a finite shift is needed in the strong
  coupling renormalisation, and transition rules have to be applied to
  convert infrared divergent amplitudes back to the HV scheme.
\end{itemize}
The two calculations are in full agreement. In the following we
provide the details, first on the general renormalisation scheme, then
on the two calculational procedures.

\subsection{One-loop diagrams and renormalisation}\label{sec:RenConst}
\begin{figure}[ht]
a)\\
\begin{center}
\begin{tikzpicture}[line width=1.0 pt, scale=1.2, arrow/.style={thick,->}]
    \node at (0,0) {};
\begin{scope}[shift={(0,-1)}]
	\node at (-1.3,0) {$\tilde{g}_L$};
	\draw[gluon] (-1,0)--(-0.4,0);
	\draw[fermionnoarrow] (-1,0)--(-0.4,0);
	\node at (1.3,0) {$\tilde{g}_L$};
	\draw[gluon] (0.4,0)--(1,0);
	\draw[fermionnoarrow] (0.4,0)--(1,0);
	\draw[fermion] (-0.4,0) arc (180:360:0.4);
	\draw[scalar] (0.4,0) arc (0:180:0.4);
	\node at (0,0.7) {$\tilde{q}_R$};
	\node at (0,-0.7) {$q$};
\begin{scope}[shift={(3,0)}]
	\draw[gluon] (-1,0)--(-0.4,0);
	\draw[gluon] (0.4,0)--(1,0);
	\draw[scalar] (-0.4,0) arc (180:360:0.4);
	\draw[scalar] (0.4,0) arc (0:180:0.4);
	\node at (0,-0.7) {$\tilde{q}_L/\tilde{q}_R$};
	\node at (0,0.7) {$\tilde{q}_L/\tilde{q}_R$};
\end{scope}
\begin{scope}[shift={(6,0)}]
	\draw[scalarbar] (-1,0)--(-0.4,0);
	\draw[scalarbar] (0.4,0)--(1,0);
	\draw[fermionbar] (-0.4,0) arc (180:0:0.4);
	\draw[fermionnoarrow] (0.4,0) arc (360:180:0.4);
	\draw[gluonloop] (0.4,0) arc (360:180:0.4);
	\node at (0,0.7) {$u$};
	\node at (-1.3,0) {$\tilde{u}$};
	\node at (1.3,0) {$\tilde{u}$};
\end{scope}
\begin{scope}[shift={(9,0)}]
%	\draw[gluon] (-1,0)--(-0.4,0);
%	\draw[fermionnoarrow] (-1,0)--(-0.4,0);
%	\draw[scalar] (0.29,0.29)--(0.707,0.707);
%	\draw[fermionbar] (0.29,-0.29)--(0.707,-0.707);
%	\draw[scalarbar] (0.29,0.29) --(0.29,-0.29);
%	\draw[fermionnoarrow] (0.29,-0.29) -- (-0.4,0);
%	\draw[gluon] (0.29,-0.29) -- (0.29,0.29);
%	\draw[gluon] (0.29,0.29) -- (-0.4,0);
	\node at (-1.3,0) {$\tilde{g}_L$};
	\draw[gluon] (-1,0)--(-0.4,0);
	\draw[fermionnoarrow] (-1,0)--(-0.4,0);
	\draw[scalar] (0.29,0.29)--(0.707,0.707);
	\node at (.85,.85) {$\tilde{u}_L$};
	\node at (.95,-.85) {$\overline{u}$};
	\draw[fermionbar] (0.29,-0.29)-- (0.707,-0.707);
	\draw[scalarbar] (-0.4,0) --node[label=below:$\tilde u_R$]{} (0.29,-0.29);
	\draw[fermionnoarrow] (0.29,-0.29) -- (0.29,0.29);
	\draw[gluon] (0.29,-0.29) -- (0.29,0.29);
	\draw[fermionbar] (0.29,0.29) --node[label=above:$u$]{} (-0.4,0);
\end{scope}
\end{scope}
\end{tikzpicture}
\end{center}
b)\\
\begin{center}
\begin{tikzpicture}[line width=1.0 pt, scale=1.2, arrow/.style={thick,->}]
    \node at (0,0) {};
\begin{scope}[shift={(0,-1)}]
	\node at (-1.3,0) {$\tilde{g}_R$};
	\draw[gluon] (-1,0)--(-0.4,0);
	\draw[fermionnoarrow] (-1,0)--(-0.4,0);
	\draw[gluon] (0.4,0)--(1,0);
	\node at (1.3,0) {$\tilde{g}_R$};
	\draw[fermionnoarrow] (0.4,0)--(1,0);
	\draw[fermionbar] (-0.4,0) arc (180:360:0.4);
	\draw[scalarbar] (0.4,0) arc (0:180:0.4);
	\node at (0,0.7) {$\tilde{q}_R$};
	\node at (0,-0.7) {$q$};
%\begin{scope}[shift={(2.5,0)}]
%
%\end{scope}
%\begin{scope}[shift={(5,0)}]
%
%\end{scope}
\begin{scope}[shift={(9,0)}]
	\node at (-1.3,0) {$\tilde{g}_R$};
	\draw[gluon] (-1,0)--(-0.4,0);
	\draw[fermionnoarrow] (-1,0)--(-0.4,0);
	\draw[scalar] (0.29,0.29)--(0.707,0.707);
	\node at (.85,.85) {$\tilde{u}_L$};
	\node at (.95,-.85) {$\overline{u}$};
	\draw[fermionbar] (0.29,-0.29)--(0.707,-0.707);
	\draw[scalarbar] (-0.4,0) --node[label=below:$\tilde u_L$]{} (0.29,-0.29);
	\draw[fermionnoarrow] (0.29,-0.29) -- (0.29,0.29);
	\draw[gluon] (0.29,-0.29) -- (0.29,0.29);
	\draw[fermionbar] (0.29,0.29) --node[label=above:$u$]{} (-0.4,0);
\end{scope}
\end{scope}
\end{tikzpicture}
\end{center}
c)\\
\begin{center}
\begin{tikzpicture}[line width=1.0 pt, scale=1.2, arrow/.style={thick,->}]
    \node at (0,0) {};
\begin{scope}[shift={(0,-1)}]
	\node at (-1.3,0) {$\tilde{g}_L$};
	\draw[gluon] (-1,0)--(-0.4,0);
	\draw[fermion] (-1,0)--(-0.4,0);
	\draw[gluon] (0.4,0)--(1,0);
	\draw[fermion] (0.4,0)--(1,0);
	\node at (1.3,0) {$\tilde{g}_L$};
	\draw[fermionnoarrow] (-0.4,0) arc (180:0:0.4);
	\draw[gluonloop] (-0.4,0) arc (180:0:0.4);
	\draw[scalarnoarrow] (0.4,0) arc (360:180:0.4);
	\node at (0,-0.7) {$O_s/O_p$};
\begin{scope}[shift={(3,0)}]
	\draw[gluon] (-1,0)--(-0.4,0);
	\draw[gluon] (0.4,0)--(1,0);
	\draw[scalarnoarrow] (-0.4,0) arc (180:0:0.4);
	\draw[scalarnoarrow] (0.4,0) arc (360:180:0.4);
	\node at (0,-0.7) {$O_s/O_p$};
	\node at (0,0.7) {$O_s/O_p$};
\end{scope}
\begin{scope}[shift={(6,0)}]
	\draw[scalarbar] (-1,0)--(-0.4,0);
	\draw[scalarbar] (0.4,0)--(1,0);
	\draw[scalarbar] (-0.4,0) arc (180:0:0.4);
	\draw[scalarnoarrow] (0.4,0) arc (360:180:0.4);
	\node at (0,-0.7) {$O_s$};
	\node at (0,0.7) {$\tilde{u}$};
	\node at (-1.3,0) {$\tilde{u}$};
	\node at (1.3,0) {$\tilde{u}$};
\end{scope}
\begin{scope}[shift={(9,0)}]
	\node at (-1.3,0) {$\tilde{g}_L$};
	\draw[gluon] (-1,0)--(-0.4,0);
	\draw[fermion] (-1,0)--(-0.4,0);
	\draw[gluon] (0.29,-0.29)-- (-0.4,0);
	\draw[fermionnoarrow] (-0.4,0)--(0.29,-0.29);
	\draw[scalar] (0.29,0.29)--(0.707,0.707);
	\node at (.95,.85) {$\tilde{u}_L$};
	\node at (.95,-.85) {$\overline{u}$};
	\draw[fermionbar] (0.29,-0.29)--(0.707,-0.707);
	\draw[scalarnoarrow] (-0.4,0) --node[label=above:$O_s$]{} (0.29,0.29);
	\draw[scalar] (0.29,-0.29) --node[label=right:$\tilde{u}_L$]{} (0.29,0.29);
\end{scope}
\end{scope}
\end{tikzpicture}
\end{center}

%\begin{center}
%\begin{tikzpicture}[line width=1.0 pt, scale=1.2, arrow/.style={thick,->}]
%    \node at (0,0) {};
%\begin{scope}[shift={(0,-1)}]
%	\draw[gluon] (-1,0)--(-0.4,0);
%	\draw[scalarbar] (-0.4,0)--(0.29,-0.29);
%	\draw[scalar] (0.29,0.29)--(0.707,0.707);
%	\draw[scalarbar] (0.29,-0.29)--(0.707,-0.707);
%	\draw[scalar] (-0.4,0) --(0.29,0.29);
%	\draw[scalarnoarrow] (0.29,-0.29) -- (0.29,0.29);
%\begin{scope}[shift={(2.5,0)}]	
%\draw[gluon] (-0.707,0.707)--(-0.4,0);
%\draw[gluon] (-0.707,-0.707)--(-0.4,0);
%\draw[scalar] (-0.4,0)--(0.29,0.29) ;
%\draw[scalarbar] (-0.4,0)--(0.29,-0.29);
%\draw[scalar] (0.29,0.29) -- (0.707,0.707);
%\draw[scalarbar] (0.29,-0.29) -- (0.707,-0.707);
%\draw[scalarnoarrow] (0.29,0.29)--(0.29,-0.29);
%\end{scope}
%\begin{scope}[shift={(5,0)}]
%\draw[fermion] (-0.707,0.707)--(-0.29,0.29);
%\draw[fermion] (-0.707,-0.707)--(-0.29,-0.29);
%\draw[scalar] (0.29,0.29) -- (0.707,0.707);
%\draw[scalar] (0.29,-0.29) -- (0.707,-0.707);
%\draw[scalar] (-0.29,0.29)--(0.29,0.29);
%\draw[scalar] (-0.29,-0.29)--(0.29,-0.29);
%\draw[scalarnoarrow] (0.29,0.29)--(0.29,-0.29);
%\draw[gluon] (-0.29,0.29)--(-0.29,-0.29);
%\draw[fermion] (-0.29,0.29)--(-0.29,-0.29);
%\end{scope}
%\end{scope}
%\end{tikzpicture}
%\end{center}
\caption{Examples of Feynman diagrams relevant for the
calculation of the counterterms. Category a) diagrams are as in the MSSM. 
For comparison category b) contains diagrams which lead to contributions in the MSSM but not the MRSSM.
Category c) are novel diagrams originating from the sgluons.}\label{fig:oneloopcts}
\end{figure}
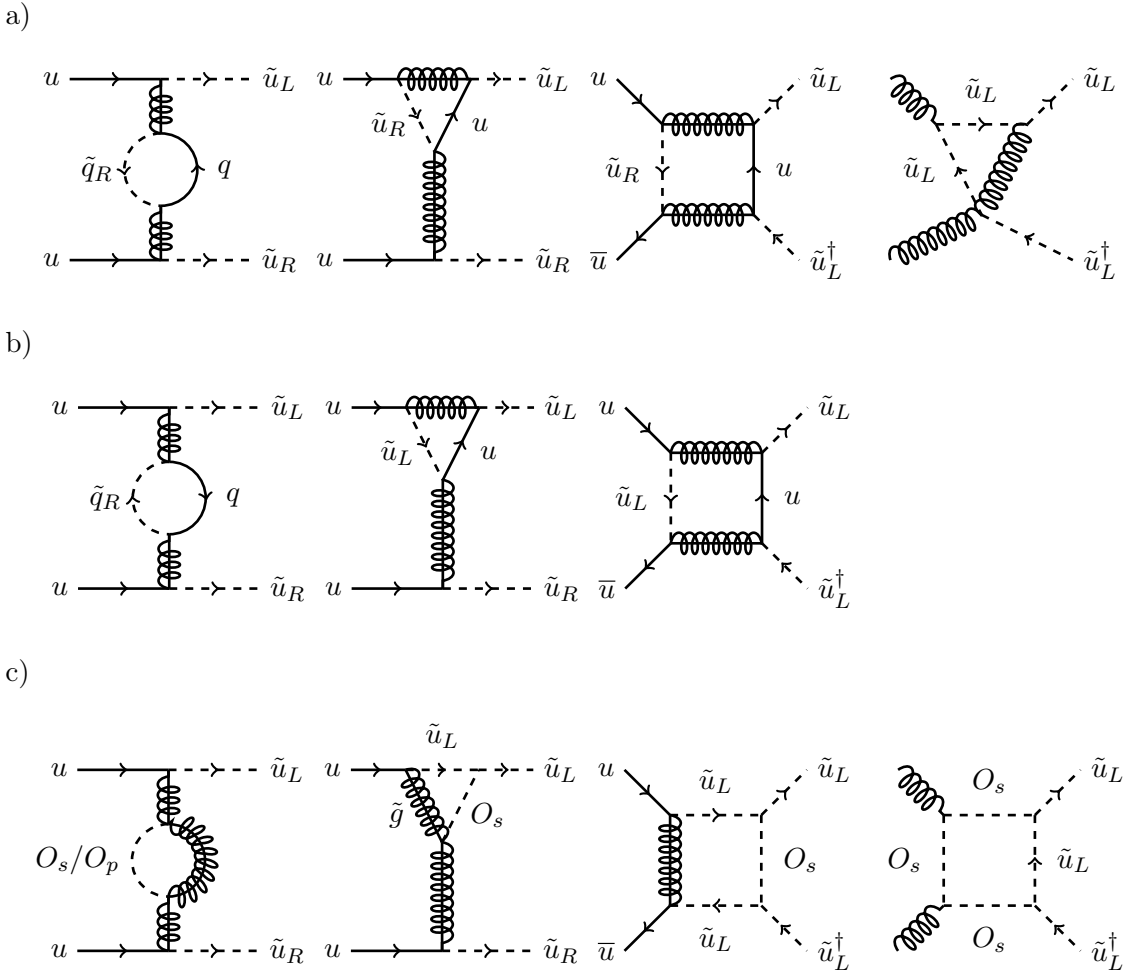
\begin{figure}[ht]
a)
\begin{center}
\begin{tikzpicture}[line width=1.0 pt, scale=1.2, arrow/.style={thick,->}]
    \node at (0,0) {};
\begin{scope}[shift={(0,-1)}]
	\draw[fermion] (-1,-1)--(0,-1);
	\draw[fermion] (-1,1)--(0,1);
	\node at (-1.2,1) {$u$};
	\node at (-1.2,-1) {$u$};
	\draw[gluon] (0,-1)--(0,-0.4);
	\draw[fermionnoarrow] (0,-1)--(0,-0.4);
	\draw[gluon] (0,0.4)--(0,1);
	\draw[fermionnoarrow] (0,0.4)--(0,1);
	\draw[fermion] (0,-0.4) arc (270:450:0.4);
	\draw[scalar] (0,0.4) arc (90:270:0.4);
	\node at (-0.7,0){$\tilde{q}_R$};
	\node at (0.7,0){$q$};
	\draw[scalar] (0,-1)--(1,-1);
	\draw[scalar] (0,1)--(1,1);
	\node at (1.3,1) {$\tilde{u}_L$};
	\node at (1.3,-1) {$\tilde{u}_R$};
\begin{scope}[shift={(3,0)}]
	\draw[fermion] (-1,-1)--(0,-1);
	\draw[fermion] (-1,1)--(-0.4,1);
	\node at (-1.2,1) {$u$};
	\node at (-1.2,-1) {$u$};
	\draw[scalar] (-0.4,1)--(0,0.2);
	\draw[fermionnoarrow] (-0.4,1)--(0.4,1);
	\draw[gluon] (-0.4,1)--(0.4,1);
	\draw[fermion] (0,0.2)--(0.4,1);
	\draw[fermionnoarrow] (0,0.2)--(0,-1);
	\draw[gluon] (0,0.2)--(0,-1);
	\node at (-0.5,0.5){$\tilde{u}_R$};
	\node at (0.5,0.5){$u$};
	\draw[scalar] (0,-1)--(1,-1);
	\draw[scalar] (0.4,1)--(1,1);
	\node at (1.3,1) {$\tilde{u}_L$};
	\node at (1.3,-1) {$\tilde{u}_R$};
\end{scope}
\begin{scope}[shift={(6,0)}]
    \draw[fermion] (-1,1)--(-0.5,0.5);
	\draw[fermionbar] (-1,-1)--(-0.5,-0.5);
	\node at (-1.2,1) {$u$};
	\node at (-1.2,-1) {$\overline{u}$};
	\draw[scalar] (0.5,0.5) -- (1,1);
	\draw[scalarbar] (0.5,-0.5) -- (1,-1);
	\node at (1.3,1) {$\tilde{u}_L$};
	\node at (1.3,-1) {$\tilde{u}_L^\dagger$};
	\draw[gluon] (-0.5,0.5)--(0.5,0.5);
	\draw[gluon] (-0.5,-0.5)--(0.5,-0.5);
	\draw[fermionnoarrow] (-0.5,0.5)--(0.5,0.5);
	\draw[fermionnoarrow] (-0.5,-0.5)--(0.5,-0.5);
	\draw[fermion] (0.5,-0.5)--node[label=right:$u$]{}(0.5,0.5);
	\draw[scalar] (-0.5,0.5)--node[label=left:$\tilde{u}_R$]{}(-0.5,-0.5);
\end{scope}
\begin{scope}[shift={(9,0)}]
	\draw[gluon] (-1,1)--(-0.5,0.5);
	\draw[gluon] (-1,-1)--(0,-0.5);
	\draw[scalar] (0.5,0.5) -- (1,1);
	\draw[scalarbar] (0,-0.5) -- (1,-1);
	\node at (1.3,1) {$\tilde{u}_L$};
	\node at (1.3,-1) {$\tilde{u}_L^\dagger$};
	\draw[scalar] (-0.5,0.5)--node[label=above:$\tilde{u}_L$]{}(0.5,0.5);
	\draw[scalarbar] (-0.5,0.5)--node[label=left:$\tilde{u}_L$]{}(0,-0.5);
	\draw[gluon] (0,-0.5) -- (0.5,0.5);
\end{scope}
\end{scope}
\end{tikzpicture}
\end{center}
b)
\begin{center}
\begin{tikzpicture}[line width=1.0 pt, scale=1.2, arrow/.style={thick,->}]
    \node at (0,0) {};
\begin{scope}[shift={(0,-1)}]
	\draw[fermion] (-1,-1)--(0,-1);
	\draw[fermion] (-1,1)--(0,1);
	\node at (-1.2,1) {$u$};
	\node at (-1.2,-1) {$u$};
	\draw[gluon] (0,-1)--(0,-0.4);
	\draw[fermionnoarrow] (0,-1)--(0,-0.4);
	\draw[gluon] (0,0.4)--(0,1);
	\draw[fermionnoarrow] (0,0.4)--(0,1);
	\draw[fermionbar] (0,-0.4) arc (270:450:0.4);
	\draw[scalarbar] (0,0.4) arc (90:270:0.4);
	\node at (-0.7,0){$\tilde{q}_R$};
	\node at (0.7,0){$q$};
	\draw[scalar] (0,-1)--(1,-1);
	\draw[scalar] (0,1)--(1,1);
	\node at (1.3,1) {$\tilde{u}_L$};
	\node at (1.3,-1) {$\tilde{u}_R$};
\begin{scope}[shift={(3,0)}]
	\draw[fermion] (-1,-1)--(0,-1);
	\draw[fermion] (-1,1)--(-0.4,1);
	\node at (-1.2,1) {$u$};
	\node at (-1.2,-1) {$u$};
	\draw[scalar] (-0.4,1)--(0,0.2);
	\draw[fermionnoarrow] (-0.4,1)--(0.4,1);
	\draw[gluon] (-0.4,1)--(0.4,1);
	\draw[fermion] (0,0.2)--(0.4,1);
	\draw[fermionnoarrow] (0,0.2)--(0,-1);
	\draw[gluon] (0,0.2)--(0,-1);
	\node at (-0.5,0.5){$\tilde{u}_L$};
	\node at (0.5,0.5){$u$};
	\draw[scalar] (0,-1)--(1,-1);
	\draw[scalar] (0.4,1)--(1,1);
	\node at (1.3,1) {$\tilde{u}_L$};
	\node at (1.3,-1) {$\tilde{u}_R$};
\end{scope}
\begin{scope}[shift={(6,0)}]
	\draw[fermion] (-1,1)--(-0.5,0.5);
	\draw[fermionbar] (-1,-1)--(-0.5,-0.5);
	\node at (-1.2,1) {$u$};
	\node at (-1.2,-1) {$\overline{u}$};
	\draw[scalar] (0.5,0.5) -- (1,1);
	\node at (1.3,1) {$\tilde{u}_L$};
	\draw[scalarbar] (0.5,-0.5) -- (1,-1);
	\node at (1.3,-1) {$\tilde{u}_L^\dagger$};
	\draw[gluon] (-0.5,0.5)--(0.5,0.5);
	\draw[gluon] (-0.5,-0.5)--(0.5,-0.5);
	\draw[fermionnoarrow] (-0.5,0.5)--(0.5,0.5);
	\draw[fermionnoarrow] (-0.5,-0.5)--(0.5,-0.5);
	\draw[fermion] (0.5,-0.5)--node[label=right:$u$]{}(0.5,0.5);
	\draw[scalar] (-0.5,0.5)--node[label=left:$\tilde{u}_L$]{}(-0.5,-0.5);
\end{scope}
\begin{scope}[shift={(9,0)}]
	\node at (-1.2,1) {};
	\node at (1.3,1) {};
\end{scope}
\end{scope}
\end{tikzpicture}
\end{center}
c)
\begin{center}
\begin{tikzpicture}[line width=1.0 pt, scale=1.2, arrow/.style={thick,->}]
    \node at (0,0) {};
\begin{scope}[shift={(0,-1)}]
	\draw[fermion] (-1,-1)--(0,-1);
	\draw[fermion] (-1,1)--(0,1);
	\node at (-1.2,1) {$u$};
	\node at (-1.2,-1) {$u$};
	\draw[gluon] (0,-1)--(0,-0.4);
	\draw[fermionnoarrow] (0,-1)--(0,-0.4);
	\draw[gluon] (0,0.4)--(0,1);
	\draw[fermionnoarrow] (0,0.4)--(0,1);
	\draw[fermionnoarrow] (0,-0.4) arc (270:450:0.4);
	\draw[gluon] (0,-0.4) arc (270:450:0.4);
	\draw[scalarnoarrow] (0,0.4) arc (90:270:0.4);
	\node at (-1,0){$O_s/O_p$};
	%\node at (0.7,0){$\tilde{g}$};
	\draw[scalar] (0,-1)--(1,-1);
	\draw[scalar] (0,1)--(1,1);
	\node at (1.3,1) {$\tilde{u}_L$};
	\node at (1.3,-1) {$\tilde{u}_R$};
\begin{scope}[shift={(3,0)}]
	\draw[fermion] (-1,-1)--(0,-1);
	\draw[fermion] (-1,1)--(-0.4,1);
	\node at (-1.2,1) {$u$};
	\node at (-1.2,-1) {$u$};
	\draw[gluon] (-0.4,1)--(0,0.2);
	\draw[fermionnoarrow] (-0.4,1)--(0,0.2);
	\draw[scalar] (-0.4,1)--node[label=above:$\tilde{u}_L$]{}(0.4,1);
	\draw[scalarnoarrow] (0,0.2)--(0.4,1);
	\draw[fermionnoarrow] (0,0.2)--(0,-1);
	\draw[gluon] (0,0.2)--(0,-1);
	\node at (-0.5,0.5){$\tilde{g}$};
	\node at (0.5,0.5){$O_s$};
	\draw[scalar] (0,-1)--(1,-1);
	\draw[scalar] (0.4,1)--(1,1);
	\node at (1.3,1) {$\tilde{u}_L$};
	\node at (1.3,-1) {$\tilde{u}_R$};
\end{scope}
\begin{scope}[shift={(6,0)}]	
	\draw[fermion] (-1,1)--(-0.5,0.5);
	\draw[fermionbar] (-1,-1)--(-0.5,-0.5);
	\node at (-1.2,1) {$u$};
	\node at (-1.2,-1) {$\overline{u}$};
	\draw[scalar] (0.5,0.5) -- (1,1);
	\draw[scalarbar] (0.5,-0.5) -- (1,-1);
	\node at (1.3,1) {$\tilde{u}_L$};
	\node at (1.3,-1) {$\tilde{u}_L^\dagger$};
	\draw[scalar] (-0.5,0.5)--node[label=above:$\tilde{u}_L$]{}(0.5,0.5);
	\draw[scalarbar] (-0.5,-0.5)--node[label=below:$\tilde{u}_L$]{}(0.5,-0.5);
	\draw[scalarnoarrow] (0.5,0.5)--node[label=right:$O_s$]{}(0.5,-0.5);
	\draw[gluon] (-0.5,0.5)--(-0.5,-0.5);
	\draw[fermionnoarrow] (-0.5,0.5)--(-0.5,-0.5);
\end{scope}
\begin{scope}[shift={(9,0)}]
	\draw[gluon] (-1,1)--(-0.5,0.5);
	\draw[gluon] (-1,-1)--(-0.5,-0.5);
	\draw[scalar] (0.5,0.5) -- (1,1);
	\draw[scalarbar] (0.5,-0.5) -- (1,-1);
	\draw[scalarnoarrow] (-0.5,0.5)--node[label=above:$O_s$]{}(0.5,0.5);
	\draw[scalarnoarrow] (-0.5,-0.5)--node[label=below:$O_s$]{}(0.5,-0.5);
	\draw[scalarbar] (0.5,0.5)--node[label=right:$\tilde{u}_L$]{}(0.5,-0.5);
	\draw[scalarnoarrow] (-0.5,0.5)--node[label=left:$O_s$]{}(-0.5,-0.5);
	\node at (1.3,1) {$\tilde{u}_L$};
	\node at (1.3,-1) {$\tilde{u}_L^\dagger$};
\end{scope}
\end{scope}
\end{tikzpicture}
\end{center}
\caption{Examples of Feynman diagrams relevant for the
NLO corrections to $\tilde{q}\tilde{q}$ and $\tilde{q}\tilde{q}^\dagger$ production. Categories as in figure~\ref{fig:oneloopcts}.}\label{fig:oneloopdiagrams}
\end{figure}
We need to consider the one-loop amplitudes for the three partonic
processes $q\overline{q} \to \tilde{q}\tilde{q}^\dagger$,
$gg \to \tilde{q}\tilde{q}^\dagger$, and $q q\to \tilde{q}\tilde{q}$.
Sample Feynman diagrams for the calculation of renormalisation constants 
 are shown in figure~\ref{fig:oneloopcts}. Diagrams contributing
 to the production of (anti)squarks are shown in figure~\ref{fig:oneloopdiagrams}.
Diagrams of type a) are exactly the same in the MRSSM and in
the conventional MSSM and give equal results. 
Diagrams of type b) can be drawn in the
MRSSM but unlike in the MSSM do not contribute. 
The result is proportional to the Majorana mass and leads
therefore to R-charge violation in the MRSSM.
Diagrams of type c) exist in the MRSSM but are absent in the MSSM because of
the appearance of sgluons, additional strongly interacting
particles.

The ultraviolet renormalisation requires the introduction of coupling,
mass, and field renormalisation. We define all SUSY masses in the on-shell
renormalisation scheme to have physical masses as input. The quark
masses are set to zero, except for the top quark mass.
For simplicity we take all squarks as mass degenerate.
We define the field renormalisation in the on-shell scheme, which leads to a
correctly normalised S-matrix but also to infrared divergent
renormalisation constants.
The strong coupling is renormalised in a decoupling scheme as
described below.

With these definitions, standard methods lead to the following results
for the quark/squark field and squark mass renormalisation constants,
computed with one-loop SUSY-QCD corrections:
\begin{align}
\label{eq:dzq}
\delta Z_q &= 2 C_F \frac{g_s^2}{16\pi^2} \Re \left[ B_1(0 ,m_{\tilde{g}}^2, m_{\tilde{q}}^2) \right]\,,\\
% newline
\delta Z_{\tilde{q}} &= \frac{g_s^2}{16\pi^2} C_F \Re \left[ 4B_1(m_{\tilde{q}}^2,0,m_{\tilde{g}}^2) + 2B_1(m_{\tilde{q}}^2,0,m_{\tilde{q}}^2) + 3B_0(m_{\tilde{q}}^2,0,m_{\tilde{q}}^2)  + 4m_{\tilde{q}}^2 B^\prime_1(m_{\tilde{q}}^2,0,m_{\tilde{g}}^2)\right.\nonumber\\
& \left. - 4m_{\tilde{g}}^2 B^\prime_0(m_{\tilde{q}}^2,m_{O_s}^2,m_{\tilde{q}}^2) + 2m_{\tilde{q}}^2 B^\prime_1(m_{\tilde{q}}^2,0,m_{\tilde{q}}^2) + 4m_{\tilde{q}}^2 B^\prime_0(m_{\tilde{q}}^2,0,m_{\tilde{q}}^2) \right]
\,,\\
% newline
\delta m_{\tilde{q}}^2 &= \frac{g_s^2}{16\pi^2} C_F \Re \left[ A_0(m_{\tilde{q}}^2) - (4A_0(m_{\tilde{g}}^2) + 4B_1(m_{\tilde{q}}^2,0,m_{\tilde{g}}^2)m_{\tilde{q}}^2) + 4 m_{\tilde{g}}^2 B_0(m_{\tilde{q}}^2,m_{O_s}^2,m_{\tilde{q}}^2) \right.\nonumber\\
&\left. -(2B_1(m_{\tilde{q}}^2,0,m_{\tilde{q}}^2) m_{\tilde{q}}^2 + 4 B_0(m_{\tilde{q}}^2,0,m_{\tilde{q}}^2)m_{\tilde{q}}^2) \right]\,.
\end{align}
We use standard Passarino-Veltman integrals and their derivatives, see
e.g.\ ref.~\cite{Denner:1991kt} for the definition.
The renormalisation constants are computed in both the HV and FDH
scheme; we denote terms only appearing in HV by the subscript
\texttt{HV}. 
We do not have to distinguish left- and right-handed squarks since we
do not take electroweak corrections into account.

Though not strictly necessary for our calculations, we also quote the
result for the gluino field renormalisation. Here the left- and
right-handed parts renormalise differently, since they are part of
different superfields, i.e. the right-handed part of the gluino %, referred to as octino, 
does not couple to (s)quarks. This is reflected by a gluino self-energy which is not left-right symmetric
and has the following basic structure:
\begin{align}
\Sigma^{\mathrm{MRSSM}}_{\tilde{g}\overline{\tilde{g}}}(p^2) &= A(p^2)P_L \slashed{p} + \hdots ,\\
\Sigma^{\mathrm{MSSM}}_{\tilde{g}\overline{\tilde{g}}}(p^2) &= A(p^2)(P_L+P_R) \slashed{p} + \hdots,
\end{align}
where the dots stand for contributions not stemming from (s)quarks.
The results read
\begin{align}
\delta Z_{\tilde{g}}^R &= \frac{g_s^2}{16\pi^2}\Re \left[C_A 
(B_1(m_{\tilde{g}}^2,m_{\tilde{g}}^2,m_{O_s}^2) + 
B_1(m_{\tilde{g}}^2,m_{\tilde{g}}^2,m_{O_p}^2)) \right.\nonumber\\
&+C_A(\left. 1\right|_{\texttt{HV}}-2(B_0(m_{\tilde{g}}^2,0,m_{\tilde{g}}^2) + 
B_1(m_{\tilde{g}}^2,0,m_{\tilde{g}}^2)))\nonumber\\
&+ 4T_F m_{\tilde{g}}^2 \left( (n_f-1)  
B^\prime_1(m_{\tilde{g}}^2,0,m_{\tilde{q}}^2) +  
B^\prime_1(m_{\tilde{g}}^2,m_t^2,m_{\tilde{q}}^2) \right)\nonumber\\
&-2 C_A m_{\tilde{g}}^2 \left( 
B^\prime_0(m_{\tilde{g}}^2,m_{\tilde{g}}^2,m_{O_s}^2)-
B^\prime_0(m_{\tilde{g}}^2,m_{\tilde{g}}^2,m_{O_p}^2) - 
B^\prime_1(m_{\tilde{g}}^2,m_{\tilde{g}}^2,m_{O_s}^2) - 
B^\prime_1(m_{\tilde{g}}^2,m_{\tilde{g}}^2,m_{O_p}^2) \right)\nonumber\\
&+4 C_A m_{\tilde{g}}^2 \left.\left( 
B^\prime_0(m_{\tilde{g}}^2,0,m_{\tilde{g}}^2) - 
B^\prime_1(m_{\tilde{g}}^2,0,m_{\tilde{g}}^2) \right)\right]\, ,\\
\delta Z_{\tilde{g}}^L &= \frac{g_s^2}{16\pi^2}\Re \left[ 4T_F\left( (n_f-1) 
B_1(m_{\tilde{g}}^2,0,m_{\tilde{q}}^2) + 
B_1(m_{\tilde{g}}^2,m_t^2,m_{\tilde{q}}^2) \right) \right] + \delta Z_{\tilde{g}}^R\, .
\end{align}
The corresponding gluino mass renormalisation constant is 
\begin{align}
\delta m_{\tilde{g}} &= \frac{g_s^2}{16\pi^2} m_{\tilde{g}}\ \Re \left[ 
-2 T_F \left( (n_f-1)B_1(m_{\tilde{g}}^2,0,m_{\tilde{q}}^2) + 
B_1(m_{\tilde{g}}^2,m_t^2,m_{\tilde{q}}^2) \right) \right.\nonumber\\
& + C_A \left( B_0(m_{\tilde{g}}^2,m_{\tilde{g}}^2,m_{O_s}^2) - 
B_0(m_{\tilde{g}}^2,m_{\tilde{g}}^2,m_{O_p}^2) - 
B_1(m_{\tilde{g}}^2,m_{\tilde{g}}^2,m_{O_s}^2) -
B_1(m_{\tilde{g}}^2,m_{\tilde{g}}^2,m_{O_p}^2) \right)\nonumber\\
& + C_A \left.\left( \left. 1\right|_{\texttt{HV}} 
- 2 B_0(m_{\tilde{g}}^2,0,m_{\tilde{g}}^2) + 2 
B_1(m_{\tilde{g}}^2,0,m_{\tilde{g}}^2) \right)\right]\,.
\end{align}
Note that the contribution stemming from the quark-squark loop is halved when 
compared to the MSSM result, due to the non-coupling right-handed gluino. 
As a consequence, the left-handed Dirac gluino allows only for a ``left-handed'' 
squark or a ``right-handed'' antisquark in the loop.
Finally, the field renormalisation constant of the gluon is given by
\begin{align}
\delta Z_G &= \frac{g_s^2}{16\pi^2} \Re \left[ T_F \left( -\frac{4}{3} B_0(0,m_t^2,m_t^2) - \frac{8}{3} m_t^2 B^\prime_0(0,m_t^2,m_t^2) +  \frac{4}{9} \right)\right.\nonumber\\
& + C_A\left( -\frac{4}{3} B_0(0,m_{\tilde{g}}^2,m_{\tilde{g}}^2) - \frac{8}{3} m_{\tilde{g}}^2 B^\prime_0(0,m_{\tilde{g}}^2,m_{\tilde{g}}^2) +  \frac{4}{9}\right)\notag\\
& + 12 T_F \left( -\frac{1}{3} B_0(0,m_{\tilde{q}}^2,m_{\tilde{q}}^2) + \frac{4}{3} m_{\tilde{q}}^2 B^\prime_0(0,m_{\tilde{q}}^2,m_{\tilde{q}}^2) - \frac{2}{9} \right) \nonumber\\
& + C_A \left( -\frac{1}{6} B_0(0,m_{O_s}^2,m_{O_s}^2) -\frac{1}{6} B_0(0,m_{O_p}^2,m_{O_p}^2) + \frac{2}{3} m_{O_s}^2 B^\prime_0(0,m_{O_s}^2,m_{O_s}^2)\right.\nonumber\\
& \left.\left. + \frac{2}{3} m_{O_p}^2 B^\prime_0(0,m_{O_p}^2,m_{O_p}^2) -\frac{2}{9} \right)\right].
\label{eq:dZG}
\end{align}

The renormalisation of the strong coupling has to be treated in a special way in order to
make use of the experimental determination of $\alpha_s$ in the SM
5-flavour scheme and for compatibility with available PDF sets. The
renormalisation has to be matched to the SM $\overline{\mathrm{MS}}$
5-flavour scheme, and contributions 
from heavy particles to the renormalisation of $g_s$ have to be
subtracted at zero momentum. In practice, we have separated the 
loop-diagrams used in  the determination of $\delta g_s$ into 
contributions from light and heavy particles. 
From loop-diagrams involving solely light particles we have kept only 
the part corresponding to the $\overline{\mathrm{MS}}$-scheme, whereas for 
diagrams involving heavy particles we have adopted zero-momentum-subtraction.
This leads to
\begin{align}
\frac{\delta g_s}{g_s} &= \frac{g_s^2}{16\pi^2} \left[ \left( \frac{2}{3}T_F n_f - \frac{11}{6} C_A + \frac{1}{3} T_F n_f + \frac{5}{6}C_A \right)\Delta_\epsilon  + (1 -\left. 1\right|_{\texttt{HV}})\frac{C_A}{6}  \right.\nonumber\\
&- \frac{2}{3} C_A \ln \frac{m_{\tilde{g}}^2}{\mu^2} - \frac{1}{3} T_F n_f \ln \frac{m_{\tilde{q}}^2}{\mu^2} - \frac{2}{3} T_F \ln \frac{m_t^2}{\mu^2}-\frac{1}{12} C_A \left.\left( \ln \frac{m_{O_s}^2}{\mu^2} + \ln \frac{m_{O_p}^2}{\mu^2} \right)\right],\label{eq:delta_gs}
\end{align}
with the typical UV-divergent constant $\Delta_\epsilon$ defined as in ref.~\cite{Denner:1991kt}.
As a result, the renormalisation constant $\delta g_s$ contains additional
$\mu$-dependent terms, where $\mu$ is the
$\overline{\mathrm{MS}}$ renormalisation scale. %\footnote{This procedure has
%  already been carried out in the corresponding calculation for the
%  MSSM \cite{Beenakker:1996ch}.}
%\marginpar{Is the footnote justified? Did they invent this? Or does
%  everybody do it? DS, PD: already done in several top qcd calculations} 
These have the effect of decoupling heavy particles from the running of $g_s$,
which is then given by the SM 5-flavour $\beta$-function. 
As a side remark, we mention that if $\delta g_s$ was defined in pure $\overline{\mathrm{MS}}$-scheme, the $\beta$-function would vanish at
one-loop-level.

\subsection{Method 1: HV and Passarino-Veltman reduction}\label{sec:virt_method1}

In our first method we have performed the calculation in HV
regularisation, i.e.\ usual dimensional regularisation for
internal particles, while particles outside of loops are kept
unregularised.

It is well known that dimensional regularisation breaks 
SUSY due to a mismatch between degrees of freedom of the gluon and the 
gluino in $D= 4-2\epsilon \neq 4$ dimensions. 
Since dimensional reduction \cite{Siegel79,Capper79} is known to preserve SUSY at the
one-loop level (for reviews of checks see
e.g.\ \cite{Jack:1997sr,Stockinger:2005gx,Jones:2012gfa}), the required SUSY-restoring
counterterms can simply be obtained from comparing renormalisation
constants in dimensional regularisation and dimensional
reduction. Appropriate transition counterterms are known for physical
parameters in generic SUSY models at one-loop level
\cite{Martin:1993yx}, for the full MSSM at one-loop level
\cite{Stockinger:2011gp}, and for SUSY-QCD at two-loop level
\cite{Mihaila:2009bn}. 
For our calculation, only one such transition counterterm is needed:
the one for the squark-quark-gluino vertex. Denoting the renormalisation
constant for this %squark-quark-gluino 
vertex by $\delta \hat{g}_s$, we find that it has to satisfy
\begin{align}
\delta \hat{g}_s & = \delta g_s + \delta g_s^{\mathrm{restore}},
\\
 \delta g_s^{\mathrm{restore}} & =
\frac{g_s^3}{16\pi^2}\left( \frac{2 C_A}{3} - \frac{C_F}{2} \right),
\end{align}
where $\delta g_s$ is given by eq.~\eqref{eq:delta_gs}.
The result for $ \delta g_s^{\mathrm{restore}}$ is the same as in the
MSSM, since SUSY-breaking is only associated with the gluon, which
has the same couplings in the MSSM and MRSSM.

The implementation of this calculation has been done with the
help of several Mathematica packages. 
The generation and processing of amplitudes has been performed by
\texttt{FeynArts}~\cite{Hahn:2000} and \texttt{FormCalc}
\cite{Nejad:2013,Hahn:1998}. The model file for the MRSSM containing the
tree-level vertices was generated by \texttt{SARAH}
\cite{Staub:2009bi,Staub:2010jh,Staub:2012pb,Staub:2013tta};
the one-loop counterterms Feynman rules were included by hand into the model file.
 The output has been passed on to a \texttt{C++} program which performs the
evaluation of loop integrals using
\texttt{LoopTools} \cite{Hahn:1998yk} and does the integration using 
the \texttt{CUBA} library \cite{Hahn:2004fe}.%\footnote{The code, called \texttt{RSymSQCD}, will become available for download from \cite{our_code} shortly after the publication of this work.} 

\subsection{Method 2: FDH and integrand reduction approach}
In our second calculation we employ the FDH regularisation scheme. We
follow the notation of ref.~\cite{Signer:2008va}, so FDH is the same
as standard dimensional reduction, but ``regular'' particles outside
of loops are kept unregularised. The FDH scheme is advantageous not
only because it preserves SUSY at one-loop level, but also because
it allows the use of powerful and efficient helicity methods for the evaluation
of loop amplitudes.

In the past, dimensional reduction was often not applied to SUSY-QCD
calculations because of an issue with QCD factorisation discussed in
refs.~\cite{Beenakker:1989,Smith:2005}. In the meantime it has been
understood that factorisation behaves differently depending on whether
FDH as defined above, or whether DRED, where also particles outside
loops are regularised, is employed
\cite{Signer:2005,Signer:2008va}.\footnote{%
In the literature, what we call FDH here is sometimes denoted as DR.}
In case of FDH, factorisation holds as expected, however the infrared
anomalous dimensions are different from the HV scheme and the
transition rules found in \cite{Kunszt:1994,Catani:1996pk} apply. In case of DRED,
factorisation is complicated by the appearance of external
$\epsilon$-scalars with separate couplings and anomalous
dimensions. The understanding of the infrared behavior of all these
schemes and their transition rules has been extended to the multi-loop
level in
refs.~\cite{Kilgore:2012tb,Gnendiger:2014nxa,Broggio:2015ata,Broggio:2015dga,Gnendiger:2016cpg}. 

The upshot of these results is that FDH can be used to regularise
ultraviolet and infrared divergences for SUSY-QCD processes. However
in order to combine the results with real corrections evaluated in HV
and to convolute with usual PDFs, we need to convert the amplitudes
from FDH to HV. The appropriate transition rules for the squared
amplitudes are, in the notation of ref.~\cite{Signer:2008va}
\begin{equation}
|\mathcal{M}_{\text{HV}}|^2_{\text{1L}}=
|\mathcal{M}_{\text{FDH}}|^2_{\text{1L}} + \frac{g_s^2}{8\pi}
|\mathcal{M}_{\text{FDH}}|^2_{\text{tree}}\times\sum_i
 \tilde\gamma^{\text{FDH}}_i
\end{equation}
with $\tilde\gamma^{\text{FDH}}_q=\tilde\gamma^{\text{FDH}}_{\bar q}=C_F/2$
and $\tilde\gamma^{\text{FDH}}_g=C_A/6$ and the sum running over all
external partons.

An alternative to Passarino-Veltman reduction for NLO calculations 
with Monte Carlo methods
is the usage of helicity methods and an integrand reduction approach, 
see e.g. ref.~\cite{Ellis:2011cr} for a review. 
For these methods, several computer codes already
exist.
They contain a general implementation of reduction methods and require the
input of model dependent information. We summarise these parts first
and then describe the implementation used for our calculation.

Needed for our calculation are the relevant Feynman rules of the MRSSM and the counterterms.
The necessary renormalisation constants of masses and wave functions are given in eqs.~\eqref{eq:dzq} to \eqref{eq:dZG}.
The renormalisation of the coupling constant $g_s$ given in 
eq.~\eqref{eq:delta_gs} includes the finite term containing the expression
$1-\left. 1\right|_{\texttt{HV}}$ which marks the 
well-known transition between $\overline{\mathrm{DR}}$ and 
$\overline{\mathrm{MS}}$ scheme needed for FDH.
Including this transition rule allows us to combine matrix elements calculated in FDH with PDFs  given in the $\overline{\text{MS}}$ scheme.
The remainder of the calculation is done automatically
as described in the following.

 For our second calculational method, we use \texttt{GoSam} 2~\cite{Cullen:2011ac,Cullen:2014yla}%
\footnote{With \texttt{GoSam} we make use of the programs 
\texttt{Qgraf}~\cite{Nogueira:1991ex},
\texttt{Form}~\cite{Vermaseren:2000nd},
\texttt{Ninja}~\cite{Mastrolia:2012bu,Peraro:2014cba,vanDeurzen:2013saa} (which
uses \texttt{OneLOop}~\cite{vanHameren:2010cp}) and 
\texttt{Golem95}~\cite{Binoth:2008uq,Cullen:2011kv,Guillet:2013msa}.
}
to calculate the virtual corrections using helicity methods and 
integrand reduction methods.
The information concerning the MRSSM is passed to \texttt{GoSam} using 
the UFO interface~\cite{Degrande:2011ua}. 
For this, the additional strongly interacting particle content of the 
MRSSM was added to the SM implementation in  \texttt{FeynRules}~\cite{Christensen:2008py}.
Numerous checks were performed to verify that the \texttt{SARAH} and \texttt{FeynRules} model files give the same tree-level results for the relevant processes.

The UV renormalisation of the MRSSM has been added by hand to \texttt{GoSam}.%
\footnote{The automatised derivation using the \texttt{NLOCT} package~\cite{Degrande:2014vpa} 
and subsequent use of \texttt{MadLoop}~\cite{Hirschi:2011pa} was not possible
as the considered sector of the MRSSM is not directly on-shell renormalisable.
This is due to the mass relation~\eqref{eq:octetmasses} 
and the Dirac mass appearing in the sgluon-squark triple vertex.
As the scalar octets do not appear at tree level, excluding it from the renormalisation procedure
would be enough to achieve the renormalisation but no such option exists in \texttt{NLOCT}.}
The counterterms are implemented using \texttt{OneLOop} for the loop functions and added to the \textsf{matrix.f90} template in the \texttt{GoSam} interface.
The exact counterterm structure has to be fixed once after generating the considered
process with \texttt{GoSam}.

Comparing the computational speed between both Methods
both implementations lead to similar running times
for the considered processes with the timings being 
within an order of magnitude of each other.
The difference are in general not relevant when combining with the
real corrections of section~\ref{sec:real} which take up the majority of the
computing time for the full calculation.
\section{Real corrections}\label{sec:real}
The singularities remaining in the virtual matrix elements, which are of infrared (IR) origin, cancel after combination with the real corrections.
The singularities left in the real emission processes are then removed by mass factorisation~\cite{Collins:1985ue,Bodwin:1984hc}.

A multitude of methods that implement the above cancellation on a numerical basis have been developed. 
In the following two subsections, we will discuss the two approaches used in this work: the two cut phase space slicing
method (TCPSS) \cite{Harris:2001sx} and the Frixione-Kunszt-Signer (FKS) subtraction~\cite{Frixione:1995ms,Frixione:1997np}. 
\subsection{Method 1: Two cut phase space slicing (TCPSS)}

A review of TCPSS can be found in ref.~\cite{Harris:2001sx}.
The method has been implement and tested by one of us in the context of the (S)QCD particle production on the process of sgluon pair production in refs.~\cite{strangethesis,Kotlarski:2016zhv}.
In this method the three-body real emission phase space is decomposed with respect to the additional parton radiation into: soft ($S$), hard collinear ($C$) and the hard non-collinear ($H\overline{C}$) regions by introducing two parameters $\delta_s$ and $\delta_c$,
taken to be numerically small.
This can be schematically written as
\begin{equation}
\sigma_R = \int d \sigma_R = \int_S d \sigma_R + \int_H d \sigma_R = \int_S d \sigma_R + \int_{HC} d \sigma_R + \int_{H\overline{C}} d \sigma_R ,
\end{equation}
where in the last step the hard part $H$ is split into collinear ($HC$) and non-collinear ($H\overline{C}$) pieces.
Within the soft and collinear approximations the divergences are then dimensionally regularised and extracted analytically.
We will now present technical details needed to carry out this calculation.

\subsubsection{Soft emissions} 
The soft phase space $S$ is defined by the condition that the energy of an outgoing gluon $E_5$, in the rest frame of colliding partons, fulfils
\begin{equation}
  \label{eq:gluon_en_soft_lim}
  E_5 < \delta_s \frac{\sqrt{\hat s}}{2},
\end{equation}
where %$\hat s$ is the partonic Mandelstam variable and 
$\delta_s \ll 1$.
We label incoming parton momenta as $p_1$ and $p_2$, such that $\hat s = (p_1 + p_2)^2$.
In the soft limit and $4-2\epsilon_{\text{IR}}$ dimensions the $2 \to 3$ amplitude can be written as
\begin{equation}
  \mathcal{M}_{3} = g_s \mu^{\epsilon} \epsilon_\mu(p_5) \,\textbf{J}^{\mu} 
  (p_5) \cdot \mathcal{M}_2 + \text{finite terms},
  \label{eq:eik}
\end{equation}
where $p_5$ is the gluon momentum and
\begin{equation}
  \textbf{J}^{\mu} (p_5) \equiv \sum_{f=1}^4 \textbf{T}_f \frac{p_f^\mu}{p_f 
  \cdot p_5}
\end{equation}
is the non-abelian eikonal current (whose sum extents over all partons except 
for the final-state gluon), which is colour-connected to the $2 \to 2$ process 
amplitude $\mathcal{M}_2$ through the colour operator $\textbf{T}$ of the particle $f$. 
Only the first term in eq.~\eqref{eq:eik} contributes to the singular part of $|\mathcal{M}_{3}|^2$, as the interference term is regular in the limit $\delta_s \to 0$.

Similarly, in the soft limit the three-body phase space factorises, with a phase space measure
\begin{equation}
  d\Phi_3^{\text{soft}} = d\Phi_2 \cdot \left(\frac{4\pi}{\hat s}\right)^\epsilon \frac{\Gamma(1-\epsilon)}{\Gamma(1-2\epsilon)} \frac{1}{2(2\pi)^2} dS
  \label{eq:eik4}
\end{equation}
with 
\begin{equation}
  dS = \frac{1}{\pi} \left (\frac{\hat s}{4}\right)^{\epsilon} E_5^{1-2\epsilon} \, d E_5 \, \sin^{1-2\epsilon} \theta_1 \, d \theta_1 \sin^{-2\epsilon} \theta_2 \, d \theta_2,
\end{equation}
where $\theta_{1/2}$ describe the direction of the gluon emission in the rest frame of colliding partons.
The gluon phase space is integrated over the solid angle, and $E_5$ is as given in eq.~\eqref{eq:gluon_en_soft_lim}.
For convenience, the necessary expressions for the D-dimensional angular integrals have been gathered in appendix D.2 of ref.~\cite{strangethesis}.

The expressions for the HV regularised real emission matrix elements have been generated using \texttt{FeynArts} and \texttt{FormCalc} with the model file described in section \ref{sec:virt_method1}.
Expanding in terms of $\epsilon_{\text{IR}}$ we find the double and single poles, and a finite part.
The double-pole terms agree with the well-known minimal structure, being proportional to the four-dimensional Born cross sections given in eqs.~\eqref{eq:qqbar_xsec}, \eqref{eq:gg_xsec} and \eqref{eq:qq_xsec}, which can be found in ref.~\cite{Signer:2008va}:
\begin{align}
 \left . \sigma_{q\bar{q}, q q}^{soft}\right |_{\text{double pole}} = & 
 \sigma_{q\bar{q}, q q}^{B} \cdot 2 \frac{\alpha_s}{2\pi} C_F \cdot \frac{1}{\epsilon_{\text{IR}}^2} , \\
\left.  \sigma_{gg}^{soft} \right |_{\text{double pole}}= & 
\sigma_{gg}^{B} \cdot 2 \frac{\alpha_s}{2\pi} C_A \cdot \frac{1}{\epsilon_{\text{IR}}^2} .
\end{align}
For numerical verification we show the cancellation of these terms %between the soft 
and the virtual matrix elements for a single phase space point in appendix~\ref{sec:pole_cancelation}.

The single-pole coefficient is not cancelled completely between virtual and soft contribution.
The remaining terms have the form
\begin{align}
\label{eq:cos1}
 \left . \sigma_{q\bar{q},qq}^{soft}\right |_{\text{soft-collinear remainder}} = & 
 - \frac{1}{\epsilon_{\text{IR}}} \,\sigma_{q\bar{q},qq}^{B} \cdot 2 \, \frac{\alpha_s}{2\pi} C_F (3/2 + 2\log \delta_s),
 \\
 \label{eq:cos2}
\left.  \sigma_{gg}^{soft} \right |_{\text{soft-collinear remainder}}= & - \frac{1}{\epsilon_{\text{IR}}} \,\sigma_{gg}^{B} \cdot 2 \, \frac{\alpha_s}{2\pi} \left[2N \log \delta_s + \frac{11 N -2 (n_f-1)}{6} \right] ,
\end{align}
where $N$ is the number of colours and $n_f-1$ is the number of massless quark flavours.
These uncancelled terms come from the phase space region where the gluon is collinear with an incoming % really only incoming?
parton, but its energy is non-zero.
They do not cancel out with the virtual contributions as they have different kinematics.
As discussed in the next subsection, these are the terms that can be absorbed by a
redefinition  of the PDF at NLO.
%will cancel out with the soft-collinear pieces of the initial state factorization %counterterms.
\subsubsection{Collinear emissions}
In the collinear limit, defined by the condition~\cite{Dawson:2003zu}
\begin{equation}
\label{eq:collinear_cut}
1 - \cos \theta_{i5} = -\frac{(p_i - p_5)^2}{\sqrt{\hat s} E_5} < \delta_c ,
\end{equation}
where $\delta_c \ll 1$ and $i = 1,2$, the real-emission cross section factorises at the level of the absolute squared matrix element.
Contrary to the definition of the collinear region in ref.~\cite{Harris:2001sx}, eq.~\eqref{eq:collinear_cut} decouples soft and collinear regions.

The double differential hadronic hard-collinear cross section is given by
\begin{multline}
\label{eq:PDFconv2}
\frac{d \sigma^{HC}}{d x_1 d x_2}  =  \sum_{ij} %\frac{2}{1+\delta_{ij}} 
\hat \sigma^B_{ij} \frac{\alpha_s}{2 \pi} \frac{\Gamma(1-\epsilon)}{\Gamma(1-2\epsilon)} \left(\frac{4 \pi \mu_R^2}{\hat s} \right)^\epsilon \left ( - \frac{1}{\epsilon} \right )\delta_c^{-\epsilon} \\ 
 \cdot \sum_k \left ( \int_{x_1}^{1-\delta_s \delta_{ik}} \frac{d z}{z} f_{k/p} \left( \frac{x_1}{z}\right ) f_{j/p} \left( x_2\right ) 
P_{ik}(z, \epsilon)  \left [\frac{(1-z)^2}{2 z} \right]^{-\epsilon} \right.
\\ 
\left. + \int_{x_2}^{1-\delta_s \delta_{jk}} \frac{d z}{z} f_{k/p} \left( x_1 \right ) f_{j/p} \left( \frac{x_2}{z} \right ) 
P_{ik}(z, \epsilon)  \left [\frac{(1-z)^2}{2 z} \right]^{-\epsilon} \right).
\end{multline}
Note that there are two possible ways in which $q\bar q$ in the initial state can be obtained from the proton-proton system.
%The factor $2/(1+\delta_{ij})$ accounts for two possible ways in which $q\bar q$ in the initial state can be obtained from the proton-proton system.
The $P_{ik}(z, \epsilon)$ are the $D$-dimensional unregulated Altarelli-Parisi (AP) splitting kernels \cite{Altarelli:1977zs}
\begin{align}
  P_{qq}(z, \epsilon) = & C_F \left[ \frac{1+z^2}{1-z} - \epsilon(1-z)\right]\,,\\
  P_{gg}(z, \epsilon) = & 2 N \left ( \frac{z}{1-z} + \frac{1-z}{z} + z(1-z) \right)\,, \\
  P_{qg} (z, \epsilon) = & \frac{1}{2} \left[ z^2 + (1-z)^2 \right] - \epsilon z (1-z)\,.
\end{align}
The $\delta_{ik}$ in the integration boundaries of eq.~\eqref{eq:PDFconv2} ensures that the
integral is taken up to $z = 1 - \delta_s$ for kernels which are singular as $z \to 1$ ($P_{qq}$ and $P_{gg}$).
The remaining $P_{gq}$ kernel is obtained by the replacement $P_{gq}(z, \epsilon) = P_{qq} (1-z, \epsilon)$.

The Bjorken variable in $f_{k/p}$ is rescaled so that in the Born configuration $\hat \sigma_B$ is taken at $\hat{s} = x_1 x_2 S$.\footnote{For quark radiation, the integral will be taken up to 1 as there is no soft singularity.}
Collinear singularities will cancel out with the renormalised PDFs. 
The first-order correction to $i$-th flavour PDF in the $\overline{\text{MS}}$ prescription is given by
\begin{equation}
\label{eq:sdPDFs}
  f_{i/p} (x, \mu_F) \equiv f_{i/p}(x) - \frac{1}{\epsilon} \left [ \frac{\alpha_s}{2 \pi} \frac{\Gamma(1-\epsilon)}{\Gamma(1-2\epsilon)} \left(\frac{4 \pi \mu_R^2}{\mu_F^2} \right)^\epsilon \right ] \sum_j \int_x^1 \frac{d z}{z} P_{ij}^+(z) f_{j/p} (x/z),
\end{equation}
where $P_{ij}^+(z)$ are the '+' regulated AP splitting kernels
\begin{align}
  P_{qq}^+(z) = & C_F \left ( \frac{1+z^2}{(1-z)_+} + \frac{3}{2} \delta(1-z)\right),\\
  P_{gg}^+(z) = & 2 N \left ( \frac{z}{(1-z)_+} + \frac{1-z}{z} + z(1-z) \right) + \frac{11 N - 2 (n_f-1)}{6} \, \delta(1-z),
\end{align}
where the associated '+' prescription is defined as 
\begin{equation}
  \label{eq:plus_distribution_definition}
  \int_{x}^1 d z \, f(z) g(z)_+  \equiv \int_{x}^1 d z \, (f(z)-f(1))g(z) - f(1) \int_0^{x} d z\, g(z). 
\end{equation}
For partonic processes which have a soft singularity there is a mismatch in the $z$ integration boundary between eq.~\eqref{eq:PDFconv2} and eq.~\eqref{eq:sdPDFs}.
To account for that, we write eq.~\eqref{eq:plus_distribution_definition} as
\begin{align}
\int_{x}^1 d z \, f(z) g(z)_+ \equiv & \int_{x}^{1-\delta_s} d z \, f(z) g(z) - f(1) \int_0^{1-\delta_s} d z\, g(z) \nonumber \\ & + \int_{1-\delta_s}^1 (f(z)-f(1))g(z),
\end{align}
where the term in the last line is of (at least) $\mathcal{O}(\delta_s)$ and can be neglected.
Eq.~\eqref{eq:sdPDFs} then reads
\begin{align}
\label{eq:sdPDFs2}
  f_{i/p} (x, \mu_F) \approx & f_{i/p}(x)\left [ 1  -   \frac{\alpha_s}{2 \pi} \frac{\Gamma(1-\epsilon)}{\Gamma(1-2\epsilon)} \left(\frac{4 \pi \mu_R^2}{\mu_F^2} \right)^\epsilon \frac{A_1^{\text{sc}, j \to i}}{\epsilon} 
  % + \mathcal{O}(\delta_s) 
  \right] 
  \nonumber \\
  & - \frac{1}{\epsilon} \frac{\alpha_s}{2 \pi} \frac{\Gamma(1-\epsilon)}{\Gamma(1-2\epsilon)} \left(\frac{4 \pi \mu_R^2}{\mu_F^2} \right)^\epsilon \int_x^{1-\delta_s \delta_{ij}} \frac{d z}{z} P_{ij}(z) f_{j/p} (x/z),
\end{align}
where now the unregularised, four dimensional AP splitting kernels appear and the soft-collinear factors $A_1^\text{sc}$ for the splittings with a soft gluon $(g)$ are given by
\begin{align}
  A_1^{\text{sc}, q \to q (g)} & = C_F ( 2 \ln \delta_s + 3/2 ) ,\\
  A_1^{\text{sc}, g \to g (g)} & = 2 N \ln \delta_s + (11 N - 2 (n_f - 1))/6 .
\end{align}
As the integral for $i \neq j$ extends up to 1, $A_1^{\text{sc}}$ is by definition 0 in that case. 

Solving eq.~\eqref{eq:sdPDFs} for $f(x)$ in the lowest order in $\alpha_s$ and convolving with the Born cross section gives 
\begin{multline}
\label{eq:PDFconv}
\frac{d \sigma^{\text{PDF}}}{d x_1 d x_2} = \sum_{ij} 
%\frac{2}{1+\delta_{ij}} 
\hat \sigma^B_{ij}  \bigg \{ f_{i/p} (x_1, \mu_F) f_{j/p} (x_2, \mu_F) \\
\cdot \left (1 
+ \frac{\alpha_s}{2 \pi} \frac{\Gamma(1-\epsilon)}{\Gamma(1-2\epsilon)} \left(\frac{4 \pi \mu_R^2}{\mu_F^2} \right)^\epsilon \sum_k \frac{A^{\text{sc}, k \to i}+A^{\text{sc}, k \to j}}{\epsilon}  \right) 
  \\ 
 + \sum_k 
\frac{\alpha_s}{2 \pi} \frac{\Gamma(1-\epsilon)}{\Gamma(1-2\epsilon)} \left(\frac{4 \pi \mu_R^2}{\mu_F^2} \right)^\epsilon \frac{1}{\epsilon}
\left [ \int_{x_1}^{1-\delta_s \delta_{ik}} \frac{d z}{z}  f_{k/p} \left( \frac{x_1}{z}, \mu_F \right )  f_{j/p} \left( x_2, \mu_F \right ) 
P_{ik}(z) \right. \\
 \left. \left. + (x_1, i) \leftrightarrow (x_2, j) \right ]  \right \}.
\end{multline}
The first term is just the Born partonic cross section convolved with the scale-dependent PDFs.
The terms $A_{\text{sc}}$ cancel out with eqs. \eqref{eq:cos1} and \eqref{eq:cos2}.
Combining now eqs. \eqref{eq:PDFconv2} and \eqref{eq:PDFconv} gives, together with  the LO cross section, a final result for the hard-collinear part
\begin{multline}
   \frac{d \sigma^{\text{HC+PDF}}}{d x_1 d x_2} 
%   = 
%   \sum_{ij} 
   %\frac{2}{1+\delta_{ij}}  
%   \hat \sigma^B_{ij} \left \{ \frac{\alpha_s}{2 \pi} \frac{\Gamma(1-\epsilon)}{\Gamma(1-2\epsilon)} \left (\frac{4 \pi \mu_R^2}{\hat s} \right)^\epsilon  \frac{1}{\epsilon} \right.
%   \\ 
%   \int_{x_1}^{1-\delta_s} \frac{dz}{z} \left[ \left( \left(\frac{\hat s}{\mu_F^2} \right)^\epsilon 
%   P_{ik}(z) - 
%   \delta_c^{-\epsilon} P_{ik}(z, \epsilon)  \left [\frac{(1-z)^2}{2 z} \right]^{-\epsilon} \right) f_{i/p} (x_1/z, \mu_F) f_{j/p} (x_2, \mu_F) + x_1 \leftrightarrow x_2 \right) \\
   = \sum_{ij} 
   %\frac{2}{1+\delta_{ij}} 
   \hat \sigma^B_{ij} \bigg \{ f_{i/p} (x_1, \mu_F) f_{j/p} (x_2, \mu_F) \\
   \cdot \left (1    +   \frac{\alpha_s}{2 \pi} \frac{\Gamma(1-\epsilon)}{\Gamma(1-2\epsilon)} \left(\frac{4 \pi \mu_R^2}{\mu_F^2} \right)^\epsilon \sum_k \frac{A_1^{\text{sc}, k \to i} + A_1^{\text{sc}, k \to j}}{\epsilon} \right ) \\
   \left. + \frac{\alpha_s}{2\pi} \sum_k \left [\int_{x_1}^{1-\delta_s \delta_{ik}} \frac{dz}{z}   \left(P_{ik}(z) \ln \left( \delta_c \frac{(1-z)^2}{2 z} \frac{\hat s}{\mu_F^2} \right) - P_{ik}'(z)\right) f_{k/p} (x_1/z, \mu_F) f_{j/p} (x_2, \mu_F)  \right. \right . \\
    \left. + (x_1, i) \leftrightarrow (x_2, j)  \bigg ] \right \}.
\end{multline}

In appendix~\ref{sec:pole_cancelation} we verify the cancellation of single poles for all partonic processes considered in this work at one phase space point.

%********************************************************************************************************
\subsection{Method 2: FKS subtraction}
As an alternative approach for the calculation of the real corrections
we make use of the FKS subtraction scheme.
Using this subtraction method, suitable subtraction terms for individual soft and collinear singularities in the squared matrix elements are constructed which allow
for a convergent numerical integration over the phase space.

This scheme is implemented as automatised method \texttt{MadFKS}~\cite{Frederix:2009yq} in
the Monte Carlo program \texttt{MadGraph5\_aMC@NLO}~\cite{Alwall:2011uj,Alwall:2014hca} (\MG).
The application of the FKS scheme is only dependent on QCD specifics
 and has been used for many different previous calculations.
The program is well tested by applications to the 
SM~\cite{Frixione:2014qaa,Wiesemann:2014ioa,Maltoni:2015ena} and
to a multitude of BSM models including models with SUSY~\cite{Degrande:2015vaa},
extra dimensions~\cite{Das:2014tva}, leptoquarks~\cite{Mandal:2015lca} 
and dark matter candidates~\cite{Neubert:2015fka,Backovic:2015soa} as well as
the Two-Higgs-Doublet Model~\cite{Hespel:2014sla,Degrande:2015vpa} and the Georgi-Machacek model~\cite{Degrande:2015xnm}.

An interface between \MG\ and \texttt{GoSAM} for SM calculations 
is already implemented~\cite{vanDeurzen:2015cga} as specified with
BLHA1~\cite{Binoth:2010xt}. To allow for calculations in BSM models we extend
this to the BLHA2~\cite{Alioli:2013nda} conventions. 
The appropriate changes are summarised in appendix~\ref{app:mg-gosam}.\footnote{
For another recent application of \texttt{GoSam} with \texttt{MadFKS} see ref.~\cite{Pruna:2016spf}.}
The correctness of the implementation has been tested thoroughly
by ensuring that all appearing divergences cancel between
the virtual and real part for all subprocesses in
various regions of phase space.

%********************************************************************************************************
\subsection{Comparison of TCPSS and FKS subtraction}
\begin{figure}
	\subfloat[]{\includegraphics[width=0.49\textwidth]{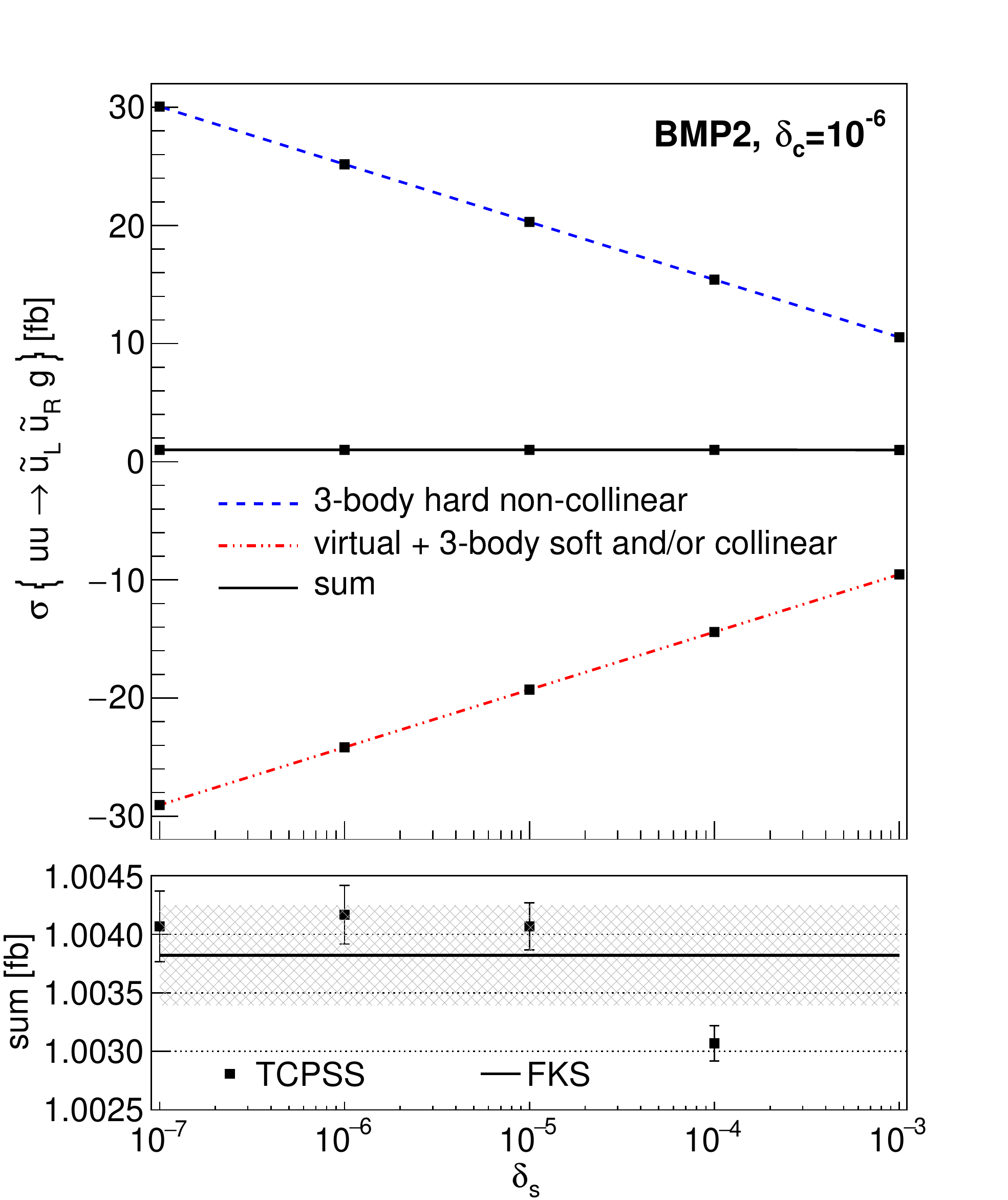} \label{img:cutcancel1}}
	\subfloat[]{\includegraphics[width=0.49\textwidth]{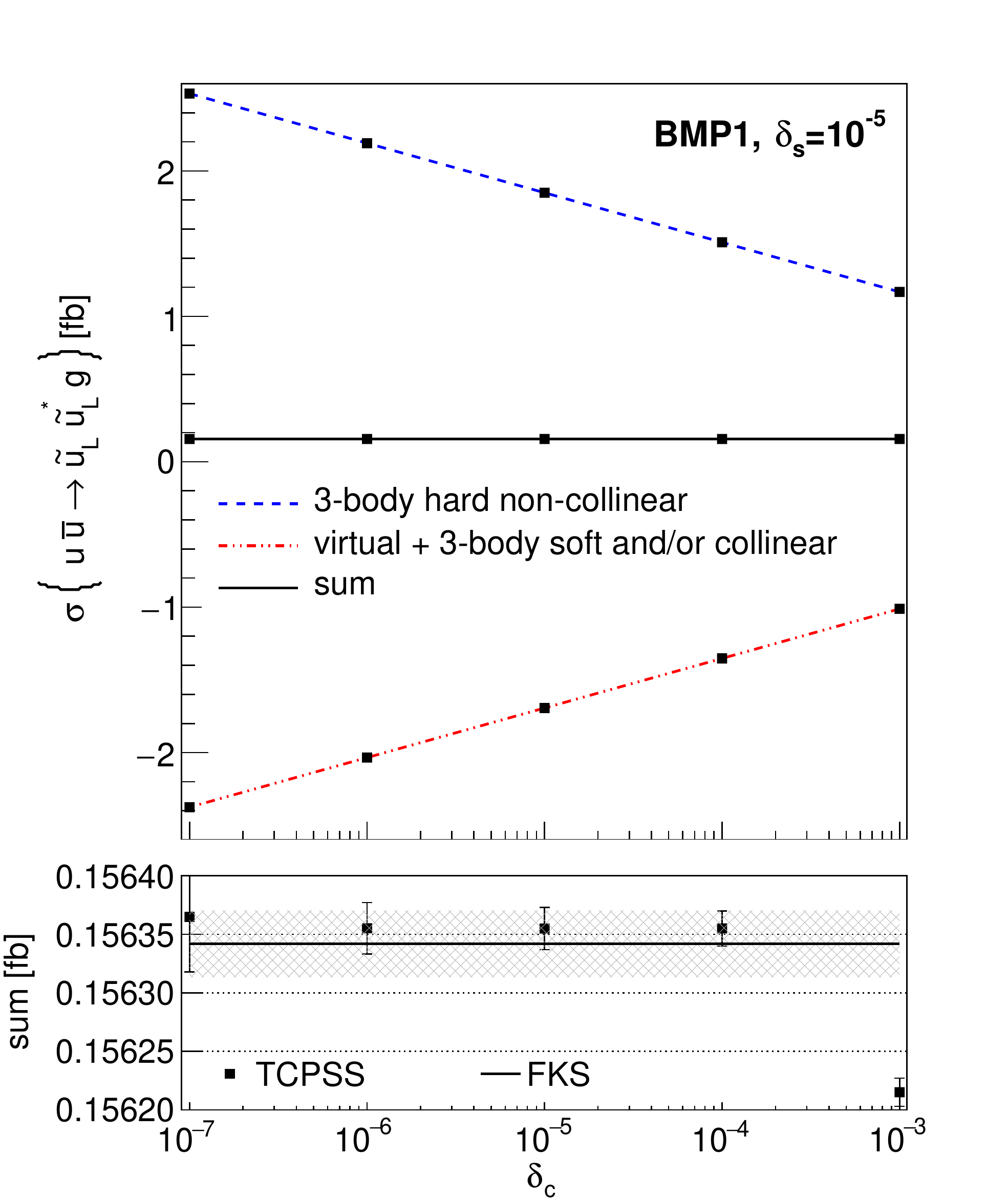} \label{img:cutcancel2}}
	\caption{
		Proof of cut parameter cancellation for the two selected subprocesses \newline
		 $uu \to \tilde{u}_L \tilde{u}_R g$ and $u\bar{u} \to \tilde{u}_L \tilde{u}_L^\dagger g$. 
		Given errors correspond to 68\% CL.
		\label{img:cutcancel}
	}
\end{figure}

In TCPSS, the dependence on the (unphysical) regulators $\delta_s$ and $\delta_c$ should vanish after adding the hard non-collinear emissions.
We verified extensively that this is indeed the case.
Examples for the cancellation of the cut parameter dependence are shown in figure~\ref{img:cutcancel} using \texttt{MMHT2014nlo68cl} PDFs \cite{Harland-Lang:2014zoa} interfaced via \texttt{LHAPDF6}~\cite{Buckley:2014ana}.
In figure~\ref{img:cutcancel1} we plot the hard non-collinear (blue) and (summed) virtual, soft and collinear parts (red) in function of the phase space slicing parameter $\delta_s$ for the $u u \to \tilde u_L \tilde u_R$ channel for BMP2 and $\delta_c = 10^{-6}$.
In the bottom subplot, the sum is then compared with the FKS result ( with a shaded band stating the statistical uncertainty for the integration given
by \MG), showing agreement within uncertainties.\footnote{We add the virtual part to the soft and collinear result as, compared to the \MG\ framework, some terms are shifted from the virtual to the soft part due to the choice of $\mathcal{O}(\epsilon)$ prefactors.}
Figure~\ref{img:cutcancel2} shows the same for the $u \bar u \to \tilde u_L \tilde u_L^\dagger$ process and BMP1 as a function of the $\delta_c$.
The comparison with the FKS subtraction reveals that a relative accuracy of $10^{-4}$ requires the cut parameters $\delta_s = 10^{-5}$ and $\delta_c = 10^{-6}$, which is what we use in our numerical analyses.

As the TCPSS method relies on a partial cancellation of two large contributions 
at the level of an integrated cross sections, it is inherently slower than 
local subtraction schemes like FKS.
For the case at hand, this means an order of magnitude slow down for the same final precision.
This directly translates into an order of magnitude speed difference between 
the standalone \texttt{C++} code and the \MG\ as for the 
standalone calculation most time is spend evaluating the $2 \to 3$ hard non-collinear part.

\subsection{Treatment of resonances in real emission diagrams using diagram removal}
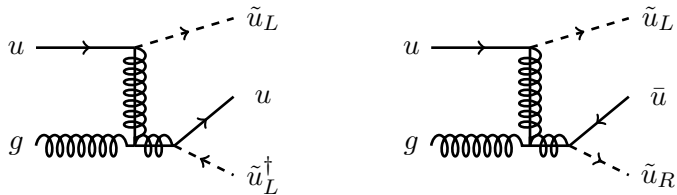
\begin{figure}
\begin{center}
\begin{tikzpicture}[line width=1.0 pt, scale=1.3, arrow/.style={thick,->,shorten >=2pt,shorten <=2pt,>=stealth}]
    \node at (0,0) {};
\begin{scope}[shift={(0,-1)}]
	\draw[fermion] (-1,0.5)--(0,0.5);
	\draw[gluon] (-1,-0.5)--(0,-0.5);
	\draw[gluon] (0,0.5)--(0,-0.5);
	\draw[fermionnoarrow] (0,0.5)--(0,-0.5);
	\draw[gluon] (0,-0.5)--(0.4,-0.5);
	\draw[fermionnoarrow] (0,-0.5)--(0.4,-0.5);
	\draw[scalar] (0,0.5)--(1,0.8);
	\draw[fermion] (0.4,-0.5)--(1,0.0);
	\draw[scalarbar] (0.4,-0.5)--(1,-0.8);
	\node at (-1.2,0.5) {$u$};
	\node at (-1.2,-0.5) {$g$};
	\node at (1.3,0.8) {$\tilde{u}_L$};
	\node at (1.3,0.0) {$u$};
	\node at (1.3,-0.8) {$\tilde{u}^\dagger_L$};
\begin{scope}[shift={(4,0)}]
	\draw[fermion] (-1,0.5)--(0,0.5);
	\draw[gluon] (-1,-0.5)--(0,-0.5);
	\node at (-1.2,0.5) {$u$};
	\node at (-1.2,-0.5) {$g$};
	\draw[gluon] (0,0.5)--(0,-0.5);
	\draw[fermionnoarrow] (0,0.5)--(0,-0.5);
	\draw[gluon] (0,-0.5)--(0.4,-0.5);
	\draw[fermionnoarrow] (0,-0.5)--(0.4,-0.5);
	\draw[scalar] (0,0.5)--(1,0.8);
	\draw[fermionbar] (0.4,-0.5)--(1,0.0);
	\draw[scalar] (0.4,-0.5)--(1,-0.8);
	\node at (1.3,0.8) {$\tilde{u}_L$};
	\node at (1.3,0.0) {$\bar u$};
	\node at (1.3,-0.8) {$\tilde{u}_R$};
\end{scope}
\end{scope}
\end{tikzpicture}
\end{center}	
\caption{
	Example of real emission Feynman diagrams with s-channel gluinos for both final states: $gu \to \tilde u_L \tilde u^\dagger_L u$ and $gu \to \tilde u_L \tilde u_R \bar u$.
	\label{fig:DS_diagrams}
	}
\end{figure}

Certain real corrections to the considered processes may contain Feynman diagrams with an intermediate massive state in an s-channel. 
As examples consider diagrams shown in figure~\ref{fig:DS_diagrams} where an intermediate s-channel gluino appears.
From a practical point of view, the region of phase space where the gluino goes on-shell should rather be classified as a Born-level gluino-squark production followed by a subsequent decay of the gluino.
This feature is not exclusive to BSM models as a similar issue appears in the case of the real corrections to the SM $tW$ production (see e.g. ref.~\cite{Frixione:2008yi}), which contain resonant contributions from the top pair production. 

A popular way of dealing with this problem is the diagram removal technique.
In this approach one either completely discards all resonant diagrams 
at the level of the amplitude or their square at the level of the squared matrix element.
In this work we employ the first of those solutions.%
\footnote{
An alternative approach would be to employ the diagram subtraction method (sometimes also referred to as the \texttt{Prospino} scheme~\cite{Beenakker:1996ed}).
We have checked that switching to this choice changes the total cross sections at a percent level.
We therefore postpone the detailed studies of this method to the publication documenting the \texttt{RSymSQCD} code.
}
This version of the DR approach is also the default way of dealing with resonant divergences when using \texttt{MadFKS} in \MG.  

The removal of a subset of Feynman diagrams in general violates gauge invariance and care has to be taken to ensure that the effect is numerically small.
For the case of the MSSM this has been studied in depth in refs.~\cite{Gavin:2013kga,Gavin:2014yga} (see also refs.~\cite{Hollik:2012rc,GoncalvesNetto:2012yt} for other implementations).
Also, it requires a careful choice of a gauge not to spoil the factorisation of collinear singularities. 

To understand this last point, consider the gluon splitting $g (p_1) \to \bar q^* (p_1-p_5) q (p_5)$ connected to some bigger amplitude through $\bar q$.
We use the $*$-symbol to emphasise that $\bar q$ is in general not on-shell.
The final state quark momentum $p_5$ can be parametrised through Sudakov decomposition as
\begin{align}
  \label{eq:sudakov_decomposition}
  p_5^\mu = (1-z) p_1^\mu - \frac{p_{5,\perp}^2}{1-z} \frac{n^\mu}{2 p_1 \cdot n} - p_{5,\perp}^\mu ,
\end{align}
where $n^\mu$ is a reference null vector, $1-z$ is the fraction of gluon energy carried by the quark, and $p_{5,\perp} \cdot p_1 = p_{5,\perp} \cdot n = 0$.
In the limit of $p_{5,\perp} \to 0$ the vectors $p_1$ and $p_5$ become spatially parallel and diagrams with this splitting develop a $(p_1-p_5)^{-2} = (1-z)/p_{5,\perp}^{2}$ singularity.
Due to the helicity conservation, physical gluons cannot decay to a pair 
of on-shell quarks.
Therefore, in the physical gauge the matrix element will always have additional power of $p_{5,\perp}$ in the numerator.
As the one particle phase space for radiation of $q$ is given by $d\Phi_1 \sim d p_{5,\perp}^2$, the collinear singularities of the interference terms are integrable since those 
terms scale as $d p_{5,\perp}^2/p_{5,\perp}$.
This is no longer true in an unphysical gauge, where longitudinally polarised gluons might appear. %in the final state.
For the full amplitude with a $ug$ initial state, which is gauge invariant, longitudinal states decouple through a Ward identity as no triple gluon vertices appear.
This is no longer true after the removal of resonant diagrams, though.

\begin{figure}
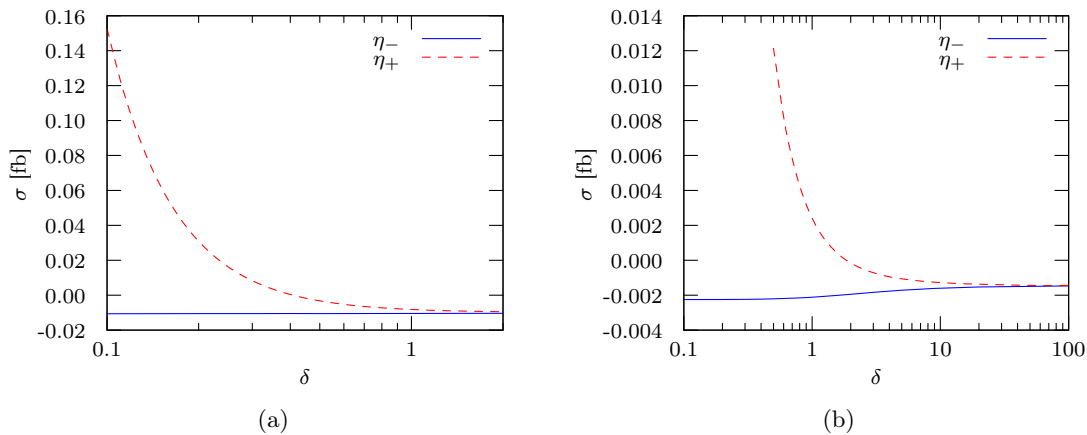

    \centering
    \subfloat[]{
      \input{img/gu_suLsuLdaggeru_DR.tex} \label{fig:DR_gauge_dep1}}
    \subfloat[]{
      \input{img/gu_suLsuRubar_DR.tex} \label{fig:DR_gauge_dep2}}
    \caption[zzz]{Gauge dependence of a diagram removed $g q \to \tilde u_L \tilde u_L^\dagger q$ with $q = u, \bar u$ (a) and $g u \to \tilde u_L \tilde u_R \bar u$ (b) matrix elements. 
    See eqs. \eqref{eq:eta-} and \eqref{eq:eta+} for the definition of $\eta_-$ and $\eta_+$ gauges.\label{fig:DR_gauge_dep}
    }
\end{figure}

We therefore calculate the real emission matrix elements in the light cone gauge.
A convenient choice of the gauge-fixing vector $\eta$ is the momentum of the other incoming parton $p_2$.
This choice allows to avoid spurious divergences present in the polarisation sum as $p_1 \cdot p_2 = \hat s/2$.\footnote{This choice is also useful if one wants to decouple interference terms between different collinear limits which, however, doesn't occur for this subprocess~\cite{Gardi:2001di}.}
To study the numerical impact of this gauge choice we consider two
families of gauge vectors (assuming the momenta $p_{1,2}$ oriented
in the $\pm$z-direction)
\begin{align}
  \eta_- \equiv & (\sqrt{1+\delta^2}, 0, \delta, -1) \label{eq:eta-},\\
  \eta_+ \equiv & (\sqrt{1+\delta^2}, 0, \delta, +1) \label{eq:eta+}.
\end{align}
For $\delta = 0$ the $\eta_-$ choice is equivalent to the choice of $\eta = p_2$, while the $\eta_+$ choice is singular as $p_1 \cdot \eta_+ = 0$.
In the limit of $\delta \gg 1$ results of both gauge choice converge as $\eta_+ \simeq \eta_-$.
Both of these features are clearly seen in figure~\ref{fig:DR_gauge_dep}, where we plot the cross sections for $g q \to \tilde u_L \tilde u_L^\dagger q$ with $q = u, \bar u$ (\ref{fig:DR_gauge_dep1}) and $g u \to \tilde u_L \tilde u_R \bar u$ (\ref{fig:DR_gauge_dep2}) in function of the $\delta$ parameter for BMP2.
For $gu \to \tilde u_L \tilde u_R \bar u$ the gauge dependence is enhanced since resonant diagrams were giving a substantial contribution to this amplitude.

We point out, that keeping the choice of the non-singular gauge $\eta_-$, the DR subtracted processes give per mille level contribution to the total cross section.
Hence, for all intents and purposes, the gauge dependence is not a problem and for studies in the following sections we use the customary choice of $\eta = p_2$.

\section{Results}
When the virtual corrections of section~\ref{sec:virt} are combined with the real 
contributions from section~\ref{sec:real} all IR divergences cancel as expected
and we are able to calculate the NLO SUSY-QCD cross section for $\tilde{q}\tilde{q}$ and $\tilde{q}\tilde{q}^\dagger$ production 
in the MRSSM.
In this section we first point out and explain the consequences of R-symmetry on NLO 
cross sections in supersymmetric QCD before exhibiting 
the results of our calculations for $\tilde{q}\tilde{q}$ and $\tilde{q}\tilde{q}^\dagger$ production. 
We conclude with remarks on the scale dependence and uncertainties of the computed 
observables.\footnote{If not noted differently, all MRSSM results
in this section are produced using method 2 and FKS subtraction including
diagram removal for possible on-shell resonances. The MSSM cross sections were calculated with \MG\ %\texttt{MadGraph5\_aMC@NLO} 
using the \texttt{UFO} model provided with ref.~\cite{Degrande:2015vaa}. PDF sets
were accessed using \texttt{LHAPDF6}.}

A useful quantity which helps to understand these different aspects
is the K-factor.
We define a K-factor $K$ as ratio of NLO over LO total cross section following 
ref.~\cite{Beenakker:1996ch}:
\begin{equation}
K(\text{NLO/LO}) = \frac{\sigma_{\text{NLO}}}{\sigma_{\text{LO}}(NLO/LO PDF)}\,,
\label{eq:kfac}
\end{equation}
where $\sigma_{\text{NLO}}$ is calculate using NLO PDFs and $\sigma_{\text{LO}}$ 
using NLO or LO PDFs, depending on the argument.
If no argument is given we use LO PDF sets as default case for $\sigma_{\text{LO}}$. 
The K-factor depends on the
process, masses of the fields in the model, and the centre-of-mass energy, as
well as the choice of renormalisation and factorisation scale. In the following, 
we set $\sqrt{S} = \unit[13]{TeV}$ and $\mu_R = \mu_F = \frac{m_1 + m_2}{2}$, 
where $m_i$ are the final state's particle masses.
%the last three parameters are chosen like stated in the caption of %fig~\ref{fig:tree_msq_1}.
For a consistent definition we identify the values of the Dirac gluino mass in the
MRSSM with the Majorana gluino mass in the MSSM, the values of
the squark masses in both models, as well as neglecting left-right squark mixing
in the MSSM.
In the MRSSM the K-factor has an additional dependency on the sgluon mass parameter.
To fix all remaining SM parameters, we set the top quark mass to  $m_t=172$~GeV and take 
the strong coupling constant $\alpha^{\overline{\text{MS}}}_s (m_Z)$ from the used 
PDF set. %If not denoted differently, we set the renormalisation and 
%factorisation scale equal to squark mass.

\subsection{Effects of R-symmetry}
The results of R-symmetry at NLO may be pinned down to two features: the presence of a 
Dirac instead of Majorana gluino and the existence of sgluons.
The conservation of R-charge, already discussed at LO in section~\ref{sec:tree-level}, also 
needs to be commented on when comparing K-factors to the MSSM.
The effect of R-symmetry is however only present in the virtual corrections, i.e. the real corrections do not differ in the MSSM and MRSSM.

The effects discussed in the following comprise the unique features of MRSSM neglected
in the study of ref.~\cite{Kribs:2012gx} by only taking MSSM NLO K-factors into account.
\begin{figure}[ht]
\begin{minipage}[t]{0.66\textwidth}
\includegraphics[width=0.49\textwidth]{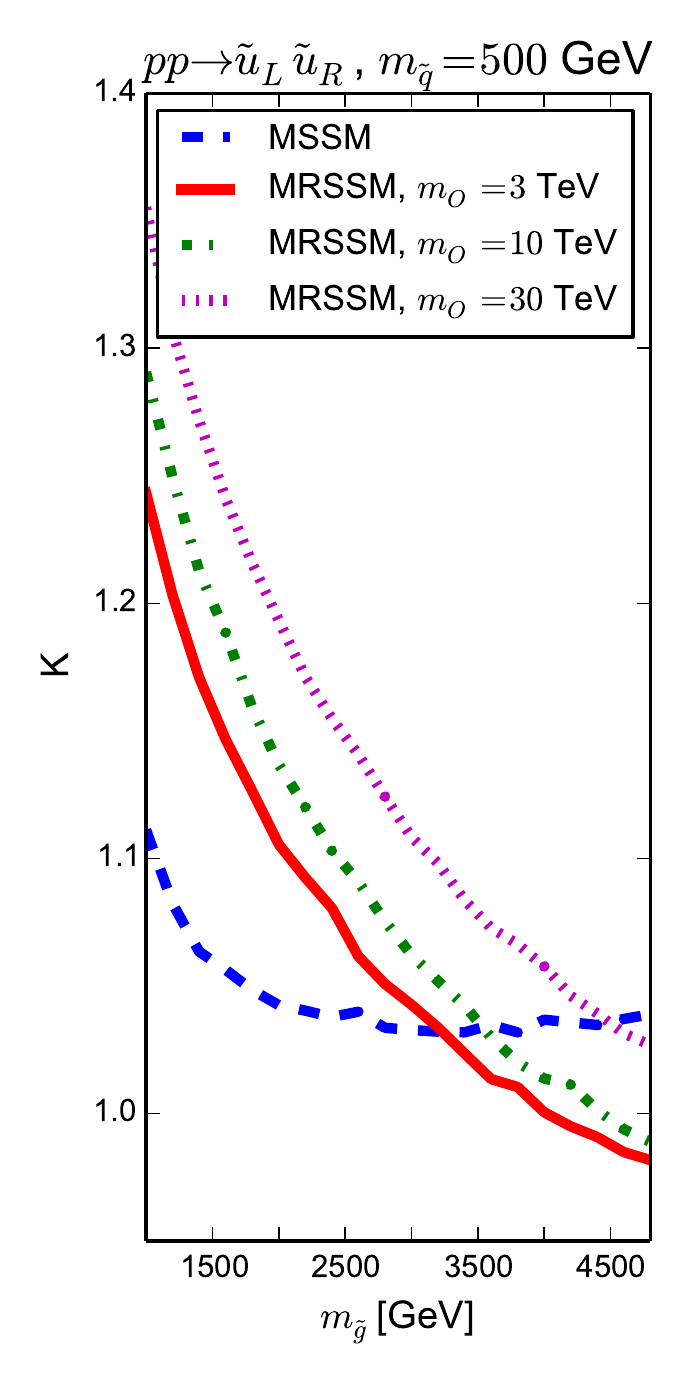}
\includegraphics[width=0.49\textwidth]{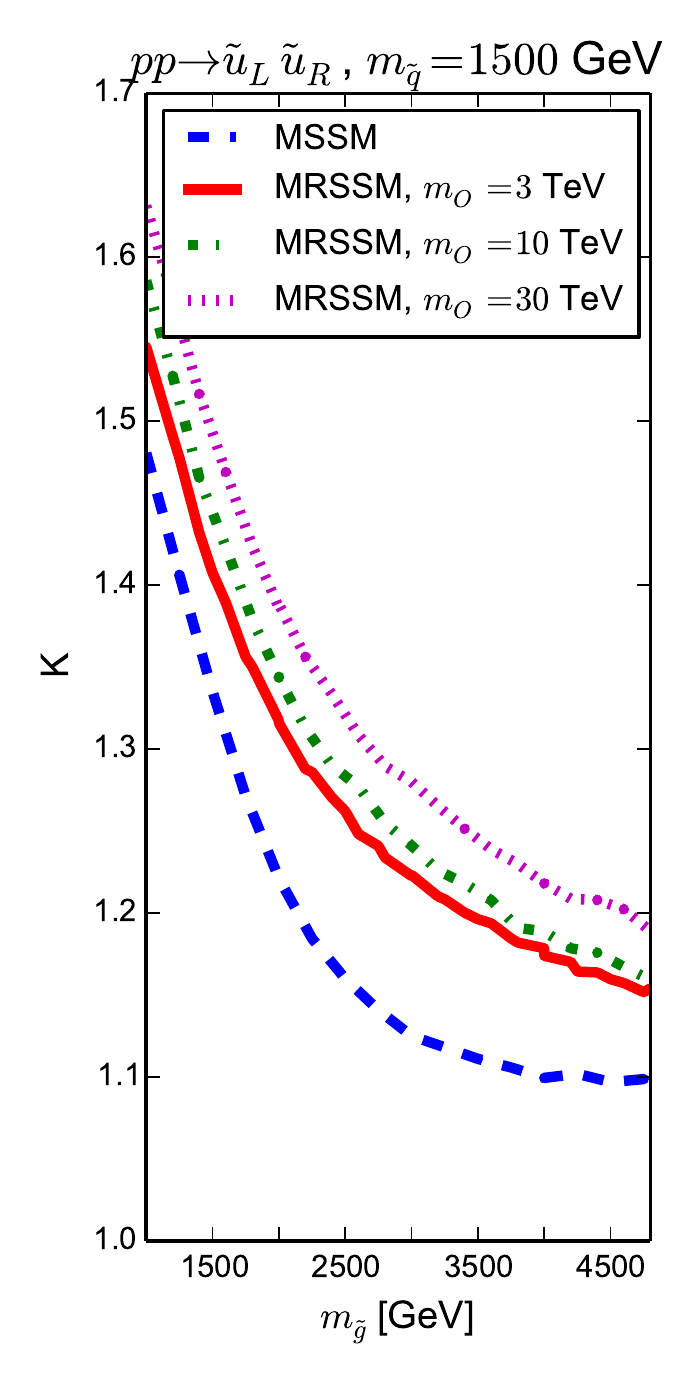}
\caption{Comparison of K-factor dependence on the gluino mass between the 
MSSM and the MRSSM at different octet masses shown for two different squark masses.}
\label{fig:mrssmeffects}
\end{minipage}
\begin{minipage}[t]{0.33\textwidth}
\includegraphics[width=\textwidth]{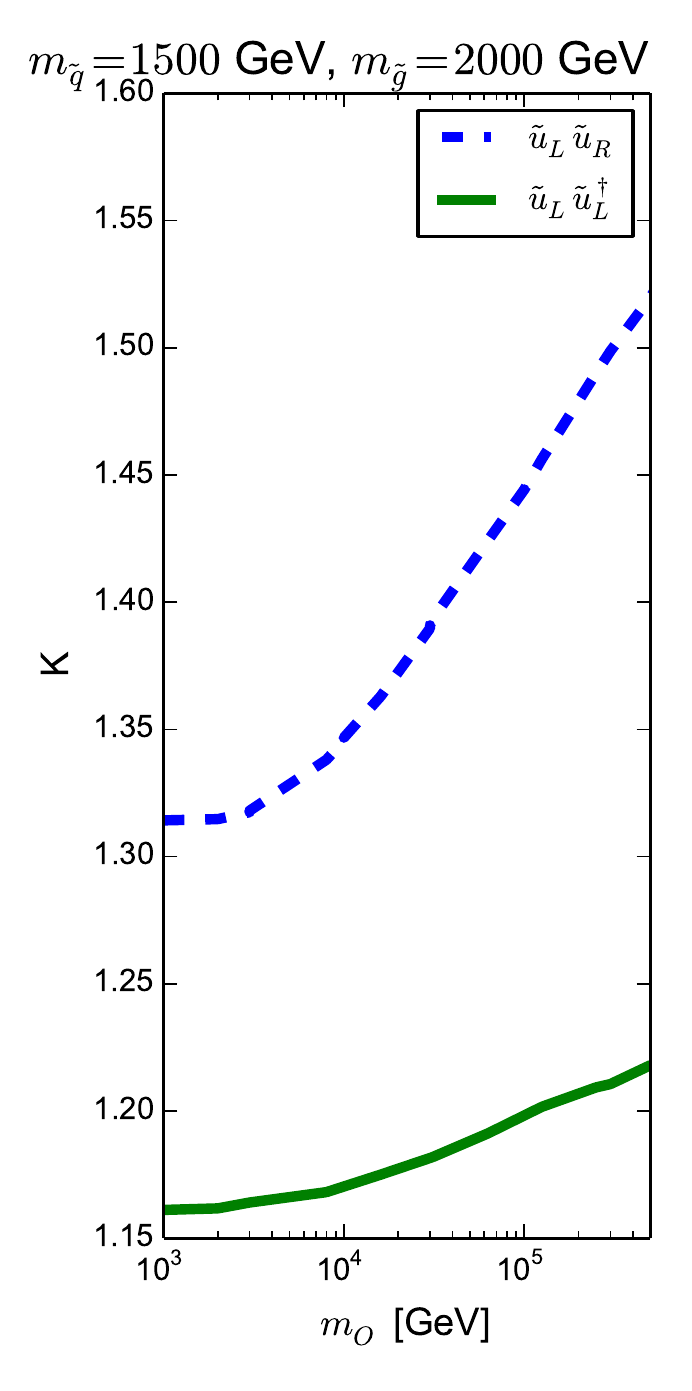}
\caption{Dependence of the K-factor on the sgluon mass for $\tilde{q}\tilde{q}$ and $\tilde{q}\tilde{q}^\dagger$ production in the MRSSM.}
\label{fig:sgluoneffects}
\end{minipage}
\end{figure}

\subsubsection{Dirac nature of the gluino}
In the presentation of renormalisation constants in 
section~\ref{sec:RenConst}, we already saw ramifications of the Dirac gluino 
whose components couple differently. Now, we study the effect of replacing 
the Majorana with a Dirac gluino in a physical process. 

To this end, consider diagram 1a) and 1b) (referring to the first diagram of category a)
and b), respectively) of 
Figure~\ref{fig:oneloopdiagrams}. The former contributes in both models, 
whereas the latter is only non-zero in the MSSM, since the gluino 
undergoes a chirality flip to its non-coupling right-handed component.
Taking only objects with spinor indices into 
account these diagrams evaluate to 
\begin{align}
\mathcal{M}_{1a)} &\propto \bar v(p_2) P_R (\slashed t+m_{\tilde g})P_L \slashed q P_R(\slashed t
+m_{\tilde g})P_L u(p_1)= \bar v(p_2)\slashed t \slashed q \slashed t P_L u(p_1),\\
\mathcal{M}_{1b)} &\propto \bar v(p_2) P_R (\slashed t+m_{\tilde g})P_R \slashed q P_L(\slashed t
+m_{\tilde g})P_L u(p_1)= m_{\tilde g}^2\bar v(p_2)  \slashed q P_L u(p_1).
\end{align}
Here $q^\mu$ is the quark's loop momentum and
$m_{\tilde g}$ is the Majorana gluino mass of the MSSM, which is zero
in the MRSSM. Hence diagram 1a) is the same in both models, while
diagram 1b) vanishes in the MRSSM. Alternatively, diagram 1b) can also be 
understood to be zero in the MRSSM as R-charge is not conserved at all vertices of
the diagram as the R-charge of the gluino flows in the opposite direction
than the one of the squark appearing in its self energy insertion.

On similar grounds we are able to understand why diagram 2b) in figure~\ref{fig:oneloopdiagrams} does not contribute in the MRSSM. Its 
matrix element has the following proportionality
\begin{align}
\mathcal{M}_{2b)} \propto \bar v P_R (\slashed t+m_{\tilde g})P_R \slashed q 
P_L(\slashed q+\slashed p_3+m_{\tilde g})P_L u=m_{\tilde g}^2\bar v\slashed q P_L u
\end{align}
in the MSSM and is therefore zero in the MRSSM. Note that its analogue, 
depicted as diagram 2a) in figure~\ref{fig:oneloopdiagrams} 
is proportional to the $u$-quark mass and therefore zero in both models.
Diagram 3b) of figure~\ref{fig:oneloopdiagrams} appears
only for $\tilde{q}\tilde{q}^\dagger$ production and is proportional to the
Majorana mass of the MSSM squared for the same reasoning as discussed before.

The number of diagrams affected by the difference between Dirac and Majorana
mass is only a subset of all contributing virtual graphs. 
Therefore, the effect is most apparent when we use the dependency on
the gluino mass and make it large compared to all other appearing scales.
This can be seen when comparing the plots of figure~\ref{fig:mrssmeffects}. 
On the right plot, where the mass scales are
not too different from each other, the MRSSM and MSSM line have a similar
dependency on the gluino mass, where the magnitude of the K-factor
is affected by the sgluon mass as described below. If, however,
the squark mass is reduced, the behaviour changes drastically for large gluino
masses. Then, the additional contributions in the MSSM lead
to a K-factor rising with gluino mass, instead of falling like in the MRSSM.

\subsubsection{Sgluon non-decoupling effects}
\label{sec:sgluon}
The sgluon necessarily appears in the MRSSM as the scalar superpartner 
of the additional component of the Dirac gluino.
The sgluon affects the MRSSM prediction in several ways. Obviously, it
enters the virtual corrections as new matter content, contributing
to the $\beta$ function of the strong coupling such that it becomes zero
at one-loop level. 
Furthermore, the mass of the pseudo-scalar (scalar) component is determined (mostly)
by the SUSY breaking mass parameter $m^2_O$, which is independent
of the gluino mass as can be seen in eq.~\eqref{eq:octetmasses}. 
It is therefore possible that the sgluon can be
much heavier than the gluino and not be directly observable at the LHC.
Still, effects of the sgluon could be experimentally accessible, even if superheavy,
as a mass splitting between superpartners in a SUSY multiplet
leads to non-decoupling, so called super-oblique, contributions~\cite{Cheng:1997sq}.

Super-oblique effects lead to a physical difference between 
the gauge coupling $g_s$ and the gaugino coupling $\hat g_s$ at loop level.
It can be understood in the context of an effective field theory (EFT) where the 
sgluon is integrated out. In the EFT, the sgluon decouples from the gauge boson and 
gaugino self-energies (first two diagrams in category c) of figure~\ref{fig:oneloopcts}) 
which changes the corresponding field renormalisation constants and therefore 
induces a change in the corresponding couplings $g_s$ and $\hat{g}_s$. 
% In an effectice theory where the sgluon is integrated out super-oblique 
% effects lead to a physical difference between 
% the gauge coupling $g_s$ and the gaugino coupling $\hat g_s$ at the loop level
This produces a one-loop difference of
\begin{equation}
\hat g_s - g_s = 
\frac{\alpha_s}{8\pi}\left(\log\frac{m_{O_s}^2}{m_{\tilde g}^2}+
\log\frac{m_{O_p}^2}{m_{\tilde g}^2}\right)
\label{eq:gaugino-gauge}
\end{equation}
in the theory without sgluon.
The production of  $\tilde u_L \tilde u_R$ at the LHC is mediated by the gluino alone
at tree level. Therefore, the super-oblique contribution
from the sgluons can be written
as product of the total tree-level cross section and the difference given
by eq.~\eqref{eq:gaugino-gauge} such that
\begin{equation}
\sigma^{\text{super-oblique part}}_{pp\to \tilde u_L \tilde u_R}=
\frac{\alpha_s}{2\pi}\left(\log\frac{m_{O_s}^2}{m_{\tilde g}^2}+
\log\frac{m_{O_p}^2}{m_{\tilde g}^2}\right)
\sigma^{\text{LO}}_{pp\to \tilde u_L \tilde u_R}\,.
\label{eq:non-decoup}
\end{equation}

The result for $\tilde{q}\tilde{q}^\dagger$ production is more complicated
as the gluino only appears in a subset of the diagrams %with quarks in the initial state 
at tree level.
Hence, the  non-decoupling correction to the cross section
affects only a part of the contributions, not allowing for a simple
factorisation as before.

The sgluon can influence the prediction of the MRSSM in another way, namely via the sgluon-squark-antisquark vertex 
induced by the gluino mass parameter, 
see \circled{9} of the Feynman rules in appendix~\ref{sec:FeynmanRules}.
This leads to additional terms in the renormalisation
constants, stemming from the third Feynman diagram in category c) of
figure~\ref{fig:oneloopcts}, as well as additional virtual corrections 
to the processes, see the last three diagrams of category c) in figure~\ref{fig:oneloopdiagrams}.
We have checked that these contributions are small 
compared to the super-oblique corrections.

In figure~\ref{fig:mrssmeffects}, the effect of super-oblique corrections
is visible when comparing lines with different sgluon masses. As expected,
their spacing follows the logarithmic dependency described 
by eq.~\eqref{eq:non-decoup} for small gluino masses.
When comparing the lines for $m_O=\unit[3]{TeV}$ and $m_O=\unit[10]{TeV}$ at large gluino masses
it can be seen that the differences between the K-factors is reduced, as the assumption that sgluons
can be integrated out is not valid. Then, also ratios of the sgluon masses to the 
Mandelstam variables become relevant.

Figure~\ref{fig:sgluoneffects} shows the effect at very large sgluon masses for both  
$\tilde{q}\tilde{q}$ as well as $\tilde{q}\tilde{q}^\dagger$ production.
For both processes we find the logarithmic enhancement for large sgluon masses,
more prominently for $\tilde{q}\tilde{q}$ production as discussed before.
Super-oblique corrections lead to a difference 
in the K-factors of roughly twenty (five) per cent when the sgluon masses changes
by two orders of magnitude for $\tilde{q}\tilde{q}$ ($\tilde{q}\tilde{q}^\dagger$) production.

\subsubsection{R-charge forbidden processes}\label{sec:apples-oranges}
For $\tilde{q}\tilde{q}$ production we have seen that in the MRSSM only the production
of left- and right-handed squarks together is allowed, while the production of
squarks with the same ``chirality'' is forbidden by R-charge conservation.
This is relevant when comparing the MRSSM K-factor for e.g.\ $pp \to
\tilde{u}_L\tilde{u}_R$ to the MSSM: should we compare it to the MSSM
K-factor for $pp \to \tilde{u}_L\tilde{u}_R$, or to the MSSM K-factor
for total squark production, $pp \to
\tilde{u}_L\tilde{u}_R,\tilde{u}_L\tilde{u}_L,\tilde{u}_R\tilde{u}_R$?
The second K-factor is commonly available via tools like
\texttt{Prospino}~\cite{Beenakker:1996ed} or
\texttt{NNLLfast}~\cite{Beenakker:2016lwe}\footnote{See also ref.~\cite{Beneke:2016kvz}
and comparable public results given via ref.~[1] therein.} and used by the experimental
collaborations for  MSSM analyses.  The first, however, allows a more
direct comparison between the models.

 \begin{figure}[ht!]
\includegraphics[width=0.5\textwidth]{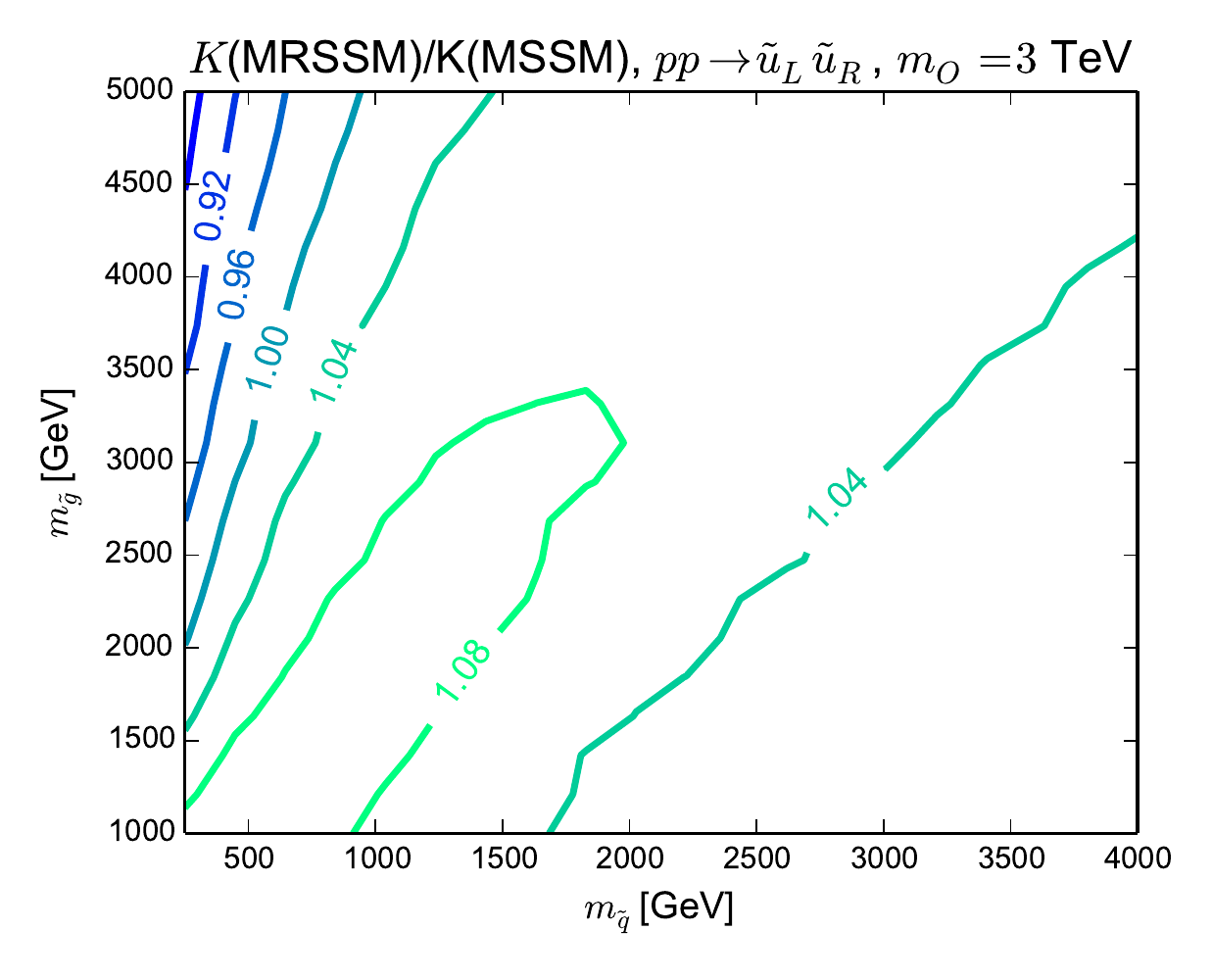}
\includegraphics[width=0.5\textwidth]{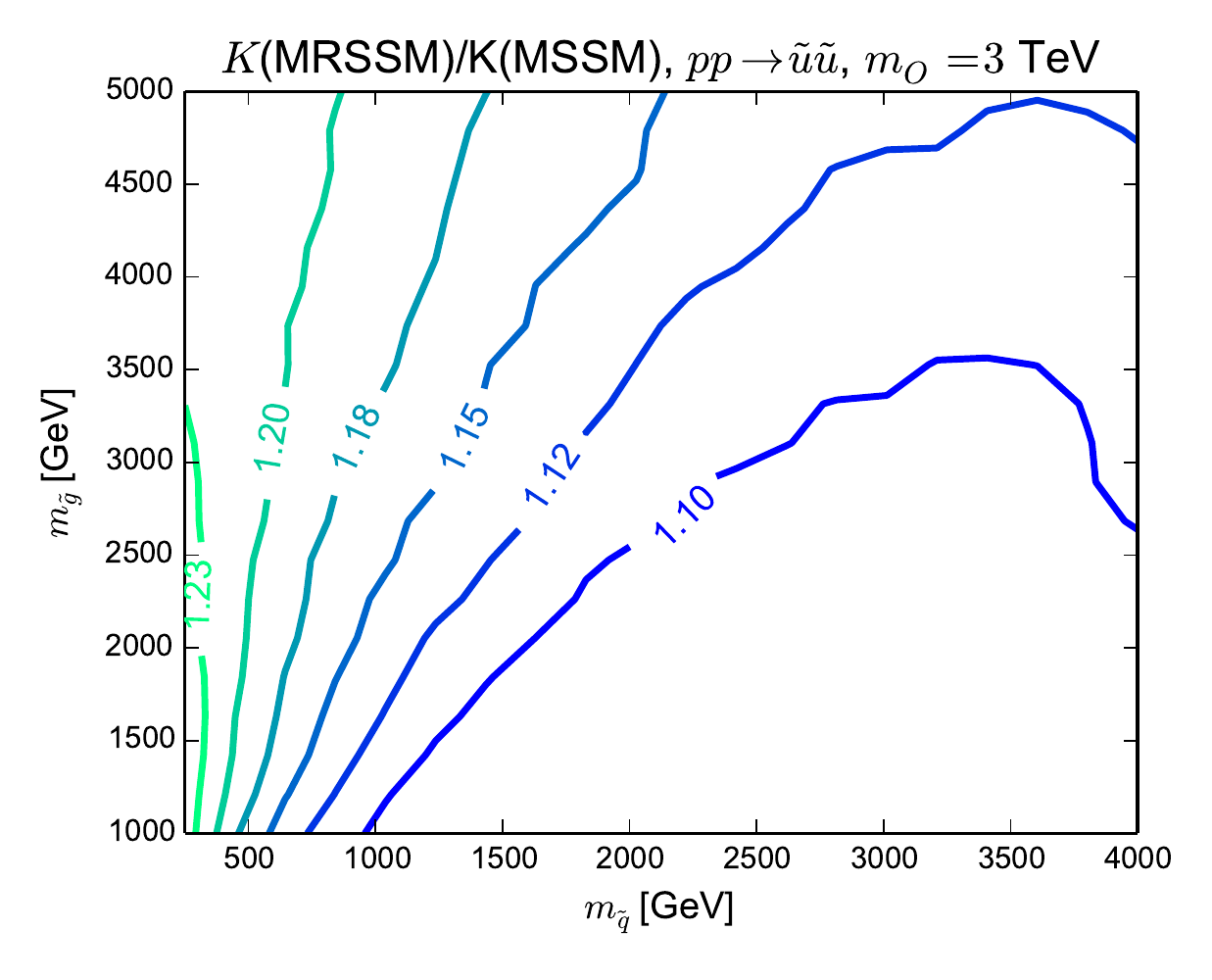}
\caption{Shown is the ratio of MRSSM over MSSM K-factors for $\tilde{q}\tilde{q}$ production for $m_O=3$~TeV. In the MSSM, only left-right $\tilde{q}\tilde{q}$ production (left) or all channels (right) are included.}\label{fig:apples_oranges}
\end{figure}

This point is depicted in figure~\ref{fig:apples_oranges}. The left-hand plot 
compares the ratios of K-factors (MRSSM over MSSM) for the same process, i.e. 
$pp \to \tilde{u}_L\tilde{u}_R$ which is equivalent to the ratio of NLO cross sections. 
It therefore allows to view the effects of the sgluons and Dirac-gluinos at NLO. On the 
other hand, the right-hand plot contrasts the K-factors for full $u$-squark production, 
i.e. $K(\mathrm{MSSM})$ is the usual K-factor including $\tilde{u}_L\tilde{u}_L$ and 
$\tilde{u}_R\tilde{u}_R$ production. Thus, this plot does not solely illustrate NLO
effects but also the conservation of R-charge, already present at LO.

On the left of figure~\ref{fig:apples_oranges} it can be seen that
the sgluon and Dirac mass effects can lead to a difference of the
K-factor of five to eight per cent
between the MRSSM and the MSSM when considering only the $\tilde{u}_L\tilde{u}_R$ 
production. The ratio is smaller than one for $m_{\tilde q}\ll m_{\tilde g}$
and larger than one in the other parts of parameter space with the maximum
at the ratio of $m_{\tilde g}/m_{\tilde q} =2$.
However, when the K-factor for the summed production in the MSSM is used
the ratio deviates significantly from one. Using the usual, 
summed K-factor of the MSSM to estimate NLO MRSSM cross sections from LO ones would consequently 
lead to a systematic underestimate between $10\%$ and $23\%$.
\FloatBarrier

\subsection{Squark production at NLO}
\begin{figure}[ht!]
\includegraphics[width=0.5\textwidth]{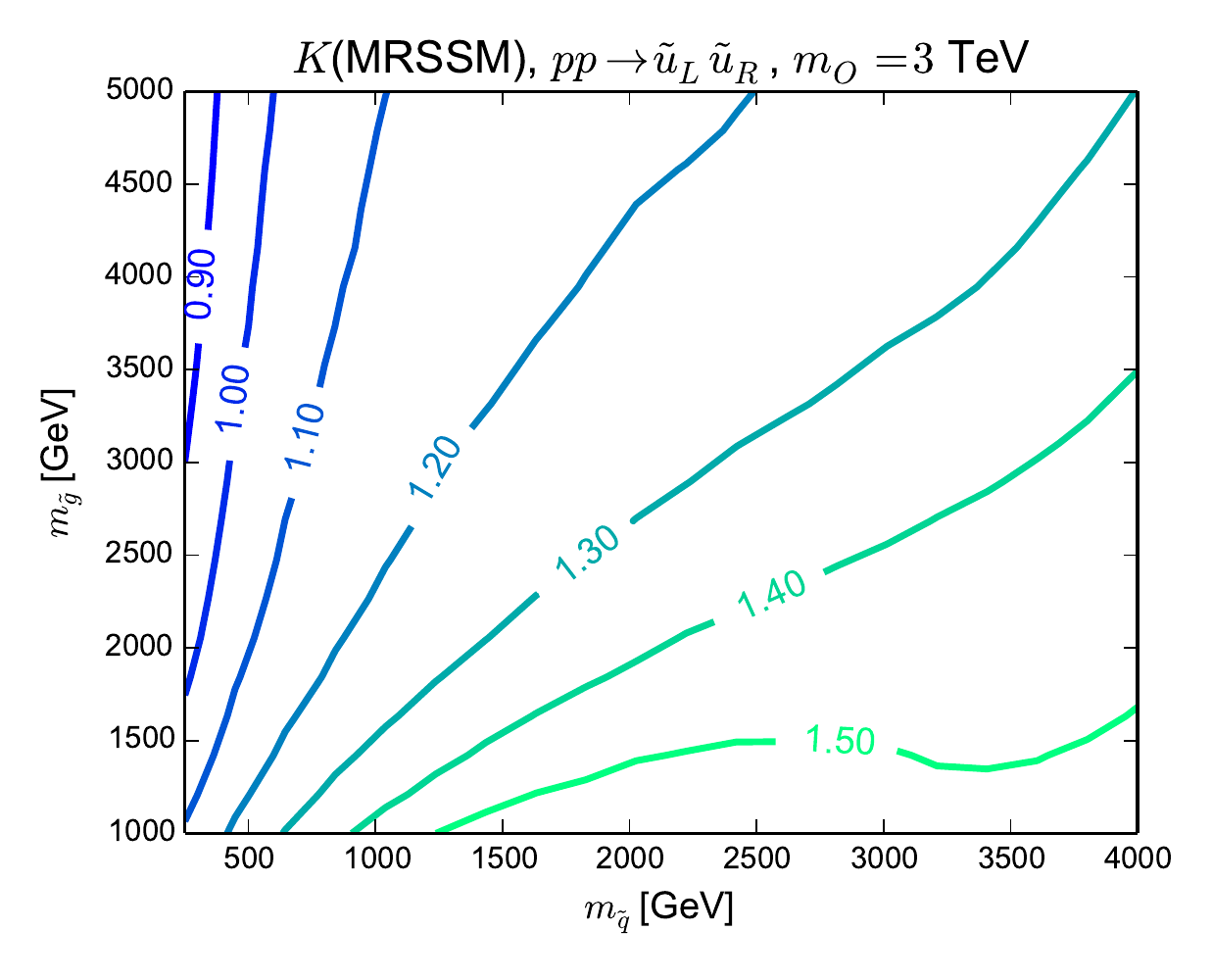}
\includegraphics[width=0.5\textwidth]{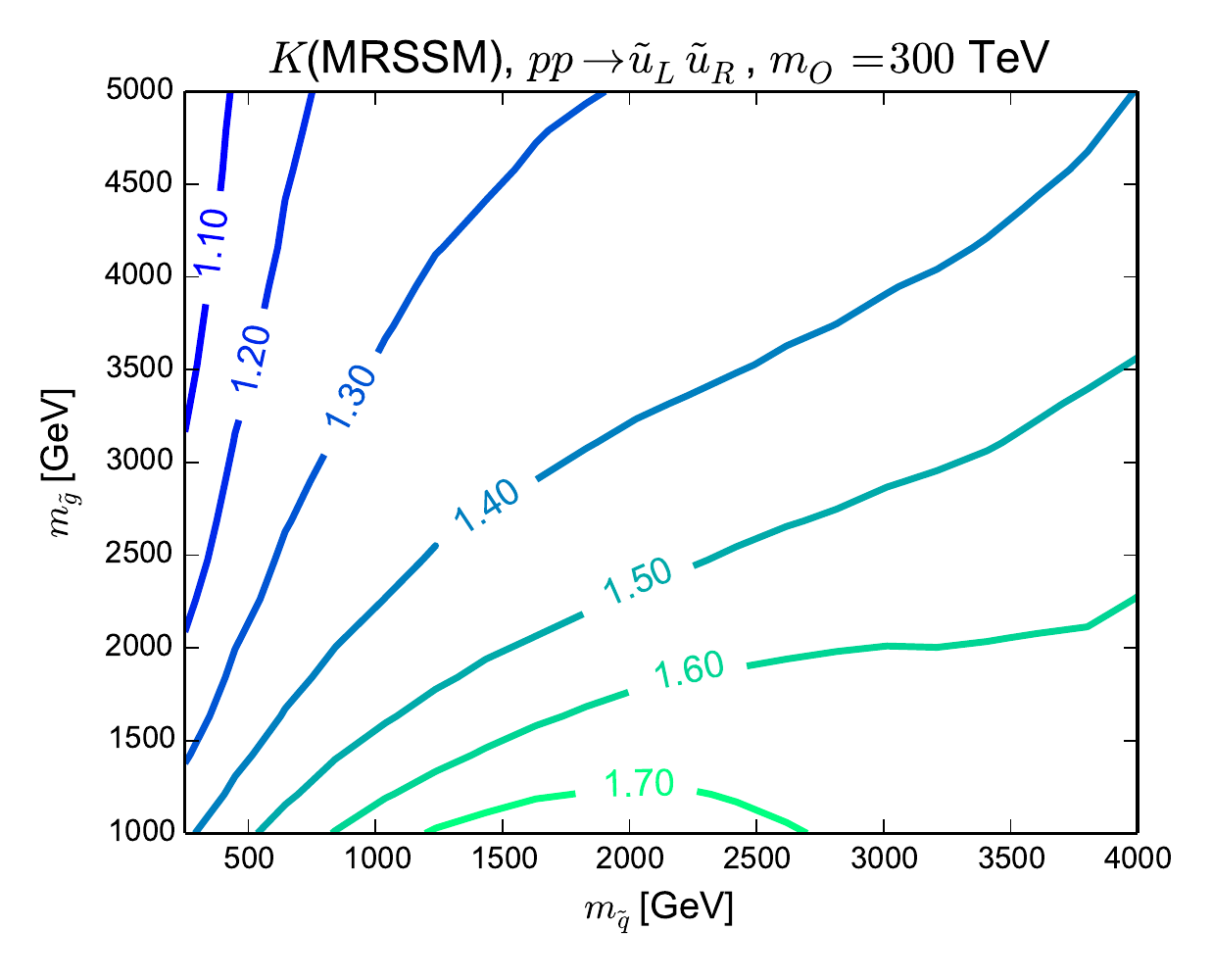}
\caption{Shown are the MRSSM K-factors for $\tilde{u}_L\tilde{u}_R$ production for two different octet masses.}
\label{fig:kfac-sqsq}
\end{figure}

\begin{figure}[ht!]
\includegraphics[width=0.5\textwidth]{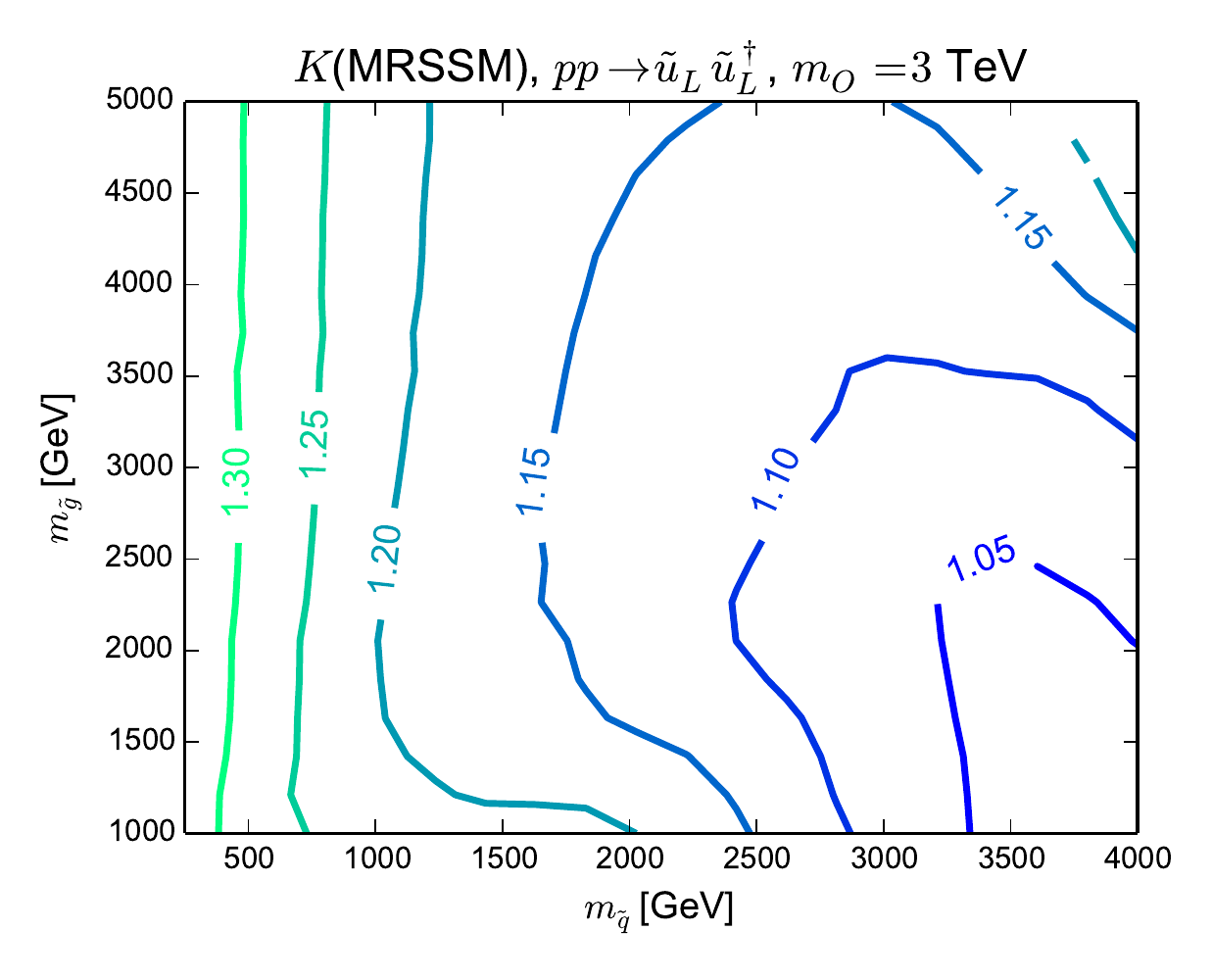}
\includegraphics[width=0.5\textwidth]{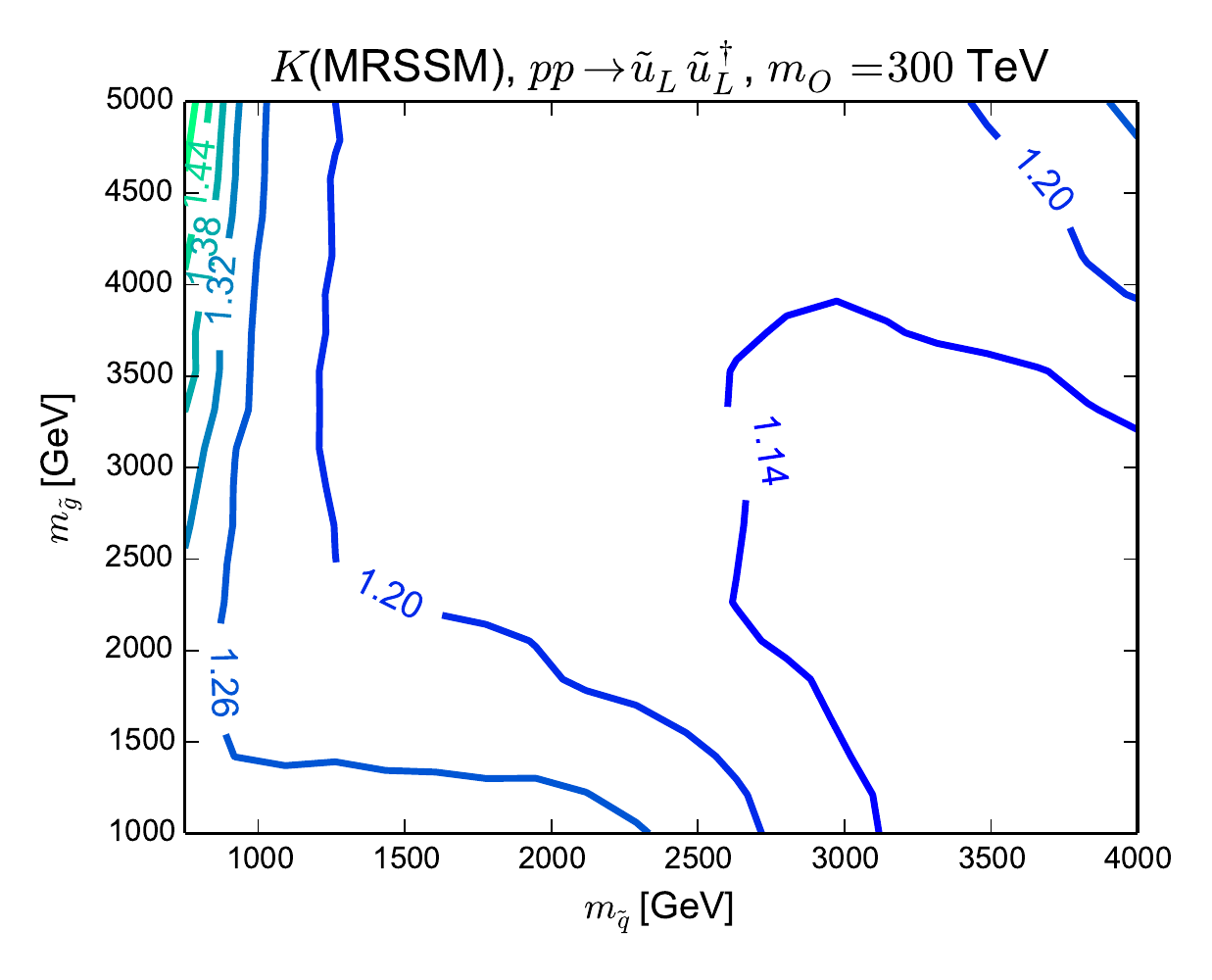}
\caption{As figure~\ref{fig:kfac-sqsq} for $\tilde{u}_L\tilde{u}_L^\dagger$ production.}
\label{fig:kfac-sqantisq}
\end{figure}

We have described the qualitative differences of the NLO corrections
between the MRSSM and MSSM in the previous section, especially highlighting
the role of the Dirac gluino mass and the appearance of the sgluon.
In the following, we give an overview of the quantitative features
of the corrections in the MRSSM analysing the variation of the K-factor.
Figures~\ref{fig:kfac-sqsq} and~\ref{fig:kfac-sqantisq} summarise the dependency
of the K-factor on the different masses of the strong sector for 
$\tilde{q}\tilde{q}$ and $\tilde{q}\tilde{q}^\dagger$ production, respectively. The left (right) plot
of each figure is given for a sgluon mass of 3~TeV (300~TeV) and shows the K-factor
depending on the Dirac gluino and common squark masses.

As discussed before, the change of the sgluon mass from 3 to 300~TeV leads to a
global enhancement of the K-factor of around twenty per cent 
in the whole parameter plane
for $\tilde{q}\tilde{q}$ production which originates from the super-oblique corrections.
The increase for the $\tilde{q}\tilde{q}^\dagger$ production is reduced to about five per cent 
as only part of the contributions receive the relevant corrections.

For $\tilde{q}\tilde{q}$ production, the total NLO contributions can decrease the LO cross 
section by $10\%$ and can increase them by more than 
$50\%$ of the LO prediction assuming sgluons are close in mass to gluino and squarks.
The relative size of the corrections falls with rising gluino mass while increasing 
squark masses lead to an enhancement. 
This feature is already present in the MSSM~\cite{Beenakker:1996ch} 
and not influenced by the Dirac 
nature of the gluino or the presence of the sgluon. 
In the scenario of interest, with a heavy Dirac
gluino and rather light squarks, the size of NLO corrections is reduced compared to 
the remainder of the parameter space.

The K-factors for $\tilde{q}\tilde{q}^\dagger$ production are in general smaller than for
$\tilde{q}\tilde{q}$ production. They yield up to $30\%$ corrections to the LO production cross 
section. The corrections are largest for small squark and gluino masses.
For small squark masses, the gluino mass does not influence the K-factor significantly.
In this region pure QCD corrections are dominant and only with a large sgluon mass
do the effects described in section~\ref{sec:sgluon} become important.
The K-factor is smallest for large squark masses and increases in this parameter
region with the gluino mass. 
\FloatBarrier

\subsection{Full results including uncertainties}
\begin{figure}[ht!]
\includegraphics[width=0.5\textwidth]{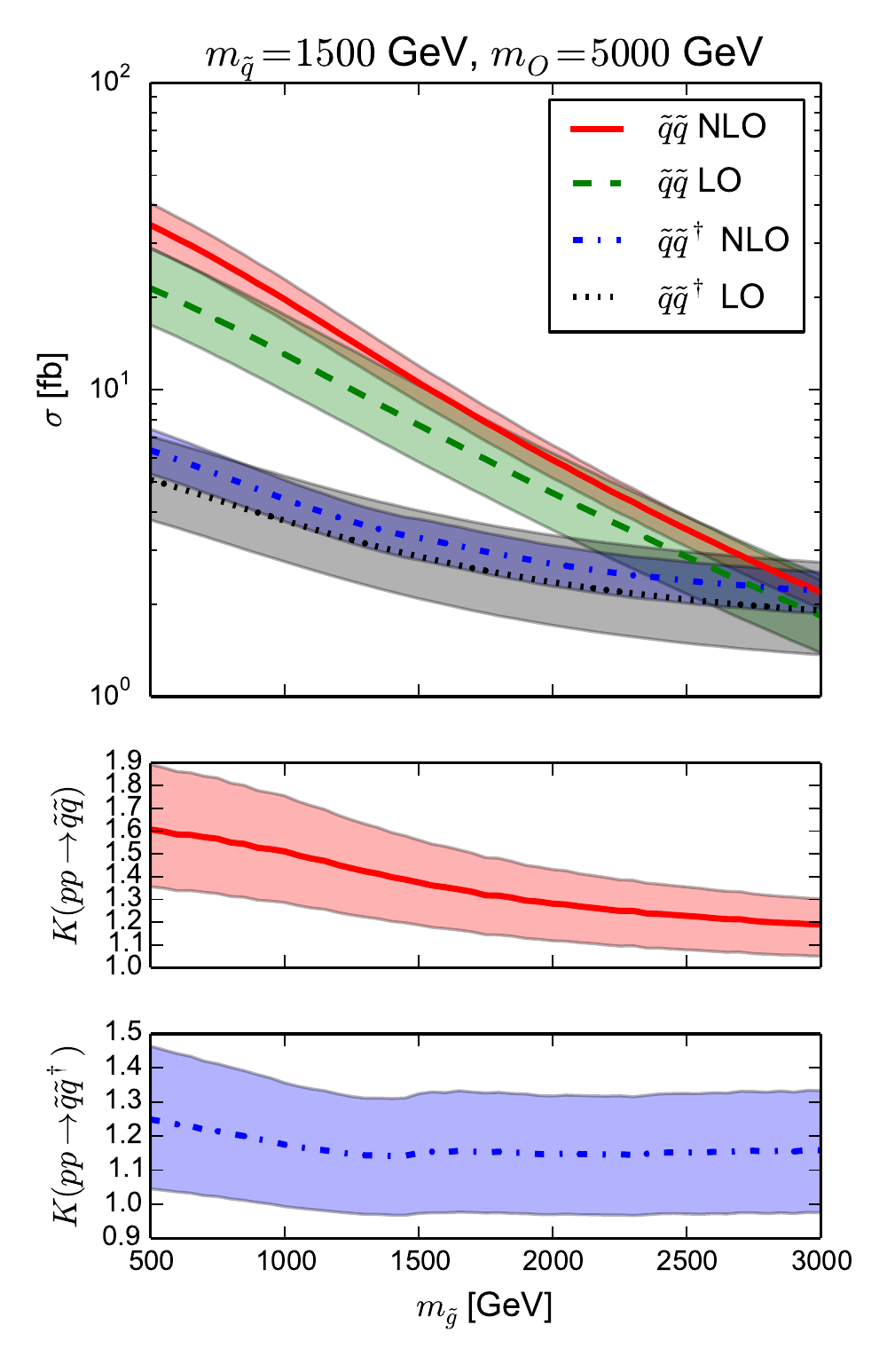}
\includegraphics[width=0.5\textwidth]{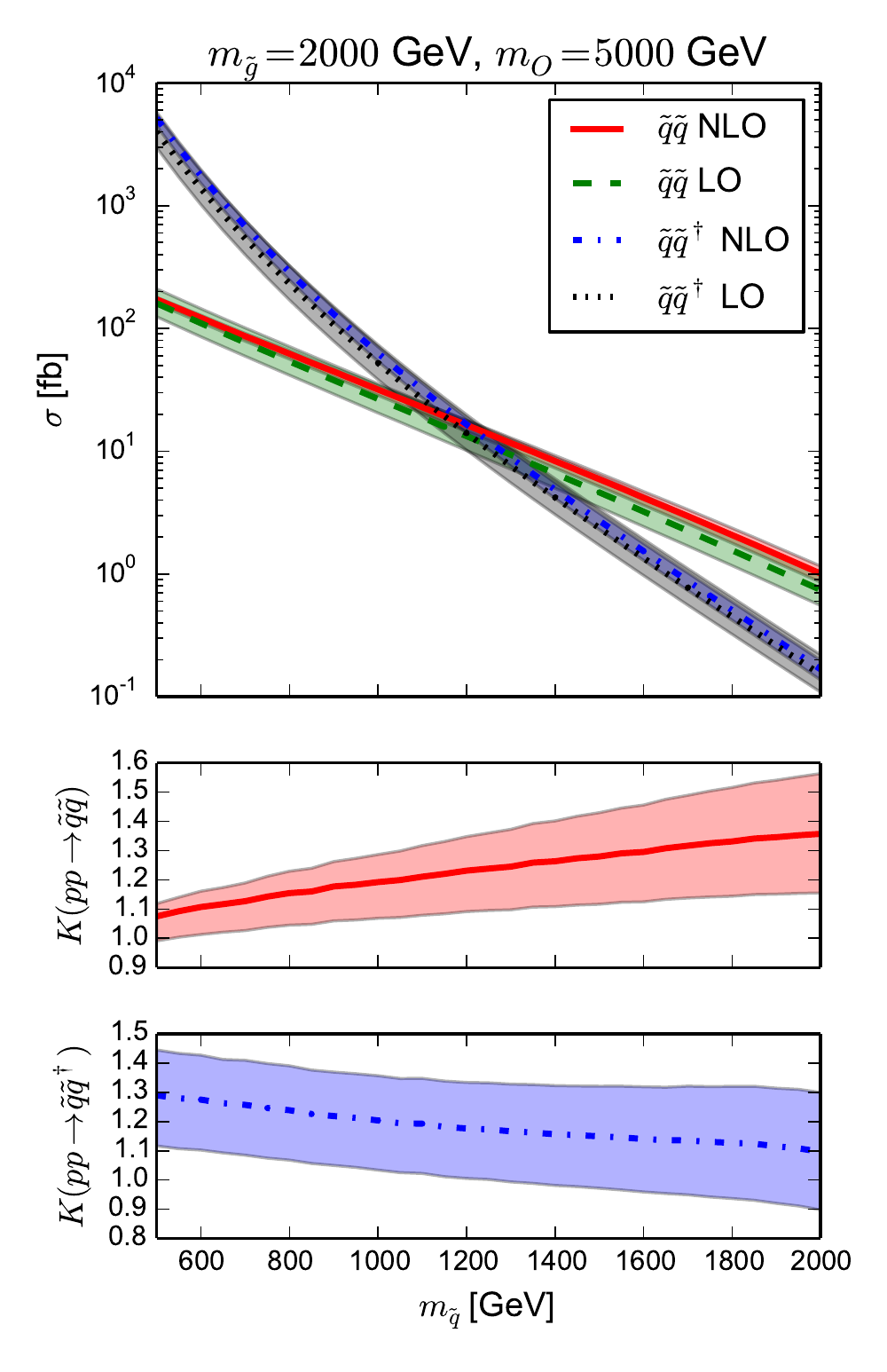}
\caption{Total LO and NLO cross section as function of the gluino (left) and common squark mass (right) with uncertainties. 
Shown are $\tilde{q}\tilde{q}$ and $\tilde{q}\tilde{q}^\dagger$ production summed over all possible flavour combinations 
as well as the corresponding K-factors. The bands give
the $68\%$~CL combined scale and pdf uncertainties as described in the text.}
\label{fig:toadd}
\end{figure}

We illustrate in figure~\ref{fig:toadd} the summarising results of our study of the 
NLO SUSY-QCD corrections for the production of squarks in the MRSSM.
The plots show LO and NLO cross sections for $\tilde{q}\tilde{q}$ and $\tilde{q}\tilde{q}^\dagger$ production
and their dependence on the squark and gluino mass as well as the corresponding
K-factors. 
The uncertainty bands are calculated by summing the uncertainty of the 
scale dependence\footnote{To estimate the scale dependence, we varied both, $\mu_F$ and $\mu_R$ like described in section \ref{sec:scale_uncer}.} and the PDF uncertainty in quadrature. A detailed
study of both sources of uncertainty is described below.

As we have seen before and has been known from the MSSM, the plots highlight that NLO 
QCD corrections are in general large and usually positive. 
Comparing the LO and NLO predictions including their respective $68\%$~CL 
uncertainty bands we see that the bands overlap for most of
the parameter space and the NLO corrections show no unexpected behaviour.

Comparing the cross sections for the MRSSM calculated with NLO precision 
to the ones of the MSSM, the distinctions between both models are
similar to the ones already discussed at tree level in section
\ref{sec:tree-level}.
These differences stem from the conservation of R-charge in the MRSSM
so that only different (same) ``chiralities'' of squarks can be
produced in $\tilde{q}\tilde{q}$ ($\tilde{q}\tilde{q}^\dagger$) production as has been discussed
in detail in section~\ref{sec:apples-oranges}.

Effects from the existence of sgluons becomes relevant only for
sgluon masses above hundreds of TeVs.
The Dirac nature of the gluino in the virtual contributions 
is of influence in the region of light
squarks and large gluino mass but does not alter the 
behaviour of the cross section significantly compared to the tree-level
differences.

The relative uncertainties of the cross 
sections are reduced from $50\%$ or more to below $20\%$ when going to NLO.
 The dominating uncertainty component comes from the scale variation.
Therefore, a further reduction of the uncertainties would require going
to next-to NLO and/or including effects from the resummation of threshold
corrections as has been done for the MSSM predictions~\cite{Beenakker:2011fu,Falgari:2012hx}.

\begin{table}
\begin{center}
\begin{tabular}{llcccc}
\toprule
& & $\sigma_{\text{LO}}(\text{LO PDF}) [\text{fb}]$  &$\sigma_{\text{LO}}(\text{NLO PDF}) [\text{fb}]$ & $K(\text{LO})$ & $K(\text{NLO})$ \\ % sigma_NLO
\midrule
BM1 & $\tilde{q}\tilde{q}$ & $13.1$ & $11.2$ & $1.46$ & $1.75$ \\%19.7
    & $\tilde{q}\tilde{q}^\dagger$  & $3.75$ & $2.72$ & $1.18$ & $1.62$ \\%4.409
BM2 & $\tilde{q}\tilde{q}$ & $4.61$ & $3.92$ & $1.28$ & $1.51$ \\ %5.913
    & $\tilde{q}\tilde{q}^\dagger$  & $2.36$ & $1.78$ & $1.15$ & $1.52$ \\ %2.712
BM3 & $\tilde{q}\tilde{q}$ & $160$  & $146$  & $1.08$ & $1.18$ \\ %172
    & $\tilde{q}\tilde{q}^\dagger$  & $3990$ & $3470$ & $1.29$ & $1.48$ \\ %5141
\bottomrule
\end{tabular}
\caption{LO cross section and corresponding K factors for the benchmark points of
table~\ref{tab:bms} using PDF sets of LO and NLO.}
\label{tab:Kfac_pdf}
\end{center}
\end{table}

We show in figure~\ref{fig:toadd} the K-factor defined in equation~\eqref{eq:kfac} using
the LO PDF sets for the LO cross section. Alternatively, it is also possible
to use the NLO PDF sets for both, NLO and LO cross section.
The difference is studied based on table~\ref{tab:Kfac_pdf}, where the LO cross sections
and NLO K factors using either LO or NLO PDF sets are given for the benchmark
points of table~\ref{tab:bms} and both squark production processes of this paper.

In general, using the NLO PDF sets for the LO cross section leads to an 
enhancement of the K factor (reduction of the LO cross section)
arising from the difference in the strong coupling constant between
the two sets. (Here, with $\alpha^{\overline{\text{MS}}}_s(m_Z)=0.135$ for 
$\mathtt{MMHT2014LO}$, $\alpha^{\overline{\text{MS}}}_s(m_Z)=0.120$ 
for \texttt{MMHT2014nlo68cl}.)
This can be seen when comparing the changes in the K-factors between
BM2 and BM3. They differ by the choice of squark mass which is used
as renormalisation and factorisation scale relevant for the extraction
of the correct $\alpha^{\overline{\text{MS}}}_s(\mu_R)$. For BM2 with a
larger squark mass the difference in K-factors is enhanced compared
to BM3.

 Additional effects due to difference in the PDF fits for quark and gluon are also present.
This leads to an enhancement of $\tilde{q}\tilde{q}^\dagger$ compared to the $\tilde{q}\tilde{q}$ production with rising squark mass.
%The relative  of both effects can be of similar size and pull the K-factor in different directions.
%This can be seen when comparing the entries for BM1 and $\tilde{q}\tilde{q}$ production
%where both effects combine in such a way that the K-factor
%with the NLO PDF set is actually smaller than with the LO PDF set for the LO cross section.

%For an meaningful prediction and comparison to possible experimental results 
%of differential distributions matching the event generator to parton shower is  advantageous.
%It is, however, not clear how much this affects exclusion bounds in the MRSSM and MSSM
%and an open question for further work.

\begin{figure}[htp]
\includegraphics[width=\textwidth]{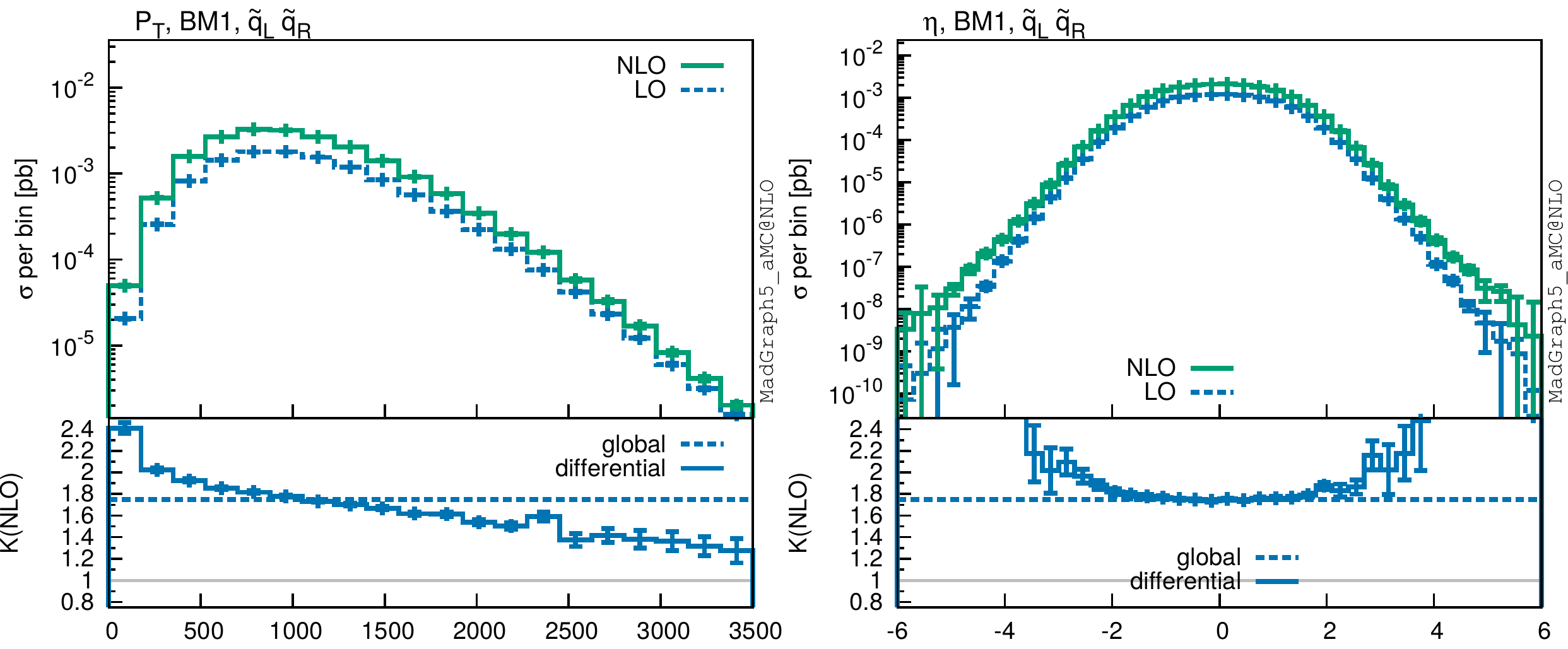}\\
\includegraphics[width=\textwidth]{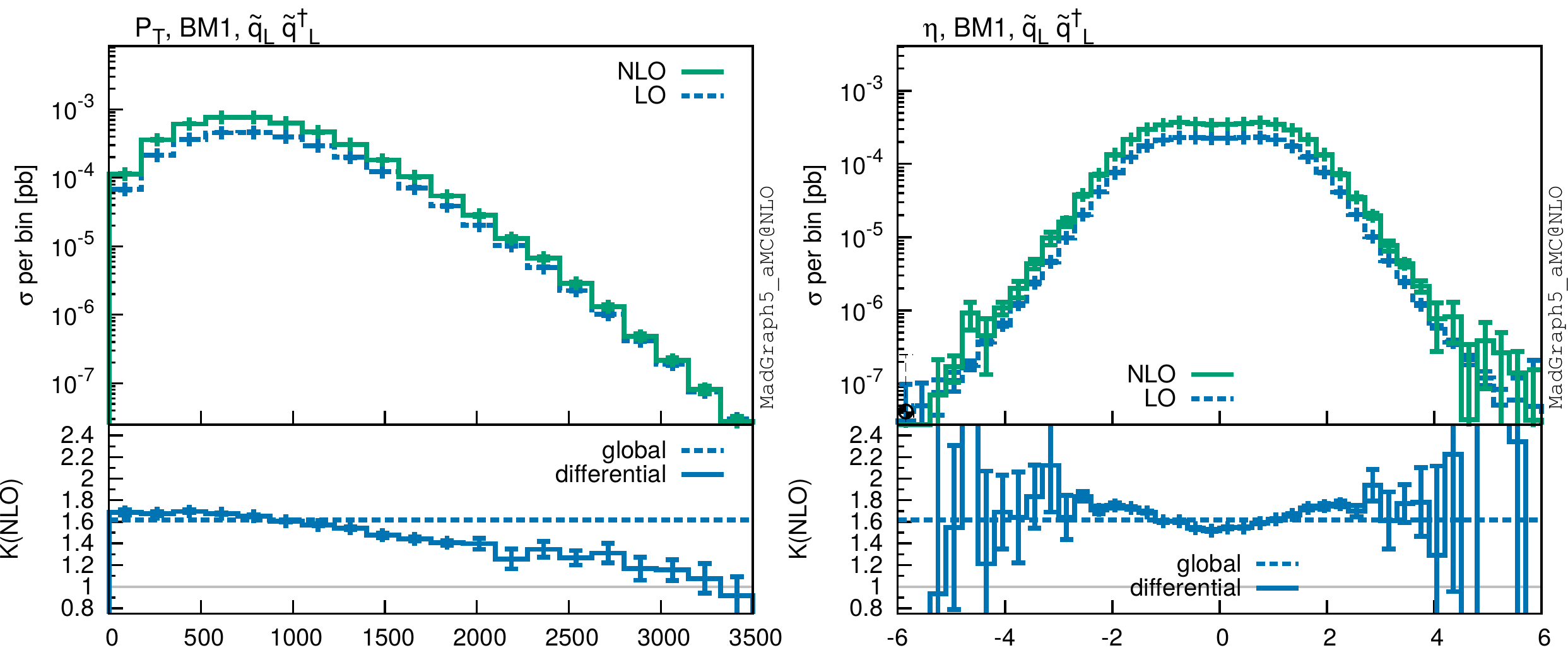}
\caption{Differential cross section of $\tilde{q}\tilde{q}$ and $\tilde{q}\tilde{q}^\dagger$ production as function of the transversal momentum $p_T$
or the pseudo-rapidity $\eta$ for BM1. The integration error is given.}
\label{fig:differential}
\end{figure}

For the results of this work we assume that all squark masses are at the same
value.
We also investigated the scenario, where the constraint of equal squark masses 
was dropped. We checked that the quantum corrections are well behaved 
in this case.
The phenomenological effects of such a scenario, especially for
final state squarks of different masses, will be studied in forthcoming 
work. 
In the following, we describe in detail the scale and PDF uncertainties as well as differential distributions of the 
NLO SUSY-QCD corrections. 
\FloatBarrier
\subsubsection{(Fixed) Renormalisation and factorisation scale dependency}\label{sec:scale_uncer}
\begin{figure}[ht!]
\includegraphics[width=0.5\textwidth]{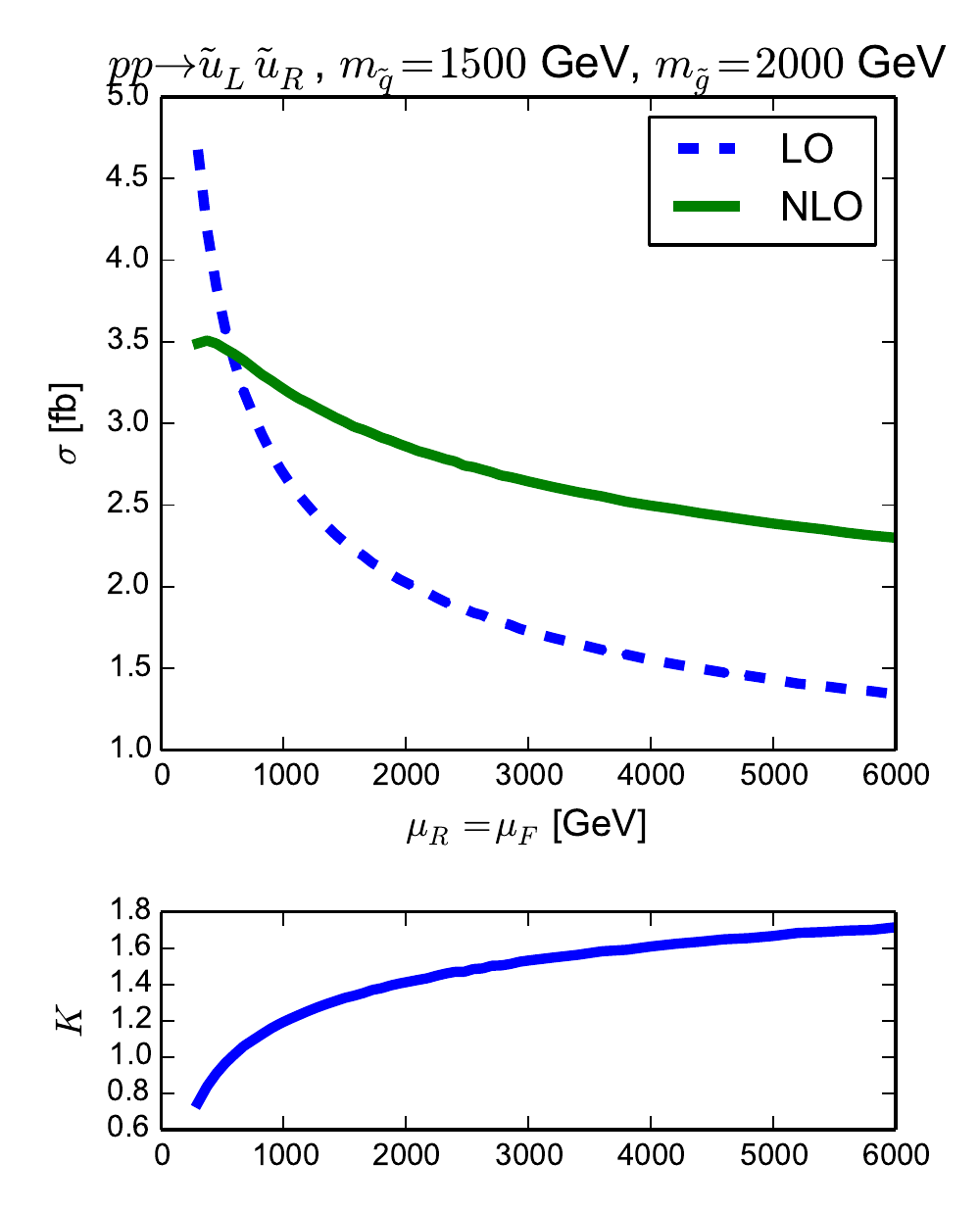}
\includegraphics[width=0.5\textwidth]{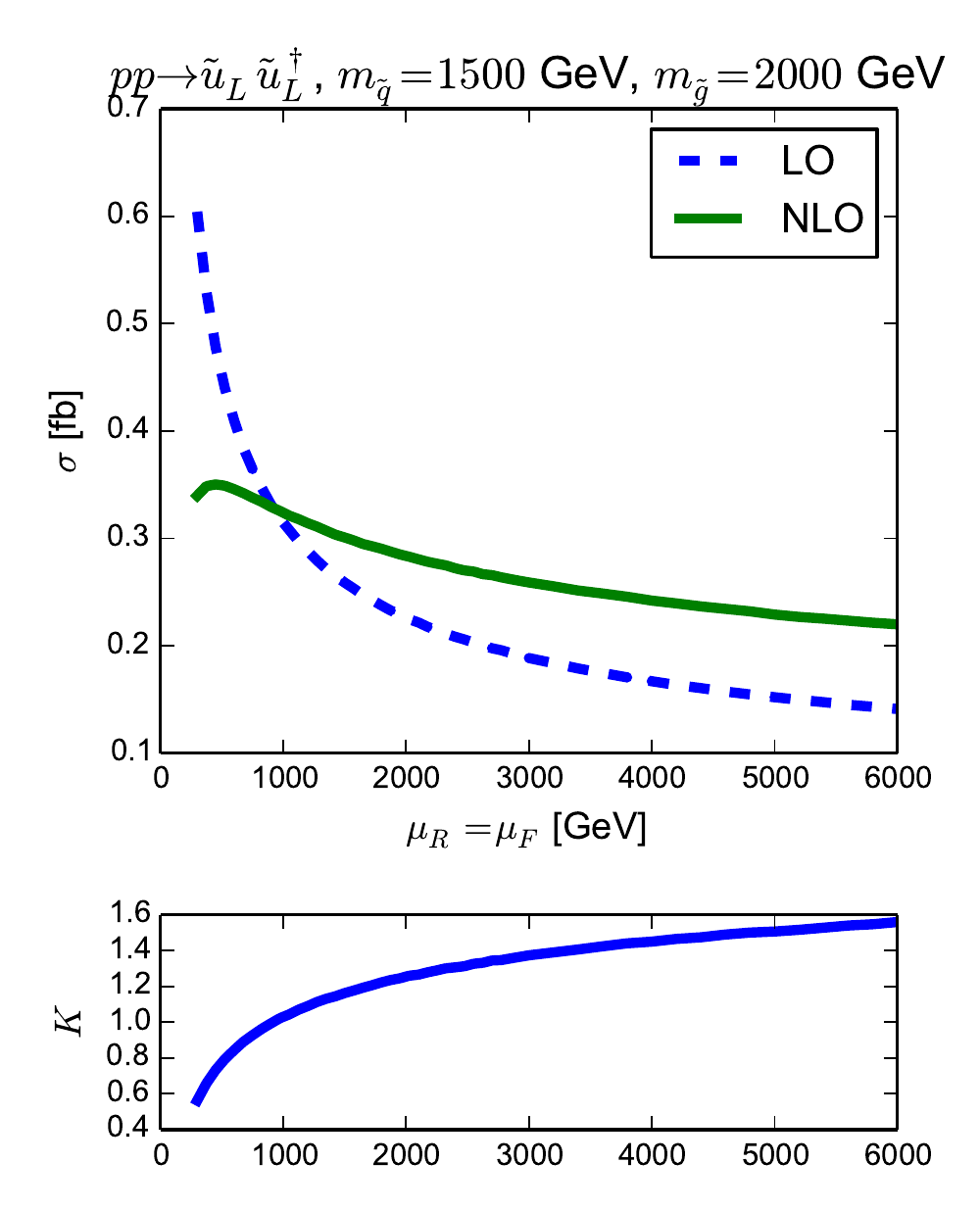}
\caption{Scale dependency of the cross section for $\tilde{u}_L\tilde{u}_R$ (left) and $\tilde{u}_L\tilde{u}_L^\dagger$ (right) production in the MRSSM at LO and NLO.}
\label{fig:scale}
\end{figure}
A major motivation to calculate perturbative corrections to the prediction of a cross
section is the reduction of the theoretical uncertainty. One way of
quantifying this is the variation of renormalisation and factorisation scale.

Figure~\ref{fig:scale} shows the variation of scales for $\tilde{u}_L\tilde{u}_R$ and $\tilde{u}_L\tilde{u}_L^\dagger$ production. To achieve a qualitative 
understanding  renormalisation and factorisation scale are set equal
and varied together.
The prediction of the cross section at LO changes by
more than $50\%$ when the
scale is varied from the reference value at $\mu_R=\mu_F=m_{\tilde q}$
by a factor of two. This variation uncertainty is reduced to below $20\%$ when
taking NLO corrections into account.
This difference in the scale dependency also leads to a strong
dependence of the K-factor, as shown in the lower plot, which can become
smaller than one for a certain choice.

The commonly accepted range for the variation of the scales is 
$[\mu_{R,F}/2,2\cdot\mu_{R,F}]$ where, for the most conservative approach, both scales 
are varied independently and one takes the envelope of those nine values as final estimate. 
This is the procedure we use to estimate the scale uncertainty in figure~\ref{fig:toadd}.

\subsubsection{PDF uncertainty}
\begin{table}
\begin{tabular}{lcccc}
\toprule
& \texttt{MMHT2014nlo68cl} & \texttt{MMHT2014nlo68clas118} & \texttt{CT14nlo} & \texttt{NNPDF30\_nlo\_as\_0118}\\
\midrule
BM1 & $9.94^{+3.5\%}_{-3.4\%}$ & $9.86^{+3.4\%}_{-3.3\%}$ & $10.1^{+3.3\%}_{-3.4\%}$ 
    & $10.1^{+3.2\%}_{-3.2\%}$ \\
BM2 & $3.01^{+3.5\%}_{-3.4\%}$ & $3.00^{+3.4\%}_{-3.3\%}$ & $3.08^{+3.4\%}_{-3.6\%}$ 
    & $3.04^{+3.3\%}_{-3.3\%}$ \\
BM3 & $64.0^{+3.0\%}_{-3.0\%}$ & $62.4^{+3.1\%}_{-2.9\%}$ & $65.0^{+2.6\%}_{-2.4\%}$ 
	& $64.0^{+2.4\%}_{-2.4\%}$ \\
\bottomrule
\end{tabular}
\caption{Cross section (in fb) of $\tilde{u}_L\tilde{u}_R$ production for the benchmark points of
table~\ref{tab:bms} using different PDF sets. The first set is the standard NLO PDF of
this paper.  The other columns compare different PDF  sets at  $\alpha_s(m_Z)=0.118$.
The factorisation and renormalisation scale are set to the common squark mass. All
uncertainties are given at $68\%$~CL. For this the uncertainties of the \texttt{CT14} set
have been rescaled by $1.642$ as the default is given at 
$90\%$~CL.~\cite{Butterworth:2015oua}}
\label{tab:xsec_uncertain_sqsq}
\end{table}
\begin{table}
\begin{tabular}{lcccc}
\toprule
& \texttt{MMHT2014nlo68cl} & \texttt{MMHT2014nlo68clas118} & \texttt{CT14nlo} & \texttt{NNPDF30\_nlo\_as\_0118}\\
\midrule
BM1 & $0.512^{+7.6\%}_{-5.4\%}$ & $0.503^{+7.5\%}_{-4.9\%}$& $0.503^{+12.9\%}_{-11.6\%}$ & $0.588^{+9.7\%}_{-9.7\%}$\\
BM2 & $0.301^{+8.1\%}_{-5.7\%}$ &$0.294^{+7.7\%}_{-5.0\%}$ & $0.300^{+14.8\%}_{-11.1\%}$ &$0.335^{+9.9\%}_{-9.9\%}$ \\
BM3 & $502^{+3.8\%}_{-3.7\%}$ & $509^{+3.9\%}_{-3.4\%}$ &$512^{+6.7\%}_{-5.4\%}$ & $533^{+3.4\%}_{-3.4\%}$ \\
\bottomrule
\end{tabular}
\caption{Cross section (in fb) of $\tilde{u}_L\tilde{u}_L^\dagger$ production for the benchmark 
points of table~\ref{tab:bms} using different PDF sets. All details as for
table~\ref{tab:xsec_uncertain_sqsq}.}
\label{tab:xsec_uncertain_sqsqc}
\end{table}

Another main source of uncertainty for the cross section prediction
at a hadron collider comes from the use of PDFs as
they can not be calculated from first principles but 
need to be extracted from data. Details on this and the application
at LHC Run II  can be found in ref.~\cite{Butterworth:2015oua}.
Here, we only aim to achieve a basic understanding of the 
relevant PDF uncertainties. For this we compare
the cross section for our three BMPs defined in table~\ref{tab:bms} and 
both processes calculated with our default set, \texttt{MMHT2014nlo68cl}, 
against one
of the same group with a different fit value for $\alpha_s(m_Z)$, 
\texttt{MMHT2014nlo68clas118}. 
Additionally, we compare to cross sections calculated using the
\texttt{CT14}~\cite{Dulat:2015mca} and \texttt{NNPDF3.0}~\cite{Ball:2014uwa}
PDF set with $\alpha_s(m_Z)=0.118$.
They are summarised in the tables~\ref{tab:xsec_uncertain_sqsq} 
and~\ref{tab:xsec_uncertain_sqsqc} .

The difference of the result  for the \texttt{MMHT2014nlo68cl} and 
\texttt{MMHT2014nlo68clas118} sets stems only from  the strong coupling constant,
$\alpha_s(m_Z)=0.120$ and $\alpha_s(m_Z)=0.118$ respectively, and gives an 
estimate for the uncertainty coming from this input parameter. For
the considered points the uncertainty is at most a few per cent.
The different PDF sets for $\alpha_s(m_Z)=0.118$ all agree within their uncertainty
as expected from the comparison done in ref.~\cite{Butterworth:2015oua}.
The uncertainty of the individual sets is
below five per cent for $\tilde{q}\tilde{q}$ production. This is understandable
as for this mainly the up quark PDF is relevant which is rather well determined,
see figure~6 of ref.~\cite{Butterworth:2015oua}. 
The PDF uncertainty for $\tilde{q}\tilde{q}^\dagger$ production on the other hand is increased
and can reach $10\%$ to $15\%$ depending on the PDF set. This originates from the
appearance of initial gluons and antiquarks as in this case the PDFs are
not as well known, see figure~5 of ref.~\cite{Butterworth:2015oua}. 

%todo: add reference
As the study of PDF uncertainties is not in the focus of this work, the PDF uncertainty in figure~\ref{fig:toadd} contains only the uncertainty of the \texttt{MMHT2014nlo68cl} set.

\subsubsection{Dynamical scales and differential distributions}
The main concern of this work was to perform the calculation of NLO corrections for the MRSSM and highlight the physical differences to the MSSM.
We addressed it by focusing on global K-factors and using fixed renormalisation and factorisation scales.
Here we give a brief comment on going beyond that.
The \MG\ framework allows in principle to use dynamical
renormalisation and factorisation scales (as opposed to fixed ones studied in section~\ref{sec:scale_uncer}) and to study differential distributions 
at the NLO. 
We have verified that these components work as expected.
The global K-factors vary by less than five per cent when choosing between a fixed (like the common squark mass) and a dynamical (like the total transverse mass in an event) scale. 
As there is no physically preferred scheme we follow the works
on the MSSM~\cite{Gavin:2013kga,Gavin:2014yga,Hollik:2012rc,GoncalvesNetto:2012yt}
and use fixed renormalisation and factorisation scales.

In figure~\ref{fig:differential}, we show sample differential distributions
in the transversal momentum $p_T$ and the pseudo-rapidity $\eta$ for the
benchmark point BM1. The top row shows it for $\tilde{q}\tilde{q}$ 
while in the bottom $\tilde{q}\tilde{q}^\dagger$ production is given.
The K-factor can actually vary substantially over the
different kinematic regions changing by 50 per cent or more. 
Still, for the regions containing
the bulk of the contributions ($\unit[500]{GeV}<p_T<\unit[1000]{GeV}$ 
and $-2 < \eta < 2$) the global K-factor is a good approximation.
\section{Conclusions}

The Minimal R-symmetric Supersymmetric Standard Model is a
well-motivated, viable model. It provides a realisation of
SUSY distinct from the familiar MSSM with a significantly modified
coloured sector: The gluino carries R-charge and is of 
Dirac instead of Majorana nature, and there are scalar and
pseudoscalar colour octets, the sgluons. Left- and right-handed
squarks carry opposite R-charge. In a motivated region of parameter
space, the squark masses are around the TeV-scale while the gluino and
sgluons are somewhat heavier.

Here we have presented an analysis of squark production at the
LHC in the MRSSM taking into account NLO corrections of the entire
strongly interacting sector. Both the tree-level results and the NLO
corrections 
show important features and differences to the MSSM:
\begin{itemize}
\item In the MRSSM only opposite-chirality squarks can be
  pair-produced, i.e.\ $\tilde{q}_L\tilde{q}_R$ production is allowed,
  while $\tilde{q}_L\tilde{q}_L$ and $\tilde{q}_R\tilde{q}_R$ are
  forbidden by R-charge conservation. For squark-antisquark
  production, the converse statement is true. Owing to this, the overall
  squark production rate in the MRSSM is lower than in the MSSM.
\item When comparing K-factors between the two models one has to be
  careful to distinguish in the MSSM between the K-factor for total
  squark production and the K-factor specific for e.g.\ $\tilde{q}_L\tilde{q}_R$
  production. The first is the one usually quoted; it differs from the
  MRSSM counterpart by $10$ to $20\%$. The second is usually not
  quoted in the MSSM, but it is more directly comparable between the
  two models.
\item Even the comparison between  $\tilde{q}_L\tilde{q}_R$ production
  in the MRSSM and MSSM reveals several differences between the two
  models. The MRSSM Dirac gluino enters the loop corrections
  differently from the MSSM Majorana gluino, and large sgluon masses
  lead to non-decoupling, logarithmic enhancements of the NLO
  corrections. To a lesser degree, these effects also lead to differences in the $\tilde{q}\tilde{q}^\dagger$ production.
\end{itemize}

Our study also has technical aspects.
During the last decade several efforts lead 
to fast and automatised ways to evaluate SM processes at NLO and beyond. 
In recent years, there has been a push to expand this also to BSM models. 
Usually, this added capability is tested in the context of well-known and 
understood models like the THDM or the MSSM.
In this paper we were able to achieve full agreement between a calculation based on a 
subset of publicly available tools and an independent calculation for NLO QCD corrections 
of the MRSSM. The latter calculation uses the techniques, described as method 1 in this 
paper and results will be available via the soon to be published \texttt{RSymSQCD} 
code~\cite{our_code}.
The MRSSM provides a valuable test of the implemented mechanisms as it contains unique 
features complementary to the previous test cases.
It therefore shows how far the latest available machinery can be pushed.
It also illustrates the possibility mentioned in ref.~\cite{Signer:2008va,Gnendiger:2017pys}, to compute
QCD corrections to SUSY processes in different renormalisation schemes.

The results obtained here are not specific for the MRSSM but can be applied
more generally also to other models with R-symmetry and a difference
in R-charge assignment. This only requires that the particle
content under consideration is the same and all the SUSY QCD vertices given in the
appendix are present.

It will be now of high interest to identify the allowed squark mass
range in the MRSSM in the light of current LHC data. In line with
simplified studies \cite{Kribs:2012gx,Kribs:2013oda} we expect that significantly lighter
squarks are allowed in the MRSSM, thanks to the fewer allowed
production channels. On the other hand, our NLO results show that the
assumption that the K-factors in the MSSM and MRSSM are equal is not
correct. The MRSSM K-factors are higher and dependent on the
hierarchies between squark, gluino and sgluon masses and should be
taken into account to obtain valid limits.

\acknowledgments

This research was supported in part by the German Research Foundation (DFG) under grant number SI 2009/1-1 and STO 876/4-1, the Polish National Science Centre under contract UMO-2015/18/M/ST2/00518 (2016-2019) and the PL-Grid Infrastructure.

%\paragraph{Note added.} This is also a good position for notes added after the paper has been written.

\appendix

%Please always give a title also for appendices.
\section{Feynman rules}\label{sec:FeynmanRules}
In SUSY models, it is often not possible to define a consistent,
conserved fermion number.\footnote{For Dirac particles like quarks, the 
direction of fermion number flow is given by the arrow on the Dirac 
propagator.} To compute Feynman diagrams in an unambiguous way, a fermion 
flow can be introduced~\cite{Denner:1992vza,Denner:1992me}, which does in general not agree 
with the flow of fermion number.

Even though there are no Majorana particles in the MRSSM, it is useful to adapt 
this procedure. The reason for this is the existence of fermion number 
violating processes like $qq \to \tilde{q}\tilde{q}$, which results in a 
clash of arrows in the associated Feynman diagrams.

In the following we list Feynman rules for the MRSSM which are new or different 
to the ones in the MSSM. We labelled the lines with the corresponding
quantum field operators of the 
respective Lagrangian term (thus this labeling does not coincide with the 
labels of external particles in section~\ref{sec:tree-level} and \ref{sec:virt}).
The fermion flow on vertices is always directed from an unbarred to a barred
spinor field and indicated by an extra arrow next to the diagrams. For
calculations involving fermions one needs to multiply the  
Feynman rules in the opposite direction of the fermion flow. 
% figure~\ref{fig:firstFeynmanRules}.

%The curved arrows indicate the fermion flow. 
The Feynman rules \circled{4b} and 
\circled{5b} are the complex conjugates of \circled{4a} and \circled{5a}, respectively. 
Applying a flipping rule to a vertex one has to reverse the curved arrow, i.e. the 
fermion flow and replace $\Psi$ with $\overline{\Psi}^C$. In addition one has to add a 
minus sign for Feynman rule \circled{1}.
\begin{center}
\begin{tikzpicture}[line width=1.5 pt, scale=1.1]
%	\node at (-1.5,0.7) {\circled{1}};
%	\draw[gluon](0.5,0)--(-1,0);
%	\draw[fermionnoarrow](-1,0)--(0.5,0);
%	\node at (-1.2,0) {$\tilde{g}^a$};
%	\node at (0.7,0) {$\overline{\tilde{g}}^b$};
%	\node at (2,0) {$\hat{=}\ i\frac{\slashed{p}+m_{\tilde{g}}}{p^2-m_{\tilde{g}}^2+i\varepsilon}\delta_{ab}$};
%	\path[line width=0.8pt,<-] (-0.7,-0.3) edge [out=0, in=180] (0.2,-0.3);
\begin{scope}[shift={(0,-2.5)}]
	\node at (-1.5,0.7) {\circled{1}};
	\draw[gluon] (0,0)--(180:1);
	\draw[gluon] (0,0)--(45:1);
	\draw[gluon] (0,0)--(-45:1);
	\draw[fermionnoarrow] (0,0)--(-45:1);
	\draw[fermionnoarrow] (0,0)--(45:1);
	\node at (45:1.4) {$\overline{\tilde{g}}^b$};
	\node at (-45:1.4) {$\tilde{g}^c$};
	\node at (180:1.5) {$G_\mu^a$};
	\node(gbar) at (0.9,0.6) {};
	\node(g) at (0.9,-0.6) {};
	\path[line width=0.8pt,->] (g) edge [out=135, in=-135] (gbar);
	\node at (2,0) {$\hat{=}  - g_s f_{abc} \gamma^\mu$};
\end{scope}
\begin{scope}[shift={(6.5,-2.5)}]
	\node at (-1.5,0.7) {\circled{2}};
	\draw[gluon] (0,0)--(180:1);
	\draw[scalar] (0,0)--(45:1);
	\draw[scalarbar] (0,0)--(-45:1);
	\node at (40:1.4) {$O_s^b(p_1)/O_p^b(p_1)$};
	\node at (-40:1.4) {$O_s^c(p_2)/O_p^c(p_2)$};
	\node at (180:1.5) {$G_\mu^a$};
	\node at (2.7,0) {$\hat{=}+ g_s (p_1-p_2)^\mu f_{abc} $}; 
	% checked, contraction G/O_s/O_p is zero
\end{scope}
\begin{scope}[shift={(0,-5)}]
	\node at (-1.5,0.7) {\circled{3}};
	\draw[scalarnoarrow] (0,0)--(45:1);
	\draw[scalarnoarrow] (0,0)--(-45:1);
	\draw[gluon] (0,0)--(135:1);
	\draw[gluon] (0,0)--(-135:1);
	\node at (45:1.4) {$O_s^c / O_p^c$};
	\node at (-45:1.4) {$O_s^d / O_p^d$};
	\node at (135:1.4) {$G_\mu^a$};
	\node at (-135:1.4) {$G_\nu^b$};
	\node at (3.5,0) {$\hat{=} +ig_s^2 g^{\mu\nu}
	(f^{aec}f^{bed}+f^{bec}f^{aed})$}; % checked
\end{scope}
\begin{scope}[shift={(0,-7.5)}]
	\node at (-1.5,0.7) {\circled{4a}};
	\draw[scalar] (0,0)--(45:1);
	\draw[fermionbar] (0,0)--(-45:1);
	\draw[fermionnoarrow] (180:1)--(0,0);
	\draw[gluon] (180:1)--(0,0);
	\node at (45:1.3) {$\tilde{q}_{Li}^\dagger$};
	\node at (-45:1.3) {$\overline{q}^C_j$};
	\node at (180:1.3) {$\tilde{g}^a$};
	\node(g) at (-0.7,-0.3) {};
	\node(u) at (0.4,-0.8) {};
	\path[line width=0.8pt,->] (g) edge [out=0, in=135] (u);
	\node at (2.3,0) {$\hat{=} -i\sqrt{2}g_s T^a_{ij}P_L$};
\end{scope}
\begin{scope}[shift={(6.5,-7.5)}]
	\node at (-1.5,0.7) {\circled{4b}};
	\draw[fermion] (0,0)--(45:1);
	\draw[scalarbar] (0,0)--(-45:1);
	\draw[fermionnoarrow] (180:1)--(0,0);
	\draw[gluon] (180:1)--(0,0);
	\node at (45:1.3) {$q^C_i$};
	\node at (-45:1.3) {$\tilde{q}_{Lj}$};
	\node at (180:1.3) {$\overline{\tilde{g}}^a$};
	\node(g) at (-0.7,0.3) {};
	\node(u) at (0.4,0.8) {};
	\path[line width=0.8pt,<-] (g) edge [out=0, in=225] (u);
	\node at (2.3,0) {$\hat{=} -i\sqrt{2}g_s T^a_{ij}P_R$};
\end{scope}
\begin{scope}[shift={(0,-10)}]
	\node at (-1.5,0.7) {\circled{5a}};
	\draw[scalarbar] (0,0)--(45:1);
	\draw[fermion] (0,0)--(-45:1);
	\draw[fermionnoarrow] (180:1)--(0,0);
	\draw[gluon] (180:1)--(0,0);
	\node at (45:1.3) {$\tilde{q}_{Rj}$};
	\node at (-45:1.3) {$\overline{q}_i$};
	\node at (180:1.3) {$\tilde{g}^a$};
	\node(g) at (-0.7,-0.3) {};
	\node(u) at (0.4,-0.8) {};
	\path[line width=0.8pt,->] (g) edge [out=0, in=135] (u);
	\node at (2.3,0) {$\hat{=} +i\sqrt{2}g_s T^a_{ij}P_L$};
\end{scope}
\begin{scope}[shift={(6.5,-10)}]
	\node at (-1.5,0.7) {\circled{5b}};
	\draw[fermionbar] (0,0)--(45:1);
	\draw[scalar] (0,0)--(-45:1);
	\draw[fermionnoarrow] (180:1)--(0,0);
	\draw[gluon] (180:1)--(0,0);
	\node at (45:1.3) {$q_j$};
	\node at (-45:1.3) {$\tilde{q}_{Ri}^\dagger$};
	\node at (180:1.3) {$\overline{\tilde{g}}^a$};
	\node(g) at (-0.7,0.3) {};
	\node(u) at (0.4,0.8) {};
	\path[line width=0.8pt,<-] (g) edge [out=0, in=225] (u);
	\node at (2.3,0) {$\hat{=} +i\sqrt{2}g_s T^a_{ij}P_R$};
\end{scope}
\end{tikzpicture}
\begin{tikzpicture}[line width=1.5 pt, scale=1.1]
\begin{scope}[shift={(0,-12.5)}]
	\node at (-1.5,0.7) {\circled{6}};
	\draw[scalarnoarrow] (0,0)--(45:1);
	\draw[fermionnoarrow] (0,0)--(-45:1);
	\draw[gluon] (0,0)--(-45:1);
	\draw[fermionnoarrow] (180:1)--(0,0);
	\draw[gluon] (180:1)--(0,0);
	\node at (45:1.3) {$O_s^b$};
	\node at (-45:1.3) {$\overline{\tilde{g}}^c$};
	\node at (180:1.3) {$\tilde{g}^a$};
	\node(g) at (-0.7,-0.3) {};
	\node(u) at (0.4,-0.8) {};
	\path[line width=0.8pt,->] (g) edge [out=0, in=135] (u);
	\node at (2.0,0) {$\hat{=} - g_s f^{abc}$}; % checked, has other sign in 
	%SARAH convention, just as the sg-sq-q vertex
\end{scope}
\begin{scope}[shift={(6.5,-12.5)}]
	\node at (-1.5,0.7) {\circled{7}};
	\draw[fermionnoarrow] (0,0)--(45:1);
	\draw[gluon] (0,0)--(45:1);
	\draw[scalarnoarrow] (0,0)--(-45:1);
	\draw[fermionnoarrow] (180:1)--(0,0);
	\draw[gluon] (180:1)--(0,0);
	\node at (45:1.3) {$\tilde{g}^b$};
	\node at (-45:1.3) {$O_p^c$};
	\node at (180:1.3) {$\overline{\tilde{g}}^a$};
	\node(g) at (-0.7,0.3) {};
	\node(u) at (0.4,0.8) {};
	\path[line width=0.8pt,<-] (g) edge [out=0, in=225] (u);
	\node at (2.0,0) {$\hat{=} +i g_s f^{abc}$};
\end{scope}
%\end{tikzpicture}
%\end{center}
%\label{fig:firstFeynmanRules}

%\begin{center}
%\begin{tikzpicture}[line width=1.5 pt, scale=1.3]
%\begin{scope}[shift={(-1,0)}]
%	\node at (-0.5,0.7) {\circled{8}};	
%	\draw[scalarnoarrow](0,0)--(1.5,0);
%	\node at (180:0.4) {$O_p^a$};
%	\node at (0:1.9) {$O_p^b$};
%	\node at (3.4,0) {$\hat{=}\ \frac{i}{p^2-m_{O_p}^2+i\varepsilon}\delta_{ab}$};
%\begin{scope}[shift={(7,0)}]
%	\node at (-0.5,0.7) {\circled{8}};
%	\draw[scalarnoarrow](0,0)--(1.5,0);
%	\node at (180:0.4) {$O_s^a$};
%	\node at (0:1.9) {$O_s^b$};
%	\node at (3.4,0) {$\hat{=}\ \frac{i}{p^2-m_{O_s}^2+i\varepsilon}\delta_{ab}$};
%\end{scope}
%\end{scope}
\begin{scope}[shift={(0,-15)}]
	\node at (-1.5,0.7) {\circled{8}};
	\draw[scalarnoarrow] (0,0)--(45:1);
	\draw[scalarnoarrow] (0,0)--(-45:1);
	\draw[scalar] (0,0)--(135:1);
	\draw[scalarbar] (0,0)--(-135:1);
	\node at (45:1.4) {$O_s^b$};
	\node at (-45:1.4) {$O_p^c$};
	\node at (135:1.4) {$\tilde{q}^\dagger_{Aj}$};
	\node at (-135:1.4) {$\tilde{q}_{Ai}$};
	\node at (3.3,0) {$\hat{=} \ -ig_s^2 T_{ij}^a f^{abc}(\delta_{AL} - \delta_{AR})$ }; % checked, O_s/O_s or O_p/O_p is zero
\end{scope}
\begin{scope}[shift={(0,-17.5)}]
	\node at (-1.5,0.7) {\circled{9}};
	\draw[scalar] (0,0)--(45:1);
	\draw[scalarbar] (0,0)--(-45:1);
	\draw[scalarnoarrow] (180:1)--(0,0);
	\node at (45:1.3) {$\tilde{q}^\dagger_{Ai}$};
	\node at (-45:1.3) {$\tilde{q}_{Aj}$};
	\node at (180:1.4) {$O_s^a$};
	\node at (3.3,0) {$\hat{=} -i2g_s m_{\tilde{g}} T^a_{ij}(\delta_{AL}-\delta_{AR})$}; % checked
\end{scope}
\begin{scope}[shift={(0,-20)}]
	\node at (-1.5,0.7) {\circled{10}};
	\draw[scalarnoarrow] (0,0)--(45:1);
	\draw[scalarnoarrow] (0,0)--(-45:1);
	\draw[scalarnoarrow] (0,0)--(135:1);
	\draw[scalarnoarrow] (0,0)--(-135:1);
	\node at (45:1.4) {$O_p^d$};
	\node at (-45:1.4) {$O_p^e$};
	\node at (135:1.4) {$O_s^b$};
	\node at (-135:1.4) {$O_s^c$};
	\node at (3.3,0) {$\hat{=} \ -ig_s^2 (f^{ace}f^{abd} + f^{abe}f^{acd})$};
	% checked, rest of contraction, e.g. Os/Os/Os/Op are zero
\end{scope}
\end{tikzpicture}
%Feynman rules involving only scalar fields.
%\label{fig:secondFeynmanRules}
\end{center}
\section{The MRSSM in MadGraph and GoSam}
\label{app:mg-gosam}
To allow the use of \texttt{MadGraph5\_aMC@NLO} (\MG) together with \texttt{GoSam}
for the MRSSM, several changes are necessary in the programs.
For \MG:
\begin{itemize}
\item Replacing the path to the model in the template \texttt{gosam.rc} of 
\MG,
\item Modifying the \texttt{write\_lh\_order} method in the 
\texttt{madgraph/iolibs/export\_fks.py} file to produce a LH file conforming to BLHA2
instead of BLHA1.
\item Adapting the \texttt{SubProcesses/BinothLHA\_OLP.f} also from BLHA1 to BLHA2 by
using  the \texttt{OLP\_SetParameter} function to pass $\alpha_s$ from
\MG\ to \texttt{GoSam} and the \texttt{OLP\_EvalSubProcess2}
function to get the virtual matrix element from \texttt{GoSam}.
\end{itemize}
For \texttt{GoSam}:
\begin{itemize}
\item Adjusting the naming of the strong coupling in the \texttt{olp\_module.f90}
template for the \texttt{aMC} interface to the UFO model name so that it compiles.
\item The default SM renormalisation of the virtual matrix element is switched off
and replaced with a model and subprocess specific renormalisation in \texttt{matrix.f90}.
\end{itemize}
\section{Validation \label{sec:pole_cancelation}}

%\subsection*{Validation on parton level}
The selected parameter and phase space point is the same for all MRSSM processes:
%fix convention for octet scalar/pseudoscalar names
\begin{center}
\begin{tabular}{ccc}
\toprule
$m_{\tilde{g}} = 1$~TeV & $m_{\tilde{q}} = 1.5$~TeV & $\mu_R = \mu_F = m_{\tilde{q}}$\\
$m_{O_p} = 5$~TeV &  $m_t = 172$ GeV\\
$\sqrt{\hat s}=6~\text{TeV}$ & $t=-22208172$~GeV${}^2$& $\alpha_s=0.1184$\\
\bottomrule
\end{tabular}
\end{center}
The mass $m_{O_s}$ is fixed by eq.~\eqref{eq:octetmasses}.
All matrix elements are compared in the t'Hooft--Veltmann scheme (HV). The LO
matrix elements |$\mathcal{M}(\text{LO})|^2$ are given in units of GeV${}^{-2}$. 
The NLO matrix elements $2\mathrm{Re}\left[\mathcal{M}(\text{NLO})\mathcal{M}^\ast(\text{LO})\right]$ are given as
\begin{align}
\frac{2\pi}{\alpha_s}\frac{2\mathrm{Re}\left[\mathcal{M}(\text{NLO})\mathcal{M}^\ast(\text{LO})\right]}{|\mathcal{M}(\text{LO})|^2}.
\end{align}
Real parts contain soft-collinear mass factorisation counterterms. The agreement of the finite parts are not as precise as the one of the poles. This is due to a limited precision of our \texttt{LoopTools} installation. If however double precision is replaced with quadruple precision the agreement is improved.

\subsubsection*{Squark squark production}
\myparagraph{$u  u\rightarrow\tilde u_L \tilde u_R (+g)$}
\begin{tabular}{S S[table-format=2.13] S[table-format=2.13] S[table-format=2.13]}
\toprule
 & {Virtual part -- Method~1} & {Virtual part -- Method~2} & {Real part -- TCPSS} \\ 
\midrule
{tree level}  & 1.1187184131205 & 1.1187184131205 & \\
{double pole} & -2.6666666666666 & -2.6666666666666 & 2.6666666666666 \\
{single pole} & 6.3424494456738 & 6.3424494456738 & -6.3424494456738 \\ 
{finite}      & 36.720472180005 & 36.720472181755 & \\
\bottomrule
\end{tabular}
% DO NOT JUST REMOVE - this is not needed for this publication but IS IMPORTRANT
%\myparagraph{$u  d\rightarrow\tilde u_L \tilde d_R$}
%\begin{tabular}{cccc}
%%\toprule
% & Virtual part -- own & Virtual part -- GoSam & Real part -- phase space slicing \\ 
%%\midrule
%tree level  & 0.18384647031257 & 0.1838464703125703 & \\
%double pole & -2.66666666666667 & $-2.666666666666667$ & 2.666666666666676\\
%single pole & 2.99539839823747 & 2.995398398237460 & $-2.995398398237452$\\ 
%finite      & 40.1995156796408 & 40.19951568021342 & \\
%%\bottomrule
%\end{tabular}
\subsubsection*{Squark anti-squark production}
\myparagraph{$gg\rightarrow\tilde u_L \tilde u _L^* (+g)$}
\begin{tabular}{S S[table-format=2.14] S[table-format=2.14] S[table-format=2.14]}
\toprule
 & {Virtual part -- Method~1} & {Virtual part -- Method~2} & {Real part -- TCPSS} \\ 
\midrule
{tree level}  & 0.11114746957753 & 0.11114746957753& \\
{double pole} & -5.9999999999999 & -6.0000000000000 & 6.0000000000000\\
{single pole} & 8.4161500386713 & 8.4161500386713& -8.4161500386713\\ 
{finite}      & -3.7965233237193 & -3.7965105147158 & \\
\bottomrule
\end{tabular}
\myparagraph{$u \bar u\rightarrow\tilde u_L \tilde u_L^* (+g)$}
\begin{tabular}{S S[table-format=2.14] S[table-format=2.14] S[table-format=2.14]}
\toprule
 & {Virtual part -- Method~1} & {Virtual part -- Method~2} & {Real part -- TCPSS} \\ 
\midrule
{tree level}  & 0.41567383692610 & 0.41567383692610 & \\
{double pole} & -2.6666666666666 & -2.6666666666666 & 2.6666666666666\\
{single pole} & 4.8560767580444& 4.8560767580444& -4.8560767580444\\ 
{finite}      & 18.078757030780 & 18.078757100653& \\
\bottomrule
\end{tabular}
\myparagraph{$d \bar d\rightarrow\tilde u_L \tilde u_L^* (+g)$}
\begin{tabular}{S S[table-format=2.14] S[table-format=2.14] S[table-format=2.14]}
\toprule
 & {Virtual part -- Method~1} & {Virtual part -- Method~2} & {Real part -- TCPSS} \\ 
\midrule
{tree level}  & 0.15281365356525 & 0.15281365356525& \\
{double pole} & -2.6666666666666 & -2.6666666666666 & 2.6666666666666\\
{single pole} & 4.8369561733416 & 4.8369561733417 & -4.8369561733416\\ 
{finite}      & -4.7136022244356 & -4.7136020740789 & \\
\bottomrule
\end{tabular}

\bibliographystyle{tex/JHEP}
\bibliography{bibliography}

\end{document}